%% file: main.tex
\documentclass[aps,prd,reprint,superscriptaddress,nofootinbib]{revtex4-2}
\usepackage[english]{babel}
\usepackage[utf8]{inputenc}
\usepackage[colorinlistoftodos, color=green!40, prependcaption]{todonotes}

\input{preamble}

\usepackage[pdftex, pdftitle={Article}, pdfauthor={Author}]{hyperref} 
\usepackage{graphicx}
\usepackage{color}
\usepackage{natbib}
\usepackage{float}
\usepackage{xcolor}
\usepackage{soul}
\usepackage{comment}
\usepackage{amsmath,amssymb,amsfonts}
\usepackage{subfigure}

\begin{document}

\title{The impact of hyperons on neutron star mergers: gravitational waves, mass ejection and black hole formation}

\author{Hristijan Kochankovski}
\affiliation{Departament de F\'{\i}sica Qu\`antica i Astrof\'{\i}sica and Institut de Ci\`encies del Cosmos, Universitat de Barcelona, Mart\'i i Franqu\`es 1, 08028, Barcelona, Spain}

\author{Georgios Lioutas}
\affiliation{GSI Helmholtzzentrum f\"ur Schwerionenforschung, Planckstra{\ss}e 1, 64291 Darmstadt, Germany}
\affiliation{Heidelberger Institut für Theoretische Studien (HITS), Schloss-Wolfsbrunnenweg 35, 69118 Heidelberg, Germany}

\author{Sebastian Blacker}
\affiliation{Institut f\"ur Kernphysik, Technische Universit\"at Darmstadt, 64289 Darmstadt, Germany}
\affiliation{GSI Helmholtzzentrum f\"ur Schwerionenforschung, Planckstra{\ss}e 1, 64291 Darmstadt, Germany}

\author{Andreas Bauswein}
\affiliation{GSI Helmholtzzentrum f\"ur Schwerionenforschung, Planckstra{\ss}e 1, 64291 Darmstadt, Germany}
\affiliation{Helmholtz Research Academy Hesse for FAIR (HFHF), Campus Darmstadt, 64291 Darmstadt, Germany}

\author{Angels Ramos}
\affiliation{Departament de F\'{\i}sica Qu\`antica i Astrof\'{\i}sica and Institut de Ci\`encies del Cosmos, Universitat de Barcelona, Mart\'i i Franqu\`es 1, 08028, Barcelona, Spain}

\author{Laura Tolos}
\affiliation{Institute of Space Sciences (ICE, CSIC), Campus UAB, Carrer de Can Magrans, 08193 Barcelona, Spain}
\affiliation{Institut d'Estudis Espacials de Catalunya (IEEC), 08860 Castelldefels (Barcelona), Spain}

\date{\today} 

\begin{abstract}

We study the influence of hyperons in binary neutron star (NS) mergers considering a total of 14 temperature dependent equations of state (EoSs) models which include hyperonic degrees of freedom and partly delta resonances. Thermally produced hyperons induce a higher heat capacity and a lower thermal index, i.e.~a reduced thermal pressure for a given amount of thermal energy, compared to purely nucleonic models. We run a large set of relativistic hydrodynamics simulations of NS mergers to explore the impact on observables of these events. In symmetric binaries, we describe a characteristic increase of the dominant postmerger gravitational-wave (GW) frequency by a few per cent, which is specifically linked to the occurrence of hyperons and can thus be potentially used as a discriminator between purely nucleonic and hyperonic systems. We corroborate that this effect occurs similarly for asymmetric binaries and becomes more prominent with increasing total binary mass. Hyperonic models tend to stick out in relations between the dominant postmerger GW frequency and the tidal deformability of massive stars providing a signature to identify the presence of hyperons. Distinct secondary postmerger GW spectral features are differently affected by the presence of hyperons, in the sense that one feature exhibits a characteristic frequency shift due to the specific thermal properties of hyperonic EoSs while the other does not. The average temperature of the remnant is reduced for hyperonic models. The dynamical mass ejection of mergers is tentatively enhanced for hyperonic models in comparison to nucleonic EoSs which yield roughly the same stellar properties of cold NSs. This may serve useful to identify exotic degrees of freedom in these systems by kilonova observations. Also, for hyperonic EoSs the threshold mass for prompt black hole formation is reduced by about 0.05~$M_\odot$ in comparison to nucleonic systems with the same stellar parameters of cold NSs.

\end{abstract}


\maketitle

\input{sections/intro.tex} 
\input{sections/section02.tex}
\input{sections/section03.tex}

\input{sections/section04.tex}
\input{sections/acknowledgements.tex}

\bibliography{main}

\clearpage

\appendix
\input{sections/appendix1.tex}

\end{document}

%% file: preamble.tex
\usepackage{amsthm}
\usepackage{mathtools}
\usepackage{physics}
\usepackage{xcolor}
\usepackage[left=23mm,right=13mm,top=35mm,columnsep=15pt]{geometry} 
\usepackage{adjustbox}
\usepackage{placeins}
\usepackage[T1]{fontenc}
\usepackage{lipsum}
\usepackage{csquotes}

\newcommand{\apjs}{Astrophys. J. Supp.}%
\newcommand{\apjl}{Astrophys. J. Lett.}%
\newcommand{\aap}{Astron. Astrophys.}

%% file: sections/intro.tex
\section{Introduction} \label{sec:introduction}

The composition of dense matter, such as the one found in neutron stars (NSs), has been an open question for several decades. It has been speculated that different forms of exotic matter can occur with hyperons, baryons containing strange quarks, being among the most probable candidates. Their presence can have a sizable impact on the equation of state (EoS) and, therefore, on certain astrophysical observables as, for instance, the mass-radius relation of NSs which is uniquely determined by the EoS.

The role of hyperons in cold isolated NSs in $\beta$ equilibrium has been extensively studied in the literature \cite{1960SvA.....4..187A, Glendenning:1982nc, 1985ApJ...293..470G,
Glendenning:1987wb, Weber:1989qf, Glendenning:1991es,Knrren:1995rv,SCHULZE199521,Schaffner:1995th, Balberg_1997, PhysRevC.57.704,  PhysRevC.58.3688,1999ApJS..121..515B, PhysRevC.61.055801,PhysRevC.61.025802,
PhysRevC.62.035801,PhysRevC.73.058801, PhysRevC.79.034301, PhysRevC.81.035803, PhysRevC.84.035801,Bonanno:2011ch,
Weissenborn_2012, Bednarek:2012zs, Maslov:2015gxa, Tolos:2017kfa, Tolos:2017lgv,  Fortin:2017zck,PhysRevC.87.055806, Dalen:2014mda, PhysRevC.89.014314,   Drago:2016srv,
Drago:2016kpi,
Fortin:2018aa, Li:2018kyx, Li:2018qap,  Providencia:2019pny, Sedrakian:2020qja, Thapa:2021ysx,Motta:2022nlj, Leong:2023yma}.
The later works have also addressed the so-called hyperon puzzle, i.e.~the tension between the tendency of hyperons to soften the EoS and the observation of NSs with mass $M \gtrsim 2M_{\odot}$ \cite{Demorest2010ShapiroStar,Antoniadis:2013pzd,Fonseca2016,Cromartie2020RelativisticPulsar,Romani:2022jhd}, which requires matter at densities above nuclear saturation to be rather stiff.

Several ideas have been put forward to provide a solution to this issue. Depending on the model, the required stiffness is usually generated by introducing three-body forces in microscopical approaches  \cite{Haidenbauer:2016vfq,Logoteta:2019utx,Gerstung:2020ktv,Takatsuka2002,Takatsuka:2004ch,Vidana:2010ip,Yamamoto:2013ada,Yamamoto:2014jga,PhysRevLett.114.092301, tong2024ab}, or by considering an additional strange vector hidden meson in mean-field frameworks \cite{Burgio:2021vgk, Sedrakian2022,  Sedrakian:2022ata,  Bednarek:2012zs, Weissenborn_2012, Maslov:2015gxa, Tolos:2017kfa,Tolos:2017lgv, Fortin:2017zck,  Ribes:2019kno, Muto:2021jms, Thapa:2021kfo}. In some of these works \cite{Gerstung:2020ktv, PhysRevLett.114.092301}, the additional stiffness even prevents hyperons to appear in nuclear matter, predicting that strangeness can not be produced in a cold dense medium.

Identifying the presence of hyperons from observations of stellar parameters like the NS radius is not straightforward. The mass-radius relations of NSs with hyperon content can be very similar to those of purely nucleonic models. This is equivalent to the ``masquerade problem'', which was raised in the context of the hadron-quark phase transition~\cite{Alford:2004pf}. Obviously, for a hyperonic and a purely nucleonic EoS to look alike, the underlying nucleonic interactions must be, respectively, different such that the final EoSs match even if one of them contains hyperons additionally. As long as the interactions and the EoS of purely nucleonic matter are not fully settled, it may be difficult to understand if hyperons occur in NSs.

At finite temperature hyperons can be expected to be more abundant because thermal energy favors the conversion of nucleons to hyperons.
In recent years, the role of hyperons has been extensively studied in matter at finite temperature and out of $\beta$ equilibrium as those conditions can be met in violent astrophysical events (see for example Refs.~\cite{Oertel:2016xsn,Fortin:2017dsj, lenka2019properties, PhysRevD.87.043006, Raduta:2021coc,Raduta:2022elz, Sedrakian_2021, Malik:2021nas, Sedrakian_2022, Guichon:2023iev, Stone2021,PhysRevD.107.103054,tsiopelas2024finitetemperature} or some general reviews \cite{Chatterjee:2015pua,Tolos:2020aln,Schaffner-Bielich:2020psc,Burgio:2021vgk,Sedrakian2022,Sedrakian:2022ata,Logoteta:2021iuy}).  

In particular, the first unambiguous detection of gravitational waves and electromagnetic radiation from a NS merger in 2017 has highlighted the prospect to learn about the conditions in hot and dense matter from the observations of these events~\cite{LIGOScientific:2017vwq,Abbott2017em,Abbott2018GW170817:State,Abbott2019}. At the latest, the new generation of gravitational-wave detectors like the Einstein Telescope and Cosmic Explorer will provide the sensitivity to measure the postmerger gravitational wave emission, which is produced by the hot, massive rotating merger remnant~\cite{Punturo:2010zz,LIGOScientific:2016wof,Torres-Rivas:2018svp}. Since densities and temperatures increase during the merging process, this phase is particularly sensitive to the physics at high densities and finite temperature and hence may potentially reveal the presence of hyperons.

Early works on hyperons in binary neutron star (BNS) mergers have studied a very limited set of EoS models of hyperonic matter~\cite{Sekiguchi:2011mc,Radice:2016rys}. By investigating individual models, it is, however, difficult to identify observable features which can be unambiguously linked to the occurrence of hyperons. This is related to the above mentioned observation that the mass-radius relations of cold NSs of hyperonic EoS models can look very similar to those of purely nucleonic matter. Specifically, it remains unclear whether a hyperonic model yielding a mass-radius relation which closely resembles that of purely nucleonic EoS, would yield a different, hence discriminating, signature.

In Ref.~\cite{Blacker:2023opp} we have considered a large sample of hyperonic EoSs and provided a systematic and detailed comparison to purely nucleonic matter.
In particular, we pointed out that the thermal behavior of the EoS can serve as a potential discriminator between hyperonic and purely nucleonic matter in NS merger remnants. Specifically, the presence of hyperons is connected to a sizable reduction of the thermal pressure for a given amount of thermal energy. This leads to a small but systematic shift of the dominant GW oscillation frequency of the postmerger remnant. In principle, one can access the thermal regime of the EoS if its cold part is sufficiently well known through an independent observable. For a given cold EoS, one can predict the postmerger GW frequency which should result for purely nucleonic matter. Compared to this reference value, we found that a frequency shift of up to 150~Hz could be indicative of hyperon production. In practice, this signature alone remains challenging to be detected. However, it may still represent an interesting approach considering the general difficulties to identify the presence of hyperons. Also, it may be combined with additional information on the hyperon content of dense matter as the cold EoS is also modified by the presence of hypronic degrees of freedom in comparison to a fiducial EoS of purely nucleonic matter.

Clearly, the total GW postmerger frequency shift or any other effect on merger observables that is introduced by the occurrence of hyperons relative to a fiducial purely nucleonic system with identical nucleonic interactions will be a result of not only the modifications of the thermal behavior but also of the cold regime of the EoS. In other words, the impact of hyperons is in principle more significant than the frequency shift introduced by the reduction of the thermal pressure component alone. However, we focus on the thermal component because in reality a fiducial nucleonic reference system does not exist. Only one cold EoS is realized and thus measurable in nature, which is either hyperonic or purely nucleonic and as argued above may not reveal unambiguous information on the hyperon content. The issue is obviously more general and other exotic degrees of freedom as for instance deconfined quark matter may play a role as well. 

In this follow-up work we extend and deepen our study of the impact of hyperons on NS mergers and their observables. We include more hyperonic EoS models and run simulations for other binary masses confirming the occurrence of a characteristic shift in the dominant postmerger GW frequency due to the specific thermal properties of hyperonic matter. We provide a more thorough analysis of the thermal properties of hyperonic EoS models and the remnant evolution. Furthermore, we investigate the impact of hyperons on other merger observables including  secondary GW spectral features of the postmerger phase, mass ejection and the threshold binary mass for prompt black-hole formation.

The paper is organized as follows. In Section \ref{sec:EoS}, we briefly discuss the different hyperonic models that we use and we present their main features. We show the results obtained from all our equal-mass $1.4M_{\odot}-1.4M_{\odot}$ BNS merger simulations in Section \ref{sec:Simulations}. In Sections \ref{sec:moremassice} and \ref{sec:asymmetric} we extend our study to more massive and asymmetric binaries. In Section \ref{sec:threshold} the influence of hyperons on the threshold masses for prompt black-hole formation is studied. We summarize our findings and conclude in Section~\ref{sec:conclusions}. In this paper, the total binary mass $M_\mathrm{tot}=M_1+M_2$ refers to the sum of gravitational mass at infinite orbital separation. The binary mass ratio is defined as $q=M_1/M_2$ such that $q\leq1$.

%% file: sections/section02.tex
\section{Thermal behavior of the hyperonic Equation of State} \label{sec:EoS}

In this section we give a brief overview of the different hyperonic EoSs \footnote{ Note that we refer to hyperonic EoSs when considering  nucleons and hyperons and, in certain cases, also $\Delta$ baryons. In the manuscript, we do not consistently employ the term ``purely nucleonic'' but will sometimes simply use ``nucleonic'' when we refer to systems which contain only nucleons but no heavier baryons. }  that we will use in our simulations, emphasizing the features of their behavior at finite temperature. The properties of the nucleonic models used for comparison have been discussed in length in other works and will not be presented here (see for example \cite{LATTIMER2007,LATTIMER2010101, Lattimer2012, Ozel2016,LATTIMER2016,RevModPhys.89.015007,Raduta2021}). However, in the subsequent discussion we will often point out the systematic differences in the results obtained with the nucleonic and the hyperonic models.

A large number of hyperonic EoSs at finite temperature  have been developed  within energy density functional theories (DFTs), assuming weak interaction equilibrium between the hadrons \cite{Chatterjee:2015pua}. DFTs are fitted to properly reproduce the nuclear matter properties and finite nuclei data, thus being appropriate for densities around saturation density ($\rho_0\simeq 2.7\cdot 10^{17}\mathrm{g/cm^3}$) and for systems with relatively low isospin asymmetry \cite{Burgio:2020fom}. However, with a careful extrapolation they are used for studying denser systems at higher asymmetries at finite temperature.  

In the present work we consider models that belong to the broad category of covariant relativistic mean-field density-functional models  expressed in terms of a Lagrangian density. The complicated many-body hadron interactions are modeled with different meson-exchange channels  at the mean-field level. In particular, the scalar-isoscalar ($\sigma$), the vector-isoscalar ($\omega$), and the vector-isovector ($\rho$) channels are taken into account. In addition, there are also scalar-isoscalar ($\sigma^{*}$) and vector-isoscalar ($\phi$) channels that are coupled only to the particles with non-zero strangeness. Some of the models - SFHOY \cite{Fortin:2017dsj}, and the FSU2H$^*$ family \cite{Kochankovski2022,Kochankovski:2023trc} - assume a
non-linear dependence on the mean fields; while others - DD2Y \cite{Marques2017}, BHB$\Lambda \phi$ \cite{Banik2014}, DDLSY family \cite{tsiopelas2024finitetemperature} and DD2Y$\Delta$ family \cite{Raduta:2022elz} -  consider density
dependent couplings. Moreover, the FSU2H* group covers the uncertainty of the hyperon-nucleon and hyperon-hyperon interactions, while the DDLSY family explores the uncertainty with respect to the slope of the symmetry energy. The DD2Y$\Delta$ family also includes $\Delta$ resonances in addition to the usual baryon octet members. They are built under the assumption that $\Delta$ resonances have negligible widths and that their vacuum masses are not changed due to the interactions. As we will see, the appearance of $\Delta$ resonances has an analogous effect on the EoS as that produced by hyperons and, hence, we also include the DD2Y$\Delta$ family in this work.

As a subgroup of the covariant models, the SU(3) Chiral Mean Field (CMF) is also considered, on which the DNS model \cite{Dexheimer2017} is based. Using the spontaneous chiral symmetry breaking, the masses of the baryon octet are generated via the coupling to the scalar meson field. At high temperatures/densities the chiral symmetry is restored, and the effective masses of the baryons are significantly lowered.

Finally, we consider the Quark-Meson-Coupling model QMC-A \cite{Stone2021}. This energy density functional scheme differs from the other covariant models because the quark structure of the hadrons is crucial for the hadron dynamics. It is the coupling of the valence quarks to the meson fields which gives rise to the hadronic interactions.

We are interested in the specific features that make these hyperonic schemes (or, more generally, schemes with heavy baryons) distinguishable from models that consider only nucleonic matter. Later, we explore how those differences affect the observables and thermodynamics conditions of BNS mergers.

\begin{figure}
\centering \includegraphics[width=0.49 \textwidth] {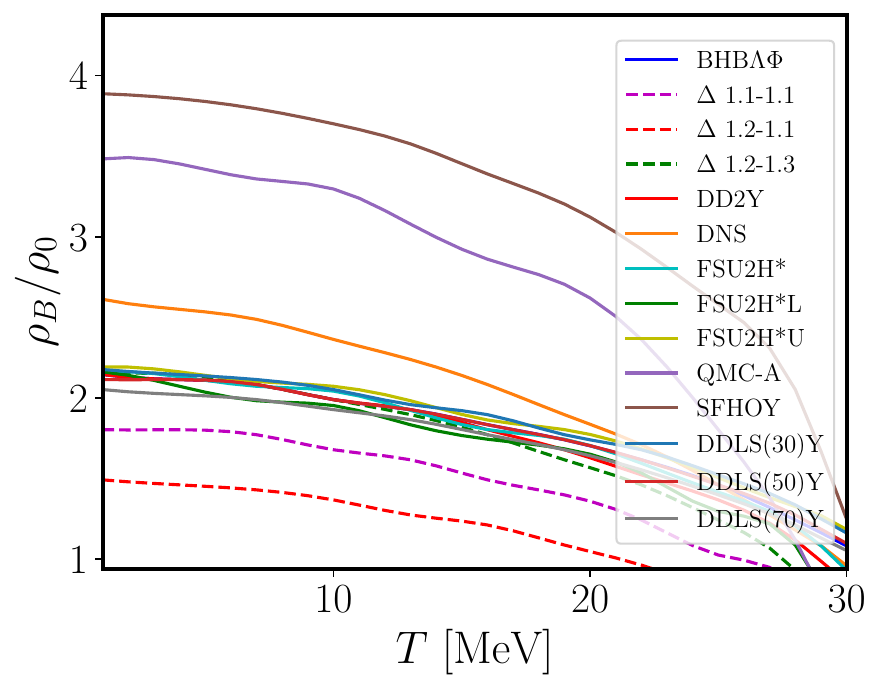}
\caption{Onset density of hyperons (and/or $\Delta$ baryons) as a function of temperature for different models considered in this work. The onset density is defined as the point at which the heavy baryon abundance in matter is larger than $1\%$.} 
\label{fig:1}
\end{figure}

In order to estimate under which conditions hyperons or $\Delta$ resonances should start playing a significant role in matter at finite temperature, in Fig.~\ref{fig:1} we show, for all models, their onset density as a function of temperature in the case of matter in $\beta$-equilibrium. Given that an amount of hyperons and $\Delta$ baryons will always be present at finite temperature, we define the onset density as the value at which the hyperon or $\Delta$ abundance in matter starts to be larger than $1\%$. 

We find that most of the models predict that hyperons appear at around $2\rho_0$ at low temperatures. 
However, some of the models, like SFHOY and QMC-A, predict the appearance of hyperons at densities larger than $3\rho_0$. On the contrary, for models that include $\Delta$ resonances, the onset density can be significantly lowered to around $1.5\rho_0$, where the fraction of negatively charged $\Delta^{-}$ resonance starts to be important. 

With increasing temperature, due to thermal excitations, the onset density is pushed towards lower values. We observe that for all models considered  the onset density is lower than $1.5\rho_0$ at a moderate temperature of $T=30$ MeV.

As we motivated in the introduction, it is important to investigate the thermal contribution to the pressure and the energy density. The thermal contribution to a quantity $X$ is defined as the difference between the quantity at some finite temperature $T$, density $\rho_B$ and charge fraction $Y_Q$ and the quantity at the same density and charge fraction but at zero temperature: $X_{\mathrm{th}} = X(T,\rho_B,Y_Q) - X(0,\rho_B,Y_Q)$.

\begin{figure*}[htbp]
    \centering
    %\subfigure[$\epsilon_{\mathrm{th}}$]
    \subfigure{
        \includegraphics[height = 2.5 in,  width=0.45\textwidth]{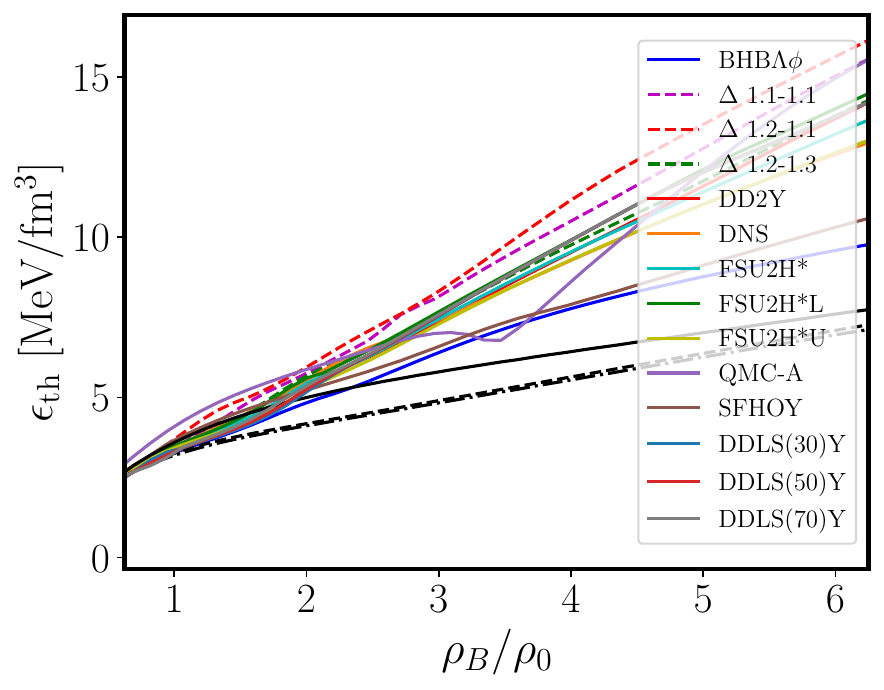}
    }
    \hspace{0.05\textwidth}
    %\subfigure[$P_{\mathrm{th}}$]
    \subfigure{
        \includegraphics[height = 2.5 in,  width=0.45\textwidth]{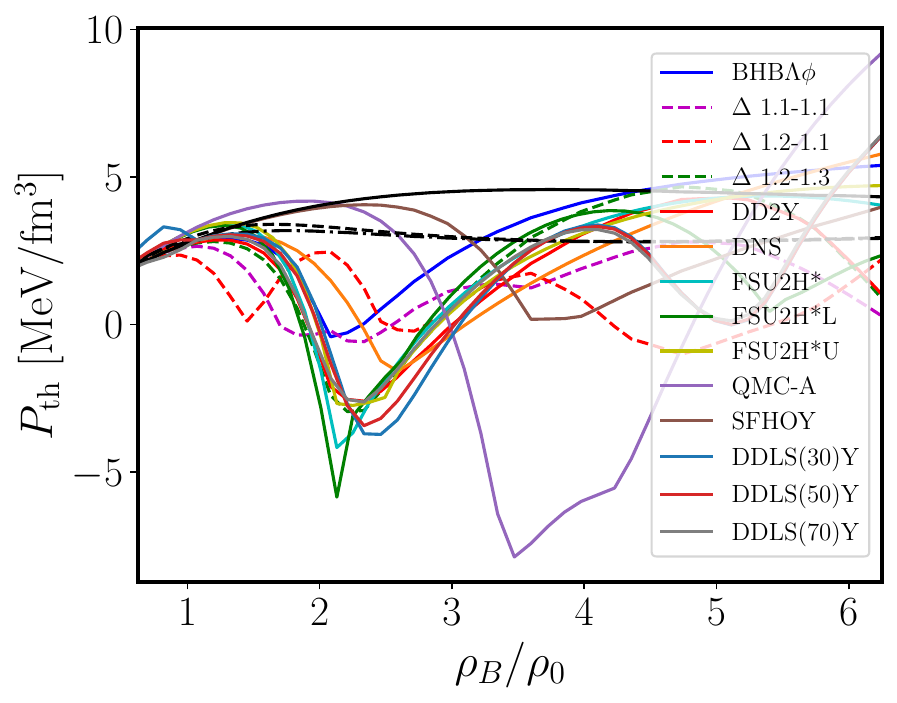}
    }
    \caption{ Thermal energy density $\epsilon_{\mathrm{th}}$ and thermal pressure $P_{\mathrm{th}}$ as functions of density at fixed charge fraction $Y_Q = 0.1$ and temperature $T = 25$ MeV. For comparison, with black curves, three different nucleonic models (solid line for SFHO, dashed line for FSU2R, dash-dotted line for DD2) are also shown.}
    \label{fig:2}
\end{figure*}

To properly compare the thermal features of the hyperonic models with respect to the nucleonic ones in the various figures, we will show results for three exemplary nucleonic models with black lines (SFHO \cite{Hempel2010a,Steiner2013} - solid line, DD2 \cite{Typel:2009sy,Hempel2010a} - dashed line, FSU2R \cite{Tolos:2017lgv} - dashed-dotted line). The choice of these three models is intentional as they are also developed in a relativistic mean-field framework and they do share many characteristics with some of the hyperonic models (DD2 with BHB$\Lambda \phi$, DD2Y and the DD2Y$\Delta$ family; SFHO with SFHOY; and FSU2R with the FSU2H* family). By comparing these sets of models we can confidently claim that the effects observed are a consequence of the appearance of hyperons (and/or $\Delta$ baryons) and are not related to the particularities of the framework to derive the EoS. 

\subsection{Density dependence of thermodynamical quantities at finite temperature}

In Fig.~\ref{fig:2} we show the density dependence of both thermal energy density\footnote{Note that in some references the symbol $\epsilon$ is used to label the specific internal energy, while in this work it refers to the energy per volume.}$\epsilon_{\mathrm{th}}$ (left panel) and thermal pressure $P_{\mathrm{th}}$ (right panel) of matter at fixed temperature $T=25$ MeV and charge fraction $Y_Q = 0.1$. As we will see later, those conditions are fairly relevant in the dynamics of BNS mergers. We find that, except for QMC-A, all other models predict a monotonically increasing behavior of $\epsilon_{\mathrm{th}}$ versus density. It is noticeable that models with more species of heavy baryons show systematically higher values of the thermal energy density. A more dramatic effect  can be seen in the behavior of the thermal pressure. While nucleonic models show a featureless behavior, all  models with heavy baryons exhibit a significant reduction of the thermal pressure, which acquires a dip at the corresponding onset density shown in  Fig.~\ref{fig:1}. Some of the models, like DD2Y$\Delta1.1-1.1$, DD2Y$\Delta1.2-1.1$ or those of the DDLSY family, for which different species appear at fairly different densities, show a few sequential dips. This peculiar drop has been discussed in some recent works \cite{Raduta:2022elz,Kochankovski2022} and has been connected with the reduction of the degeneracy pressure tied to the appearance of hyperons (or other baryons) in matter. We recall that hadrons in matter are in weak equilibrium. As a consequence, the composition of matter for a fixed charge fraction $Y_Q$ can be different at different temperatures, when heavy baryons are present. Indeed, in Figure \ref{fig:1} we already found that at large temperatures heavy baryons appear at substantially lower densities than at $T=0$.

\begin{figure*}[htbp]
    \centering
    %\subfigure[$\Gamma_{\mathrm{th}}$]
    \subfigure{
        \includegraphics[height = 2.5 in,  width=0.45\textwidth]{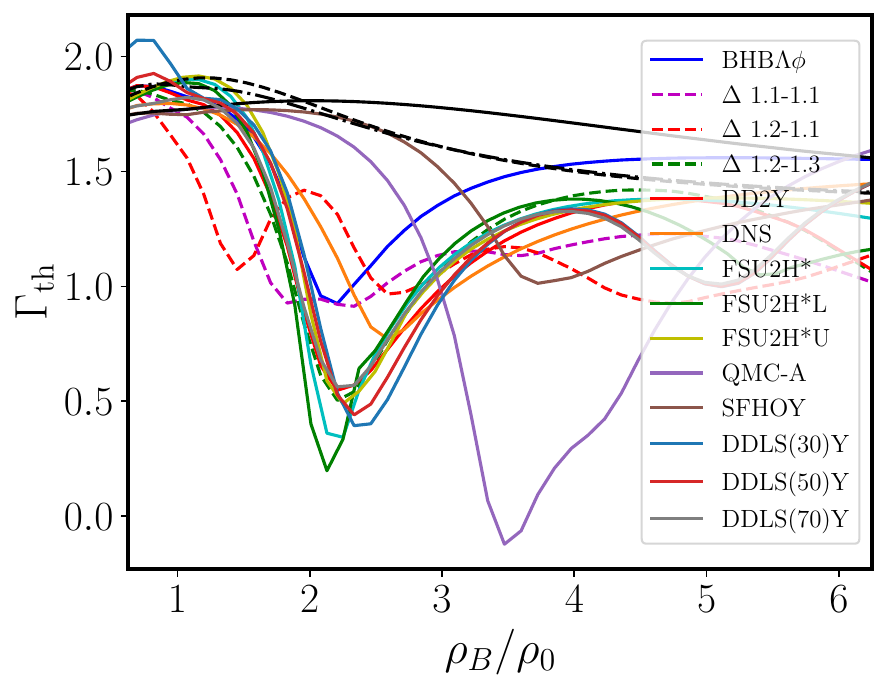}
    }
    \hspace{0.05\textwidth}
    %\subfigure[$\Delta \rho_Y $]
    \subfigure{
        \includegraphics[height = 2.5 in,  width=0.45\textwidth]{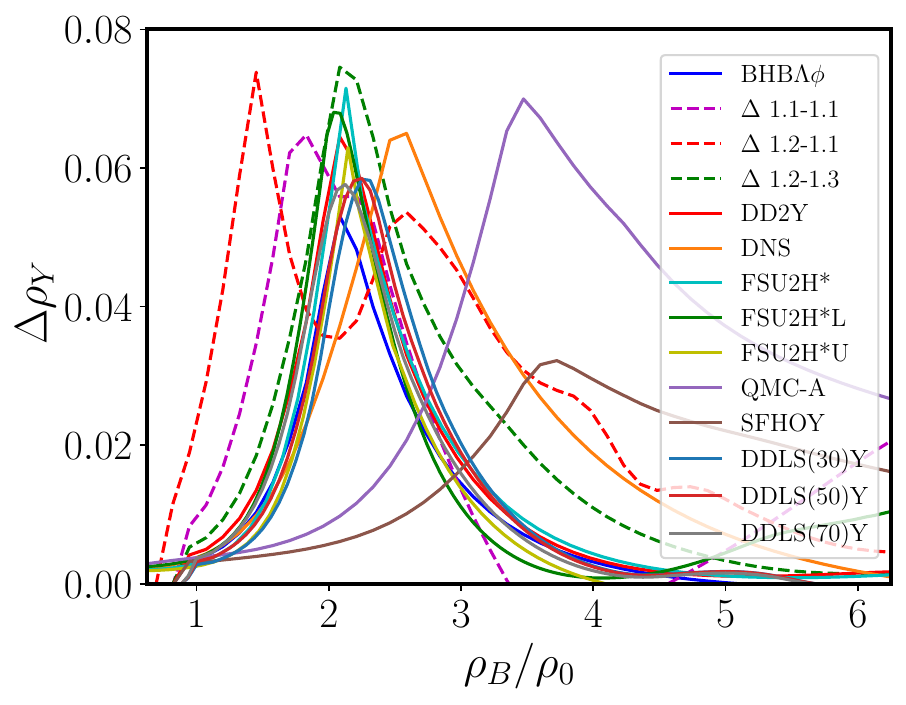}
    }
    \caption{$\Gamma_{\mathrm{th}}$ (left plot) and $\Delta \rho_Y$ (right plot) as functions of density at fixed charge fraction $Y_Q = 0.1$ and temperature $T = 25$ MeV. In the left panel, for comparison with black curves, three different purely nucleonic models (solid line for SFHO, dashed line for FSU2R, dash-dotted line for DD2) are also shown.}
    \label{fig:3}
\end{figure*}

To quantify and analyze the different thermal behavior of the thermal energy density and thermal pressure, we consider the  thermal index: $\Gamma_\mathrm{th} (\rho_B,Y_Q,T) = 1+ P_{\mathrm{th}}/\epsilon_{\mathrm{th}}$. While the thermal index of nucleonic models shows a mild dependence on the density, the behavior of the thermal index for hyperonic or $\Delta$-baryon models is characteristically different (see Fig.~\ref{fig:3}). The strong reduction of the thermal pressure leads to a significant drop of the thermal index. Note that, for most of the models, the thermal index is lower than 1.5 for a wide range of densities. In order to reinforce our conclusion that this effect is achieved by the loss of degeneracy pressure, in the same figure we display the normalized hyperon excess $\Delta \rho_Y$\footnote{Note that the hyperon excess refers to the presence of thermal hyperons and, in certain cases, also thermal $\Delta$ baryons.} as a function of density. This quantity is defined as the difference between the hyperon abundance at a given temperature $T$ and the hyperon abundance at $T=0$ for the same density and charge fraction, normalized to the total baryon density:
\begin{equation}
\Delta \rho_Y = \frac{\sum_{Y_i} (\rho_{Y_i}(T,\rho_B,Y_Q) - \rho_{Y_i}(0,\rho_B,Y_Q))}{\rho_B} ,
\end{equation}
where  $\rho_{Y_i}$ indicates the density of an arbitrary heavy baryon $Y_i$.

The different dips of the thermal index for all hyperonic models occur at densities where $\Delta \rho_Y$ shows a peak. The excess hyperons that are created with respect to $T=0$ come at the cost of highly degenerate nucleons, hence a reduction of the thermal pressure and, subsequently, a drop of the thermal index is produced. 
This effect is not significant for the thermal energy density because the loss of degeneracy is replaced by a gain in the rest mass energy as hyperons and/or $\Delta$-s are more massive than nucleons. When the hyperon excess starts to decrease, the thermal index increases again getting closer to the nucleonic one. Generally, the higher the peak in $\Delta \rho_Y$ is, the deeper the drop in the thermal index becomes.  However,  the shape of $\Delta \rho_Y$ as function of density depends on the particular model. For instance, some models like DD2Y and FSU2H$^*$ do show a sharp peak, meaning that they tend to loose the excess amount of thermal hyperons rather quickly with increasing density and, as a consequence, their thermal index increases sharply. On the contrary, models like QMC-A and SFHOY present wider peaks, so they have lower values of $\Gamma_{\mathrm{th}}$ for a wider range of densities.
It is also interesting to note the behavior of the DD2Y$\Delta1.1-1.1$ model, for which different baryonic exotic species are relevant at fairly different densities. This is the reason why one can observe three peaks in $\Delta \rho_Y$. However, although the largest peak appears at a relatively low density, the thermal index is reduced more strongly at the subsequent higher densities. This is related to the fact that the pressure of any degenerate species scales with the particle density as $\rho^{5/3}$, explaining why the reduction of the degeneracy pressure is more important at higher densities.

\begin{figure}[h]
\centering \includegraphics[height = 2.2 in ]{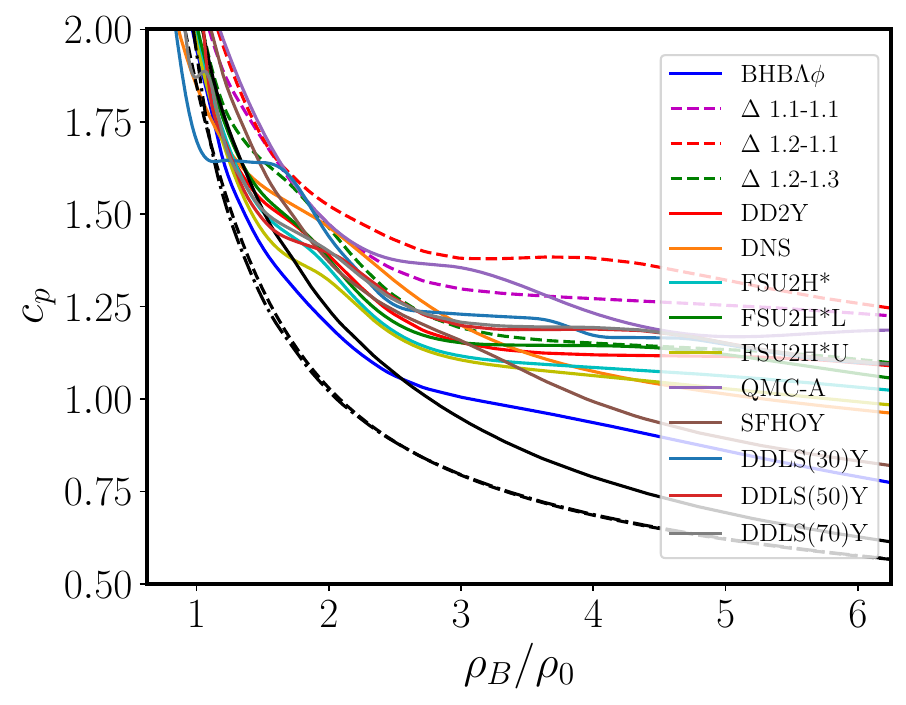}
\caption{Specific heat at constant pressure, $c_{p}$, as function of density at fixed charge fraction $Y_Q = 0.1$ and temperature $T = 25$ MeV. For comparison, with black curves, three different purely nucleonic models (solid line for SFHO, dashed line for FSU2R, dash-dotted line for DD2) are also shown.}
\label{fig:4}
\end{figure}

Finally, in Fig.~\ref{fig:4} we also show the density dependence of the specific heat at constant pressure $c_p$ for  $T=25$ MeV and $Y_Q = 0.1$.
The specific heat is defined via the temperature derivative of the entropy per particle $S/A$ 

\begin{equation}
    c_p = T\frac{d (S/A)}{dT} \big |_{P,A,Y_{q}},
\end{equation}
From the figure it is seen that at higher densities hyperonic EoSs yield substantially higher values of $c_p$ with respect to nucleonic ones. We recall that this is a direct reflection of the increase of the entropy per baryon with increasing temperature that hyperonic models can achieve compared to nucleonic ones, as more species are produced and hence less degeneracy is present \cite{Drago:2016kpi,Raduta:2022elz, Kochankovski2022}.  

%% file: sections/section03.tex
\section{BNS merger simulations} \label{sec:Simulations}
\subsection{Approach and models}
In this section, we describe BNS simulations for the hyperonic EoSs presented in the previous section, which are compared to calculations with purely nucleonic EoSs. We provide a deeper analysis of simulations for 1.4-1.4~$M_\odot$ binaries than in Ref.~\cite{Blacker:2023opp}, most of which have already been presented in that reference. The calculations are performed with a relativistic smooth particle hydrodynamics (SPH) code, which adopts the conformal flatness condition to solve the Einstein field equations~\cite{1980grg1.conf...23I,1996PhRvD..54.1317W} (see~\cite{Blacker:2023opp} for more details and~\cite{Oechslin:2001km,2007A&A...467..395O,Bauswein:2009im} for further information on the code). We use the same code to run additional simulations for other binary masses, the results of which are presented in the next sections. For all calculations we set up the stars without intrinsic spin and assume neutrino-less beta-equilibrium at zero temperature. Neutrinos and magnetic fields are not included.

The EoS set of purely nucleonic models includes APR~\cite{Akmal:1998cf,Schneider:2019vdm}, DD2~\cite{Typel:2009sy,Hempel2010a}, DD2F~\cite{Typel:2009sy,Alvarez-Castillo:2016oln},
DDLS(30)-N~\cite{tsiopelas2024finitetemperature},
DDLS(50)-N~\cite{tsiopelas2024finitetemperature},
DDLS(70)-N~\cite{tsiopelas2024finitetemperature},
FSU2R~\cite{Tolos:2017lgv}, FTNS~\cite{Furusawa:2017auz,Togashi:2017mjp}, GS2~\cite{Shen:2011kr}, LPB~\cite{Bombaci:2018ksa,Logoteta:2020yxf}, LS220~\cite{Lattimer:1991nc}, LS375~\cite{Lattimer:1991nc}, SFHo~\cite{Hempel2010a,Steiner2013}, SFHx~\cite{Hempel2010a,Steiner2013}, SRO(SLy4)~\cite{Chabanat:1997un,Schneider:2017tfi}, TM1~\cite{Sugahara:1993wz,Hempel:2011mk} and TMA~\cite{Toki:1995ya,Hempel:2011mk}, and the models 'Fiducial', 'Large Mmax', 'Large SL', 'Large R', 'Small SL' and 'Smaller R' from Ref.~\cite{Du:2021rhq}. Like the hyperonic models (see Sect.~\ref{sec:EoS}), all these purely nucleonic EoSs are available as temperature and composition dependent tables. 
\footnote{Most of these tables are available on the CompOSE website database (https://compose.obspm.fr/)~\cite{RevModPhys.89.015007, Typel:2013rza, CompOSECoreTeam:2022ddl,particles5030028}}

Following our previous analysis~\cite{Blacker:2023opp} we focus on discriminating hyperonic EoSs from purely nucleonic models by the different thermal behavior of the EoS (see also Sect.~\ref{sec:EoS}). For this purpose, we conduct two sets of simulations with all EoSs, i.e. all purely nucleonic and all hyperonic models. We first perform simulations employing the full EoS tables, which are temperature and composition dependent. For the other set of simulations, we use all EoSs at temperature $T=0$ under neutrinoless $\beta$-equilibrium conditions. These barotropic slices $P_\mathrm{cold}=P_\mathrm{cold}(\rho)$ and $\epsilon_\mathrm{cold}=\epsilon_\mathrm{cold}(\rho)$ are supplemented with an approximate thermal treatment, where the thermal pressure $P_\mathrm{th}$ is implemented via $P_\mathrm{th}=(\Gamma_\mathrm{th}-1)\epsilon_\mathrm{th}$ with $\epsilon_\mathrm{th}=\epsilon - \epsilon_\mathrm{cold}(\rho)$ (see~\cite{Janka1993,Bauswein:2010dn}). 

In the simulations with the approximate thermal treatment, the thermal index is chosen to be $\Gamma_\mathrm{th}=1.75$, although it generally depends on temperature and composition (see e.g. Fig.~\ref{fig:3}(a)). This choice reproduces well the results from fully consistent simulations for purely nucleonic EoSs, but leads to quantitative differences for hyperonic models. This implies that any arbitrary barotropic, cold EoS model can be supplemented with a ``nucleonic'' thermal behavior by using $\Gamma_\mathrm{th}=1.75$~\cite{Blacker:2023opp}.

Considering the two different sets of simulations has the following purpose. By choosing $\Gamma_\mathrm{th}=1.75$ for an actually hyperonic $T=0$ slice, we can mimic the scenario that a given cold hyperonic EoS could similarly result from a purely nucleonic model with different nuclear interactions. By comparing the fully consistent simulations with the $\Gamma_\mathrm{th}=1.75$ calculations, we thus quantify the impact of the different thermal behavior of hyperonic and purely nucleonic models.
As a long-term perspective, assuming the cold EoS is determined from future measurements with high accuracy, one could then discriminate nucleonic or hyperonic matter from its different thermal behavior.

\subsection{Dominant postmerger gravitational-wave frequency}\label{subsec:fpeak}
\subsubsection{Relative frequency shifts}
To assess the influence of the finite-temperature regime of the EoS, we focus our analysis on the post-merger phase and its GW emission. In our previous study~\cite{Blacker:2023opp} we discussed that for the same cold EoS the presence of hyperons at finite temperature yields an increase of the dominant post-merger GW frequency\footnote{In this study we determine GW frequencies by computing $h_\mathrm{eff,\times}(f)=f\cdot \tilde{h}_\times(f)$ with the Fourier transform $\tilde{h}_\times(f)$ of the strain of the cross polarization. Since the spectrum $h_\mathrm{eff,\times}(f)$ is given by discrete data points, we determine maxima of peaks by fitting a parabolic function to the three highest data points at the top of a peak. From the analytic representation we obtain the location of the maximum of the respective peak.} compared to the case when only nucleons are considered. In order to quantify this shift,  we define $\Delta f = f_\mathrm{peak} - f_\mathrm{peak}^{1.75}$ with $f_\mathrm{peak}$ being the dominant GW frequency from the simulation with the fully temperature dependent EoS table and $f_\mathrm{peak}^{1.75}$ being the dominant frequency from the $\Gamma_\mathrm{th}$-run, i.e. the calculation with the zero-temperature beta-equilibrium EoS slice of the same model and an approximate thermal treatment using $\Gamma_\mathrm{th}=1.75$. As argued above, the choice of $\Gamma_\mathrm{th} = 1.75$ is intentional since it has been shown to reproduce the thermal behavior of nucleonic models, which is fully corroborated by our analysis comparing full tables and $\Gamma_\mathrm{th}$-runs for purely nucleonic models (see below and~\cite{Blacker:2023opp}). In this sense, the quantity $\Delta f$ is a measure of how much a given model deviates from an ``idealized'' nucleonic thermal behavior.

\begin{figure*}[htbp]
    \centering
    \subfigure[$ \Delta f/f_{\mathrm{peak}} (f_{\mathrm{peak}})$]{
        \includegraphics[height = 2.5 in,  width=0.45\textwidth]{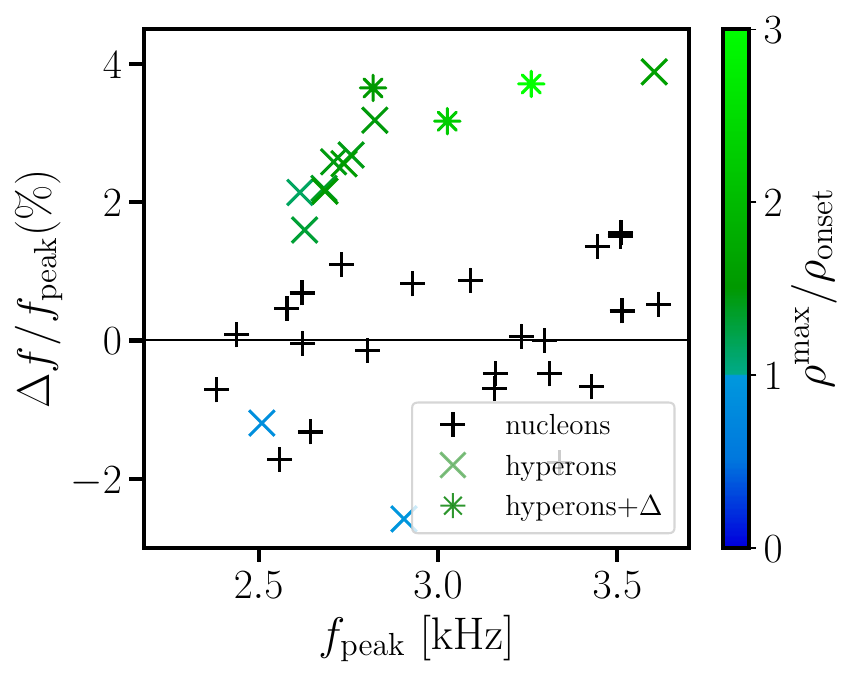}
    }
    \hspace{0.05\textwidth}
    \subfigure[$\Delta f_{\mathrm{peak}}/f_{\mathrm{peak}}(\bar{\Gamma}_{\mathrm{th}})$]{
        \includegraphics[height = 2.5 in,  width=0.45\textwidth]{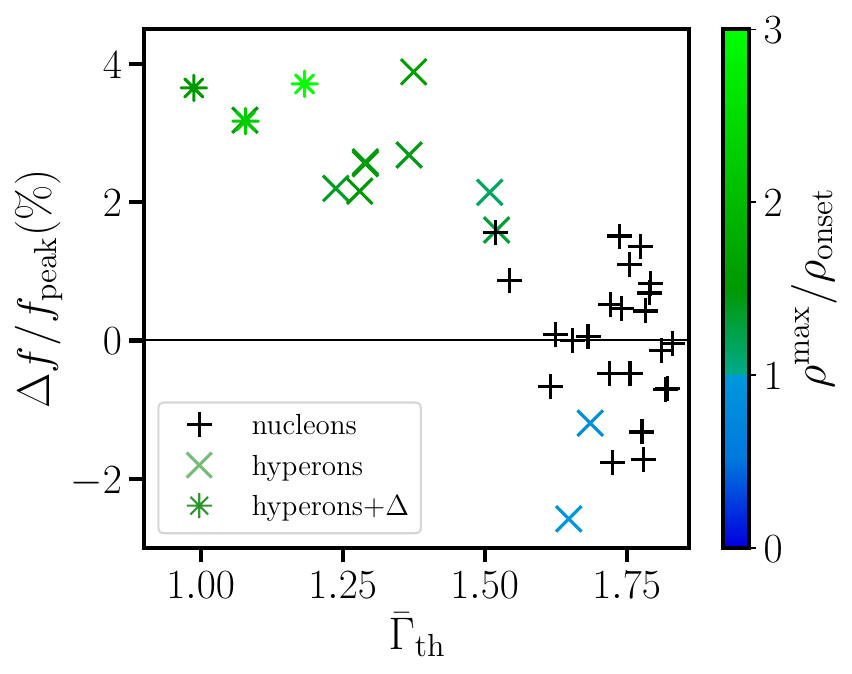}
    }
    \caption{Relative frequency shift $\Delta f/f_{\mathrm{peak}}$ as a function of the peak frequency $f_{\mathrm{peak}}$ and the average thermal index of the remnant $\bar{\Gamma}_{\mathrm{th}})$. See also similar figures in~\cite{Blacker:2023opp}.}
    \label{fig:5}
\end{figure*}

In Fig.~\ref{fig:5} we show the relative frequency shift as a function of both $f_\mathrm{peak}$ (left panel) and of an averaged thermal index $\bar{\Gamma}_{\mathrm{th}}$, which is extracted from the thermodynamical conditions in the merger remnant (right panel). Table ~\ref{tab:app-1} and Table ~\ref{tab:app-nuc} in Appendix provide the quantitative results. In order to obtain $\bar{\Gamma}_{\mathrm{th}}$, we first calculate a mass-averaged value $\Gamma^{\mathrm{av}}_{\mathrm{th}} = \sum_i m_i \Gamma_{\mathrm{th},i} / \sum_i m_i$, where $\Gamma_{\mathrm{th},i}$ is computed for an individual fluid element (SPH particle) with mass $m_i$, at a given time. If a fluid element has a temperature equal to the lowest value in the EoS table (typically 0.1 MeV), we use $\Gamma_{\mathrm{th},i} = 1$ because those SPH particles have not gained any thermal energy. We then average $\Gamma^{\mathrm{av}}_{\mathrm{th}} $ over a time window of 5 ms starting at 2.5 ms after merger (see also Subsect.~\ref{subsect:further} for the time evolution). 

In Fig.~\ref{fig:5}, models containing hyperons are labeled with crosses, models that also contain $\Delta$ baryons in addition to hyperons are labeled with asterisks, and the purely nucleonic models are labeled with plus signs. The color bar in the plots indicates the ratio between the maximum rest-mass density in the merger remnant within the first 5 ms after the merger and the density at which heavy baryons start to appear in matter in $\beta$ equilibrium at $T=0$.
It is seen that the purely nucleonic models cluster around a zero relative frequency shift, while the hyperonic models show a systematic shift towards higher values that varies in the range of $2\%-4\%$. If the density in the remnant is low, a frequency shift is not observed for hyperonic models, since the abundance of hyperons (and/or $\Delta$s) is too small. Hence, these hyperonic models result in frequency shifts similar to those of purely nucleonic ones. In the right panel the frequency shift is correlated with the average thermal index of the remnant with a smaller average thermal index corresponding to a larger frequency shift.

\begin{figure}[h]
\centering \includegraphics[height = 2.2 in ]{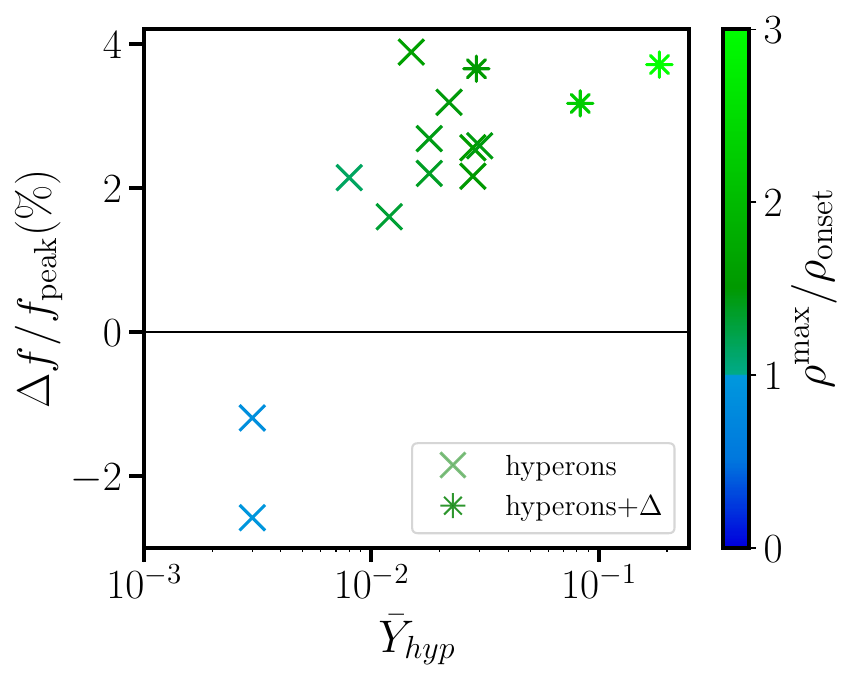} 
\caption{Relative frequency shift $\Delta f_{\mathrm{peak}}/f_{\mathrm{peak}}$ as a function of the average hyperonic fraction $\bar{Y}_{hyp}$. Note the logarithmic scale of the $x$-axis.}
\label{fig:6}
\end{figure}

In Fig.~\ref{fig:6} we display the relative frequency shift as a function of the averaged hyperon fraction $\bar{Y}_{hyp}$ inside the post-merger remnant. This quantity is defined in an analogous way as the average thermal index, namely by first computing a mass-averaged value $Y_{\mathrm{av}} = \sum_i m_i Y_{i} / \sum_i m_i$ over all fluid elements at a given time, and then averaging $Y_{\mathrm{av}}$ during a time window of 5~ms starting 2.5~ms after merging (see also Subsect.~\ref{subsect:further} for the time evolution). We observe that even a small fraction of hyperons of $\bar{Y}_{hyp}\approx0.01$ can substantially increase the GW frequency. However, the shift does not exceed four per cent even if the total number of hyperons in the remnant increases significantly. The explanation for this phenomenon lies in the behavior of the thermal index with increasing density and its correlation with the thermally produced hyperons. More specifically, we observe in the left panel of Fig.~\ref{fig:3} that the thermal index recovers to typical values for nucleonic EoSs as the density increases further beyond the onset density, although the amount of hyperons continues to grow. The right panel of Fig.~\ref{fig:3} shows that the relative abundance of {\it thermally} produced hyperons is reduced as the density becomes larger.

\subsubsection{\texorpdfstring{$f_{\mathrm{peak}}-\Lambda$}{fpeak-Lambda} relations}
It is challenging to measure a frequency shift in an actual observation because the reference value has to be independently determined (see also discussion in~\cite{Blacker:2023opp}). In~\cite{Blacker:2023opp} we thus explored the impact of heavy baryons on relations between the tidal deformability $\Lambda$ and $f_\mathrm{peak}$, which are both directly measurable. Note that we here refer to the dominant GW frequency $f_\mathrm{peak}$ from the simulations with the fully temperature dependent EoS tables alone. In such relations, hyperonic models tend to yield peak frequencies which are slightly higher than those resulting from purely nucleonic EoSs (cf. Fig.~4 in~\cite{Blacker:2023opp}). This frequency increase compared to nucleonic models, which ultimately originates from the particular thermal behavior, is thus a potentially useful signature of heavy baryons. However, the models with heavy baryons stick out most prominently if one relates $f_\mathrm{peak}$ to the tidal deformability $\Lambda_M=\Lambda(M)$ of a binary which is more massive than the one considered in the simulations. In this case the $f_\mathrm{peak}-\Lambda_M$ relations become particularly tight for purely nucleonic models (see also~\cite{PhysRevD.104.043011}). This is understandable since the densities increase during merging and $\Lambda_M$ of a more massive star with mass $M>M_\mathrm{tot}/2$ may be more representative for the EoS at these densities.

\begin{figure}[h]
\centering \includegraphics[height = 2.2 in ]{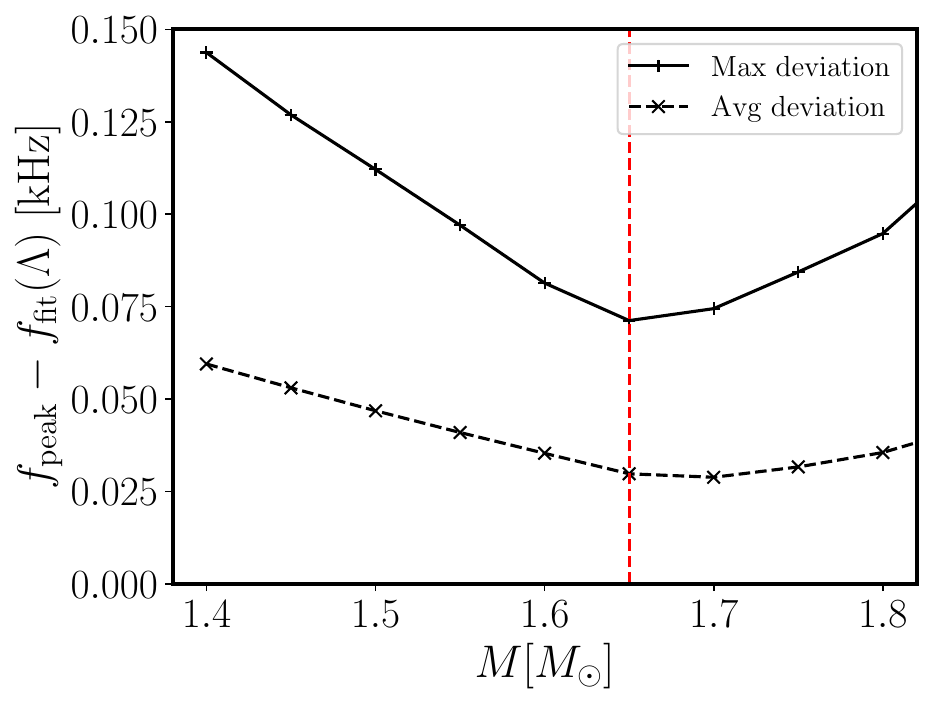} 
\caption{Average and maximum deviation of our symmetric binary data data for purely nucleonic EoSs from quadratic least-squares fits $f_\mathrm{peak}-\Lambda_M$. The fits have been performed for different reference masses $M$ at which $\Lambda$ is evaluated ($\Lambda_M=\Lambda(M)$). The dashed vertical line indicates at which $M$ the maximum deviation reaches a minimum.}
\label{fig:fpeaklambda_refef_mass_symm}
\end{figure}

We now consider $f_\mathrm{peak}-\Lambda_\mathrm{M}$ relations for the set of EoS models considered here, which is larger than the sample in~\cite{Blacker:2023opp}.  Following the approach of earlier works \cite{Blacker:2023opp,PhysRevD.104.043011}, we first identify the stellar mass $M$ for which the $f_{\mathrm{peak}}-\Lambda_M$ relation of purely nucleonic models becomes particularly tight.  
In Fig.~\ref{fig:fpeaklambda_refef_mass_symm} we show the maximum and the average deviation of the data from a quadratic fit $f_\mathrm{peak}(\Lambda_M)$, for purely nucleonic EoSs, as functions of the mass $M$\footnote{Considering all purely nucleonic EoSs, we perform a quadratic least-squares fit $f_\mathrm{peak}(\Lambda_M)$ and compute the residual between the fit and the true value as $r_i = |f_\mathrm{peak} - f_\mathrm{fit}|$ for each data point. For every fixed reference mass $M$ we compute the average value of the residuals $r_\mathrm{ave} = \sum r_i/N$ with $N$ being the number of data points and determine the maximum of all residuals $r_\mathrm{max} = \max{(r_i)}$.}. The average scatter with respect to a quadratic fit is almost constant for different reference masses $M$. The maximum deviation shows a clear minimum at $M=1.65~M_\odot$, which we thus choose as a reference value\footnote{Note that in~\cite{Blacker:2023opp} the reference mass was 1.75~$M_\odot$ for equal-mass binaries, whereas the larger set of EoS models considered here suggests a slightly lower reference mass.}.

\begin{figure}[h]
\includegraphics[width=0.9\linewidth]{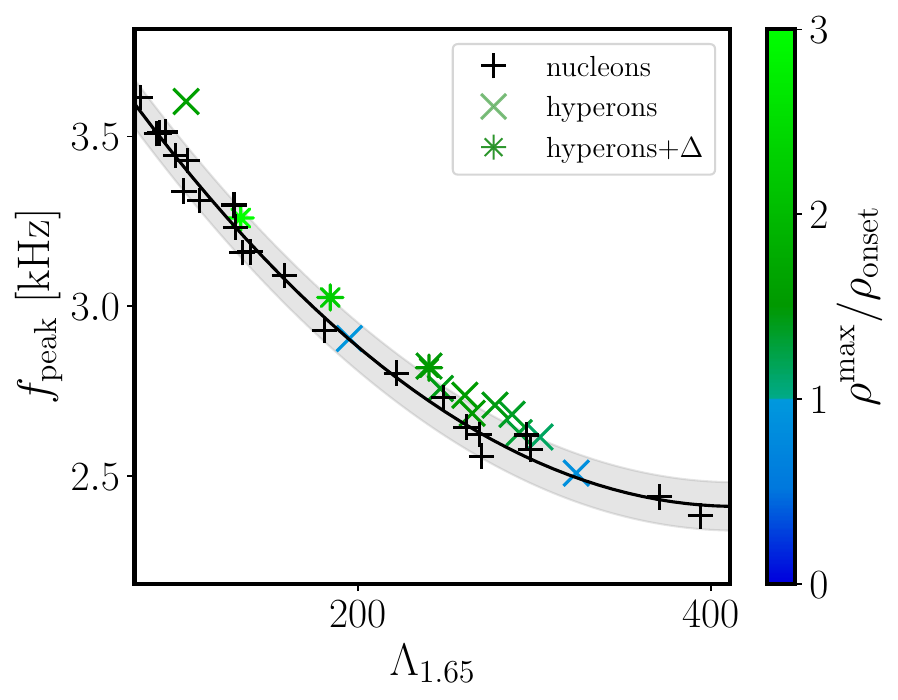}
\caption{Dominant postmerger GW frequency as function of tidal deformability of a 1.65~$M_\odot$ NS for symmetric mergers.
Black curve is a least-squares quadratic fit to purely nucleonic models. Gray band indicates maximum residual of purely nucleonic models from the fit.}
\label{fig:fpeaklambda_symm}
\end{figure}

In Fig. \ref{fig:fpeaklambda_symm} we plot the corresponding $f_{\mathrm{peak}}-\Lambda_{1.65}$ relation for all full EoS models considered in this work. 
The notation of the symbols is the same as before. The solid line is the quadratic fit to the nucleonic data, while the gray band indicates the maximum residual of the nucleonic models with respect to the fit. For a given tidal deformability, the dominant frequency of most of the models with heavy baryons is higher compared to that of the nucleonic models. In particular, some of the models with heavy baryons also lie outside of the gray band, which clearly distinguishes them from purely nucleonic EoSs.

\begin{figure*}

\begin{centering}
    
\subfigure[]{
\includegraphics[  width=0.48\textwidth]{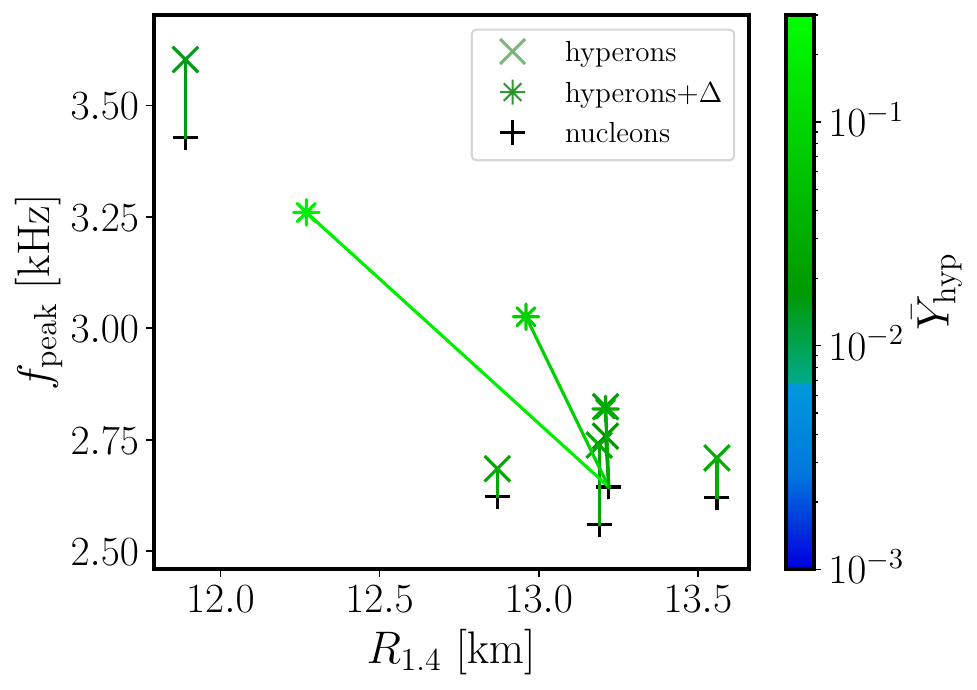}
    }
    \hfill    
\subfigure[]{
\includegraphics[  width=0.48\textwidth]{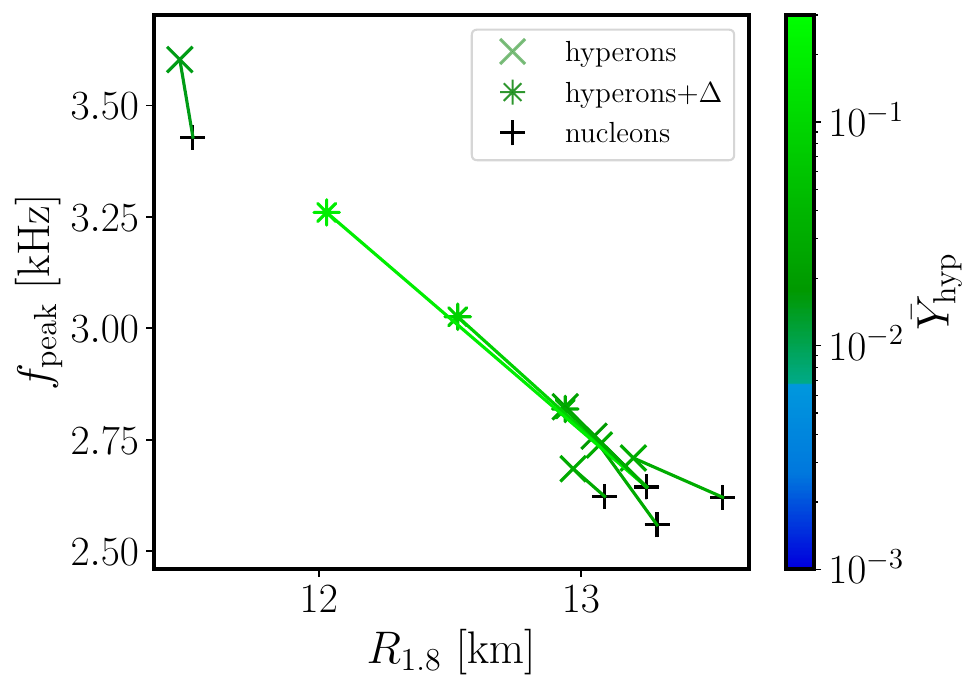}}

\subfigure[]{
\includegraphics[  width=0.48\textwidth]{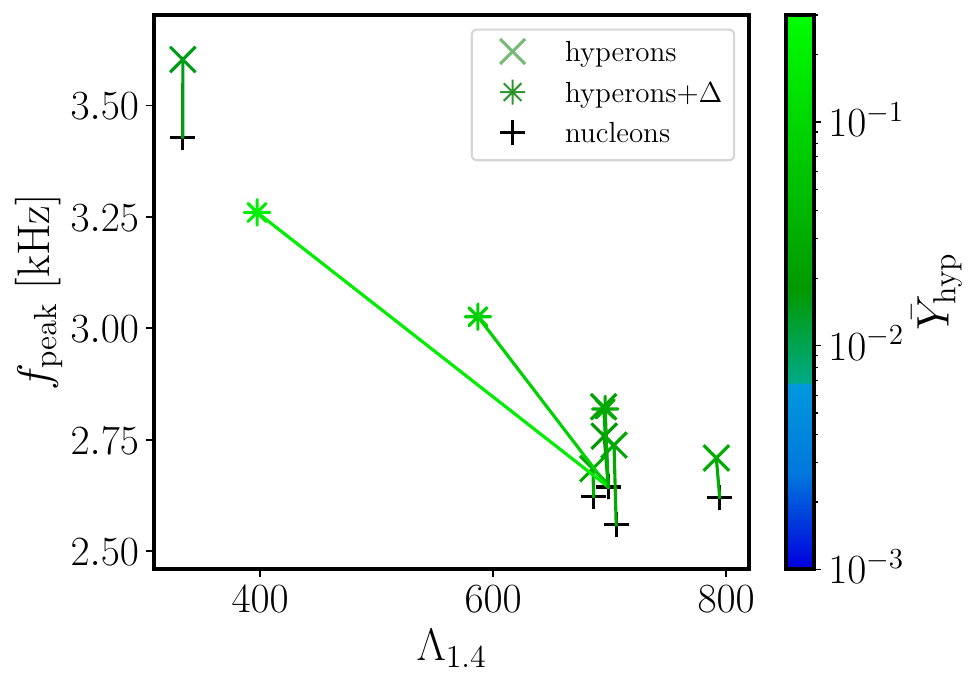}
    }
    \hfill    
\subfigure[]{
\includegraphics[  width=0.48\textwidth]{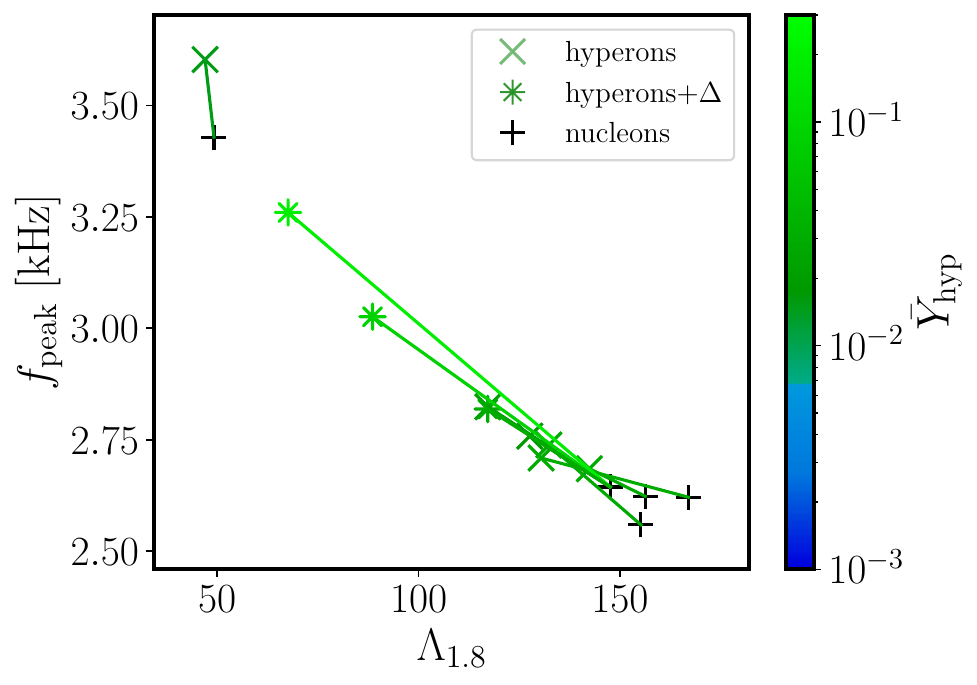}}

\caption{Dominant postmerger GW frequency $f_{\mathrm{peak}}$ as a function of stellar parameters of cold nonrotating NS, $R_{1.4}$, $R_{1.8}$, $\Lambda_{1.4}$ and $\Lambda_{1.8}$, for several pairs of nucleonic and hyperonic models which share the same nucleonic interaction. Colorcoding indicates the fraction of hyperons in the remnant.}
\label{fig:add_fpeak_coldpar}

\end{centering}

\end{figure*}

\subsubsection{Full impact of hyperons}
For completeness we illustrate the specific impact of hyperons for a given nucleonic model. Within our set of EoSs, we identify those models for which a purely nucleonic models is available as well as an EoS which includes heavy baryons and employs the very same nucleonic interactions. In Fig.~\ref{fig:add_fpeak_coldpar} we display the dominant postmerger GW frequency as function of different stellar parameters for the aforementioned subset of EoS models. Using a line we indicate those models which rely on the same nucleonic interaction. This gives a visual impression of the quantitative impact of hyperons on $f_\mathrm{peak}$ and the stellar properties. There are several models associated with the purely nucleonic DD2 EoS (DD2Y, BHB$\Lambda \phi$, DD2Y$\Delta1.1-1.1$, DD2Y$\Delta1.2-1.1$ and DD2Y$\Delta1.2-1.3$) which differ by the included heavy species and their interactions. Other pairs include SFHO and the hyperonic counterpart SFHOY, and the family of DDLSN nucleonic models with the corresponding hyperonic extensions.
The coloring of the points indicates the time averaged fraction of heavy baryons inside the remnant, 5 ms after the merging, in a time window of 2.5 ms.

Generally the impact of hyperons or heavy baryons is to shift $f_\mathrm{peak}$ to higher frequencies, which can amount to more than 500~Hz in some models. Also the stellar parameters tend to decrease by the inclusion of heavy baryons, which can exceed 1~km in the radius (see panels b) and d) in Fig.~\ref{fig:add_fpeak_coldpar}). However, for $R_{1.4}$ and $\Lambda_{1.4}$ the effect may also be absent or marginal, which is explained by the fact that heavy baryons only occur in sizable quantities at densities exceeding the central density of the respective star. In all shown models, however, $f_\mathrm{peak}$ is affected because temperature and density increase during merging.

We stress that the direct comparison between nucleonic and hyperonic models in Fig.~\ref{fig:add_fpeak_coldpar} is mostly academic and only helpful to understand the impact and the importance of hyperons in high-density matter. It is likely not useful for inferring the presence of hyperons on NS mergers because, as already detailed above, a respective model to compare with and quantify a shift does not exist in nature. One might optimistically envision a scenario where theoretical work, which may be based on experimental data, specifies the properties of a  hypothetical purely nucleonic system. A comparison to the actually observed $f_\mathrm{peak}$ or stellar parameters may then reveal a mismatch which may be compatible with the occurrence of heavy baryons.

\subsubsection{Maximum density}
\begin{figure}[h]
\centering \includegraphics[height = 2.2 in ]{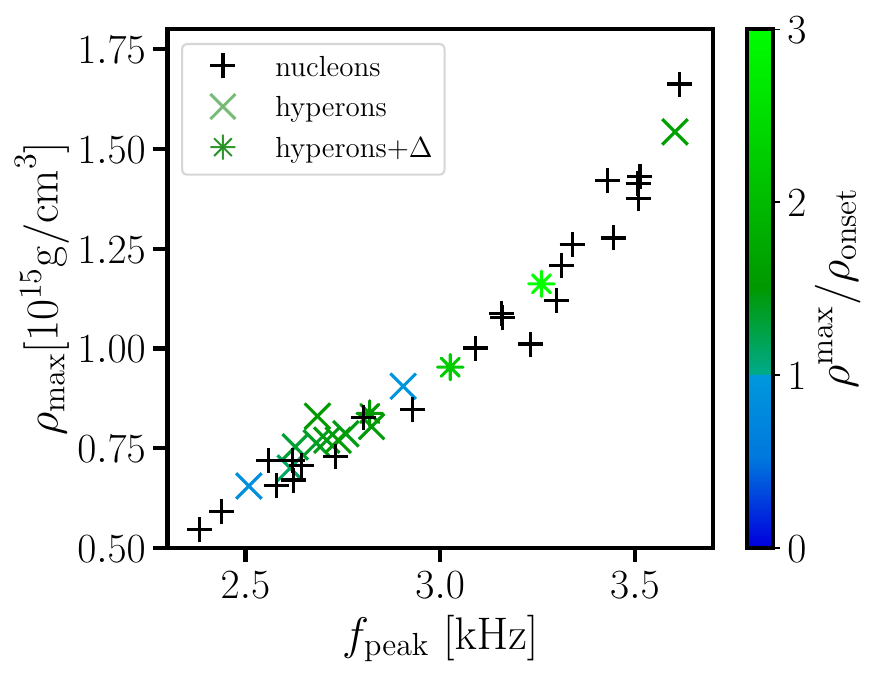} 
\caption{Maximum rest-mass density as a function of the dominant frequency peak $ f_{\mathrm{peak}}$ in the merger remnant within the first 5 ms after the merger.}
\label{fig:7}
\end{figure}

We next investigate whether the frequency shift observed in hyperonic models modifies 
a correlation between the maximum rest-mass density in the merger remnant within the first 5 ms after the merging ($\rho_{\mathrm{max}}$) and the dominant post-merger GW frequency (or, equivalently, the combined tidal deformability)~\cite{PhysRevLett.122.061102,Blacker:2020nlq}. As the correlation does not hold very well for EoSs that feature a strong first order phase transition to quark matter, it is important to investigate this relation for our large set of EoSs with heavier baryons. In Fig.~\ref{fig:7} we display the dependence of $\rho_{\mathrm{max}}$ versus the dominant post-merger GW frequency for all models employed in the present work. We do not observe quantitative differences between the relation for the nucleonic models and the hyperonic one. Hence, we confirm the conclusion that all hadronic models show the aforementioned correlation and that this universal relation can be used to estimate the maximum rest mass density in the remnant from the observed dominant frequency, even if hyperons are present. This fact may be exploited in order to put an upper (or lower) limit on the density at which hyperons start to occur (or not) using future GW observations. If a GW event suggests that there are no hyperons in the remnant, then using the universal relation $\rho_{\mathrm{max}}(f_{\mathrm{peak}})$ one can put a lower bound on the density at which hyperons are not present (at least in significant amounts). If a future measurement provides evidence for the presence of hyperons in the remnant, one can determine an upper bound for the onset density at which hyperons start to be significantly produced.

\subsubsection{Numerical robustness}
Considering that the GW frequency shifts caused by thermally produced hyperons are relatively small (a few per cent), it is important to cross-check that this effect is not an artifact of the numerical tool. To this end, we perform simulations with the moving-mesh code AREPO~\cite{2010MNRAS.401..791S,2016MNRAS.455.1134P} employing the relativistic version of the code~\cite{Lioutas2022GeneralRM}. The AREPO code implements a very different hydrodynamics solver, where the computational cells move with the fluid flow, which minimizes numerical artifacts from advecting matter across grid cells in finite-volume schemes. The code also employs the conformal flatness condition for the Einstein equations~\cite{1980grg1.conf...23I,1996PhRvD..54.1317W}. 
For the DD2 and DD2Y EoSs we rerun simulations with the relativistic moving-mesh version of the AREPO code. Again, we conduct simulations with the fully temperature dependent EoS tables and with the approximate thermal treatment choosing a constant thermal index of $\Gamma_{\mathrm{th}} = 1.75$. The results of the dominant post-merger frequency are given in Table \ref{table:AREPO}, where, for comparison, the results obtained with the SPH scheme are also provided. For the DD2 model, the resulting frequency values agree very well. For the DD2Y case, the codes yield slightly different frequencies, which may not be surprising considering that the numerical schemes are very different. However, it is important to note that also with the moving-mesh tool there is a significant frequency shift of roughly the same magnitude between the run with $\Gamma_\mathrm{th}=1.75$ and the fully consistent simulation. This corroborates our previous findings that the change of the thermal behavior of the EoS by hyperons leads to a small but characteristic increase of the post-merger GW frequency.

\begin{table}[h]
\begin{tabular}{c|cc|cc|}
\cline{2-5}
& \multicolumn{2}{c|}{DD2}         & \multicolumn{2}{c|}{DD2Y}        \\ \cline{2-5} 
 & \multicolumn{1}{c|}{$f_{\mathrm{peak}}$ (kHz)} & $f^{1.75}_{\mathrm{peak}}$ (kHz) & \multicolumn{1}{c|}{$f_{\mathrm{peak}}$ (kHz)} & $f^{1.75}_{\mathrm{peak}}$ (kHz) \\ \hline
\multicolumn{1}{|c|}{AREPO} & \multicolumn{1}{c|}{2.63} & 2.67 & \multicolumn{1}{c|}{2.71} & 2.64 \\ \hline
\multicolumn{1}{|c|}{SPH}   & \multicolumn{1}{c|}{2.64} & 2.68 & \multicolumn{1}{c|}{2.82} & 2.73 \\ \hline
\end{tabular}
\caption{Results of dominant GW frequency peak using AREPO and SPH for two EoSs, DD2 and DD2Y. $f_\mathrm{peak}$ refers to the results with the full 3D tables, while $f^{1.75}_{\mathrm{peak}}$ denotes the frequencies for the simulations based on the cold $\beta$ equilibrium slices supplemented with an approximate thermal treatment adopting $\Gamma_{\mathrm{th}} = 1.75$ to mimic the thermal behavior of nucleonic matter.}
\label{table:AREPO}
\end{table}

Frequency shifts also occur consistently for simulations with different resolution, which we test with the SPH code running models with about 150k, 300k, and 600k SPH particles, respectively, for a subset of EoSs. Within this set the absolute frequencies change by at most a few 10~Hz. We find that purely nucleonic models with a consistent temperature treatment yield very stable results (with increasing resolution 3.435~kHz, 3.429~kHz and 3.430~kHz for SFHO, and 2.635~kHz, 2.644~kHz, 2.623~kHz for DD2). Hyperonic models show a slightly stronger resolution dependence with the tendency to yield lower absolute frequencies for higher SPH particle numbers ($\sim$50~Hz difference between the runs with the highest and lowest resolution for DD2Y, SFHOY and DD2Y$\Delta$1.2-1.1). Hyperonic models might be more resolution sensitive since already the cold models feature some pronounced softening of the EoS (cf. Fig.~\ref{fig:add_fpeak_coldpar}), which is why the frequencies may be more susceptible to the exact thermodynamic conditions in the remnant. Comparing simulations with $\Gamma_\mathrm{th}=1.75$ and the full thermal treatment, we emphasize that shifts in GW frequencies of the order of 100~Hz due to the presence of hyperons occur consistently across these runs with different resolution (for DD2Y, SFHOY and DD2Y$\Delta$1.2-1.1). This indicates that the impact of hyperons as discussed in this study is a relatively robust feature although on the verge of the resolving power.

\subsection{Secondary gravitational wave peaks}

\begin{figure*}[htbp]
    \centering
    \subfigure[]{
        \includegraphics[height = 2.5 in,  width=0.45\textwidth]{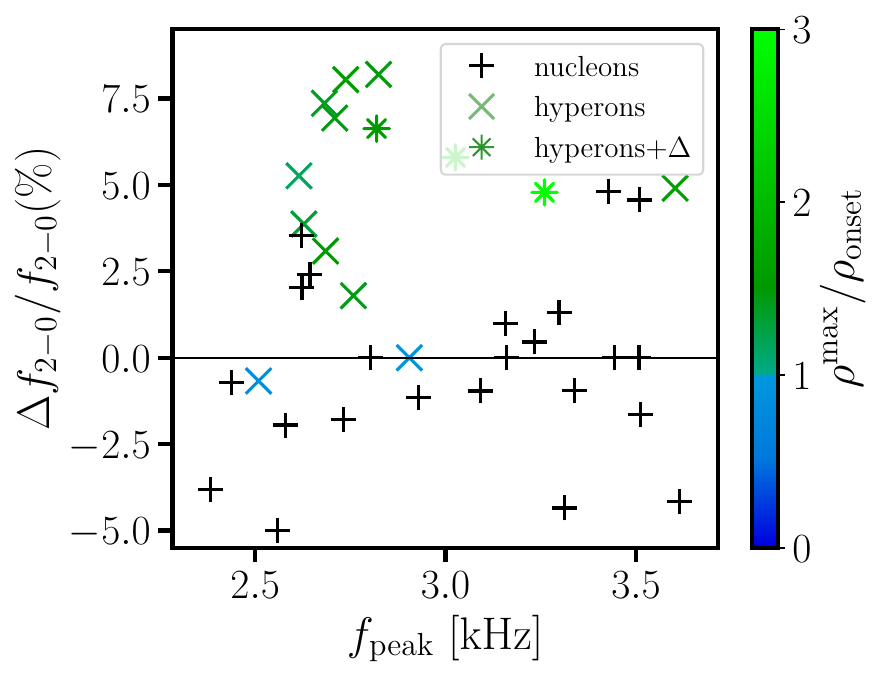}
    }
    \hspace{0.05\textwidth}
    \subfigure[]{
        \includegraphics[height = 2.5 in,  width=0.45\textwidth]{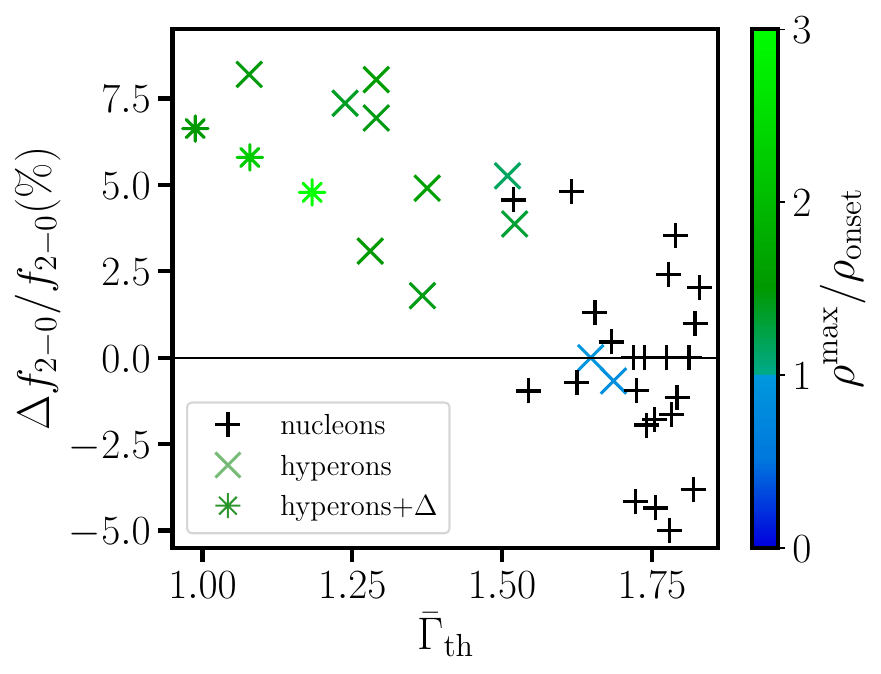}
    }
    \caption{Relative frequency shift of the secondary peak $\Delta f_{2-0}/f_{2-0}$ as a function of the dominant frequency peak $f_{\mathrm{peak}}$ and the average thermal index $\bar{\Gamma}_{\mathrm{th}}$.   
    }
    \label{fig:12}
\end{figure*}

Apart from the dominant frequency peak, additional secondary spectral features occur in the kHz range of the GW spectra from BNS merger events and may be detected in the future (e.g.~\cite{Stergioulas:2011gd,PhysRevD.88.044026,PhysRevD.91.124056,Bauswein_2016,  PhysRevD.91.064001,PhysRevD.96.063011}). 
Secondary peaks at frequencies below the dominant frequency are most relevant for detection. Specifically, we discuss the $f_{2-0}$ and $f_{\mathrm{spiral}}$ features, which are generated by different mechanisms (see~\cite{PhysRevD.91.124056,Bauswein_2019} for nomenclature and further discussion \footnote{Note that in some simulations it is not straightforward to clearly identify or distinguish the different features. However, based on experience from other simulations one can associate the correct peaks.}).
We are interested to understand if thermally produced hyperons leave an imprint in these observables. Hence, in an analogous way as for our analysis of the dominant frequency peak, for each EoS in our simulation campaign, we define relative frequency shifts:
\begin{equation}
    \frac{\Delta f_{2-0}}{f_{2-0}} = \frac{f_{2-0}-f_{2-0}^{1.75}}{f_{2-0}}; \frac{\Delta f_{\mathrm{spiral}}}{{f_{\mathrm{spiral}}}} = \frac{f_{\mathrm{spiral}}-f_{\mathrm{spiral}}^{1.75}}{f_{\mathrm{spiral}}}.
\end{equation}
These ratios quantify by how much the secondary frequency peaks obtained with an approximate thermal treatment mimicking an idealized ``nucleonic'' thermal behavior deviate from the ones which result from the simulation with fully temperature-dependent EoS tables.
The dependence of $ \Delta f_{2-0}/f_{2-0}$ on $f_{\mathrm{peak}}$ and $\bar{\Gamma}_{\mathrm{th}}$ is shown in Fig.~\ref{fig:12}. Similar to the dominant frequency, hyperonic models with sizable hyperon fractions exhibit systematic frequency shifts. The magnitude of the shift is model dependent reaching up to 8\%. Nucleonic models are scattered around zero shift with an overall larger scatter than that of the dominant frequency peak $f_{\mathrm{peak}}$, that is $\Delta f/f_\mathrm{peak}$. 
As for the dominant peak, there is a relatively strong  correlation between the shift of $f_{2-0}$ and the average thermal index. 

\begin{figure*}[htbp]
    \centering
    \subfigure[]{
        \includegraphics[height = 2.5 in,  width=0.45\textwidth]{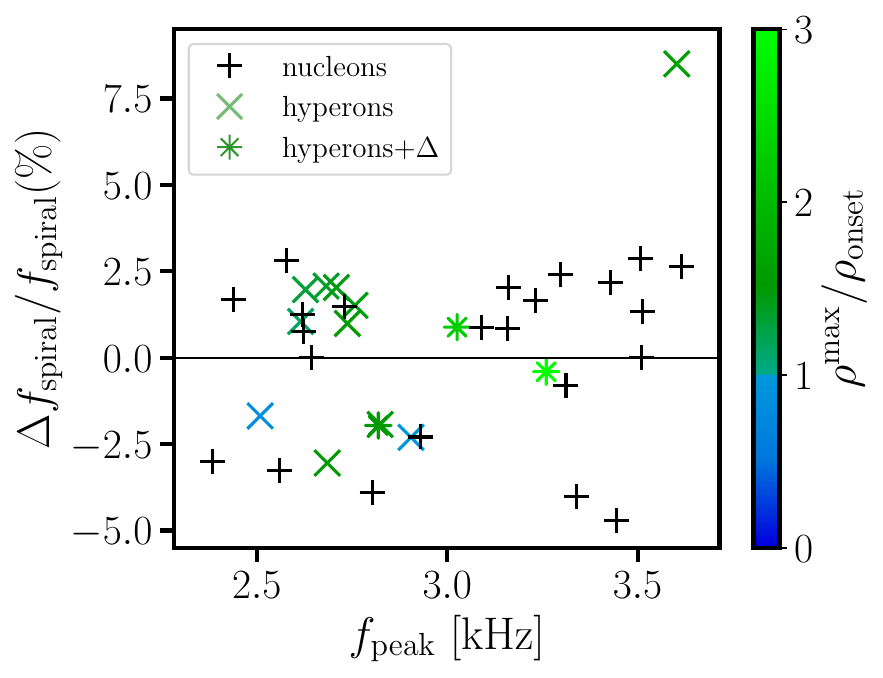}
    }
    \hspace{0.05\textwidth}
    \subfigure[]{
        \includegraphics[height = 2.5 in,  width=0.45\textwidth]{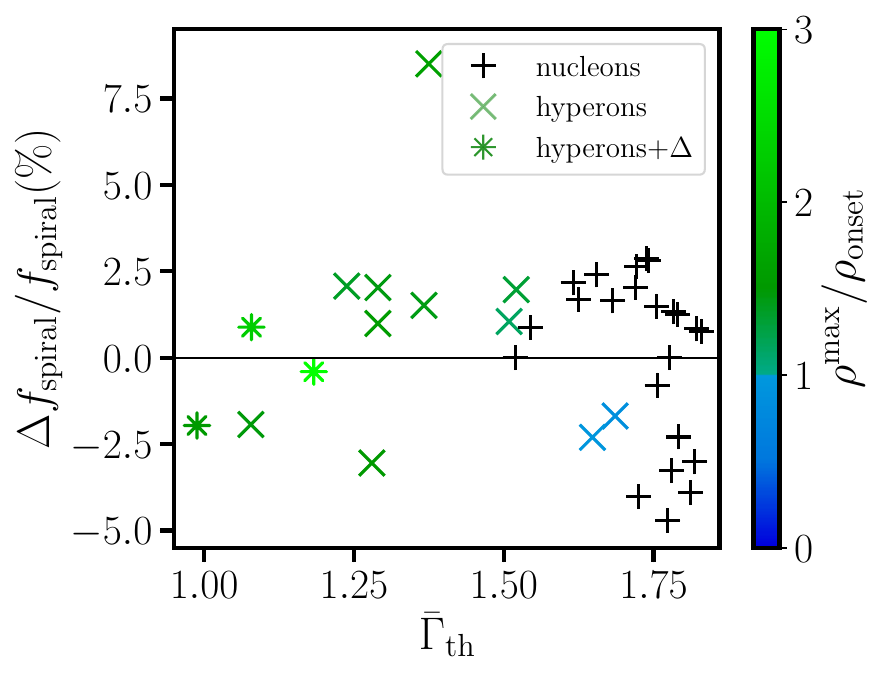}
    }
    \caption{Relative frequency shift of the secondary peak $\Delta f_{\mathrm{spiral}}/f_{\mathrm{spiral}}$ as a function of the dominant frequency peak $f_{\mathrm{peak}}$ and the average thermal index $\bar{\Gamma}_{\mathrm{th}}$. 
    %The labeling is same as in Fig. \ref{fig:12}}
    }
    \label{fig:spiral}
\end{figure*}

Interestingly, thermally produced hyperons and $\Delta$ baryons do not have a pronounced impact on $f_\mathrm{spiral}$, and, besides one outlier, frequency shifts for hyperonic models are comparable to those of purely nucleonic models (see left panel Fig.~\ref{fig:spiral}). The right panel of Fig.~\ref{fig:spiral} demonstrates that $f_\mathrm{spiral}$ is largely unaffected by the average thermal index. The very different behavior of $f_\mathrm{peak}$ and $f_{2-0}$ as opposed to $f_\mathrm{spiral}$ might be explained by the different mechanisms producing these features. $f_{2-0}$ is a combination frequency of the quandrupolar and the quasi-radial oscillation mode of the remnant. $f_\mathrm{spiral}$ is generated by two orbital bulges, which originate from the tidal stretching during merging and move on the surface of the remnant in the very early phase of the post-merger evolution.

\begin{figure*}[htbp]
    \centering
    \subfigure[]{
        \includegraphics[height = 2.5 in,  width=0.45\textwidth]{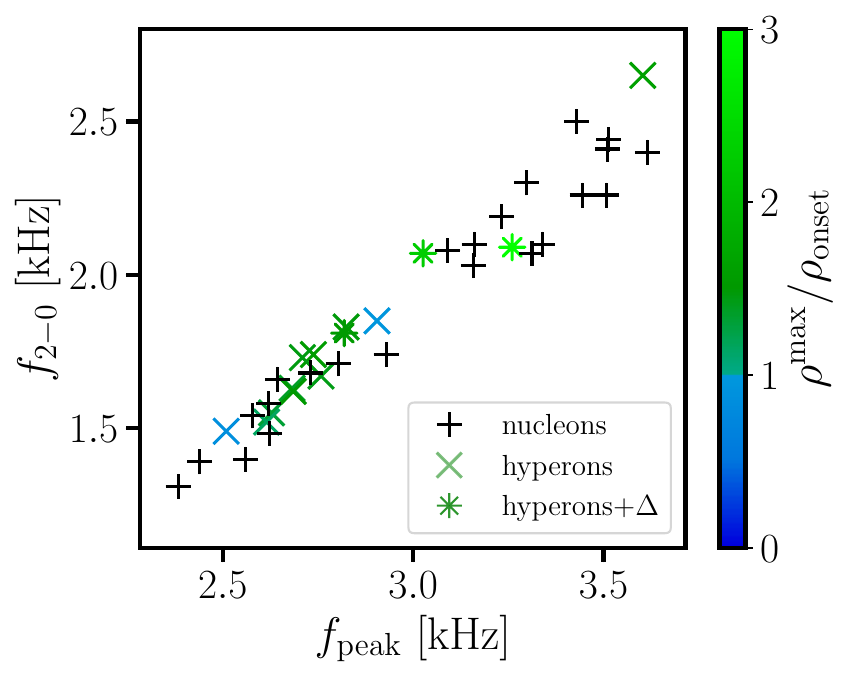}
    }
    \hspace{0.05\textwidth}
    \subfigure[]{
        \includegraphics[height = 2.5 in,  width=0.45\textwidth]{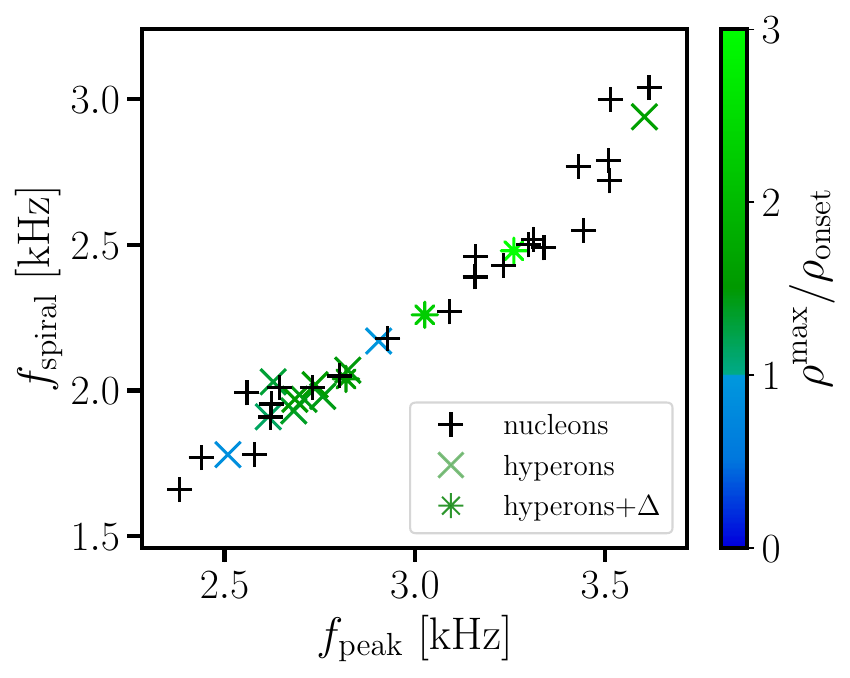}
    }
    \caption{Secondary peaks $f_{2-0}$ (left panel) and $f_{\mathrm{spiral}}$ (right panel) as functions of the dominant peak $f_{\mathrm{peak}}$.}
    %Crosses display results obtained for hyperonic models, where the coloring indicates the ratio between the maximum rest-mass density in the postmerger remnant and the onset rest-mass density of hyperon production at zero temperature. Models that also include $\Delta$-baryons are labeled with asterisk.}
    \label{fig:extra-11}
\end{figure*}

In Fig. \ref{fig:extra-11} we also show $f_{2-0}$ and $f_{\mathrm{spiral}}$ as functions of $f_{\mathrm{peak}}$. Comparing two frequencies of the same system could potentially yield an observational signature of hyperons if hyperonic models for instance showed some deviations from the relations of purely nucleonic models. Unfortunately, the hyperonic models closely follow the trends of the purely nucleonic models. Only for $f_\mathrm{spiral}(f_\mathrm{peak})$ one might recognize a slight tendency of models with significant hyperon content to have a relatively low $f_\mathrm{spiral}$ for a given $f_\mathrm{peak}$, which is in line with the fact that hyperonic models feature a positive shift in $f_\mathrm{peak}$ (left panel of Fig.~\ref{fig:5}) but not in $f_\mathrm{spiral}$ (left panel of Fig.~\ref{fig:spiral}).

\subsection{Further analysis}
\label{subsect:further}
While at the beginning of Sec.~\ref{sec:Simulations} we have focused on certain observables, we provide here a deeper analysis of the simulation data from hyperonic models.
Whenever applicable, we compare the results from the hyperonic EoSs with those from three purely nucleonic, namely, DD2, SFHO and FSU2R.

\begin{figure}[h]
\centering \includegraphics[height = 2.2 in ]{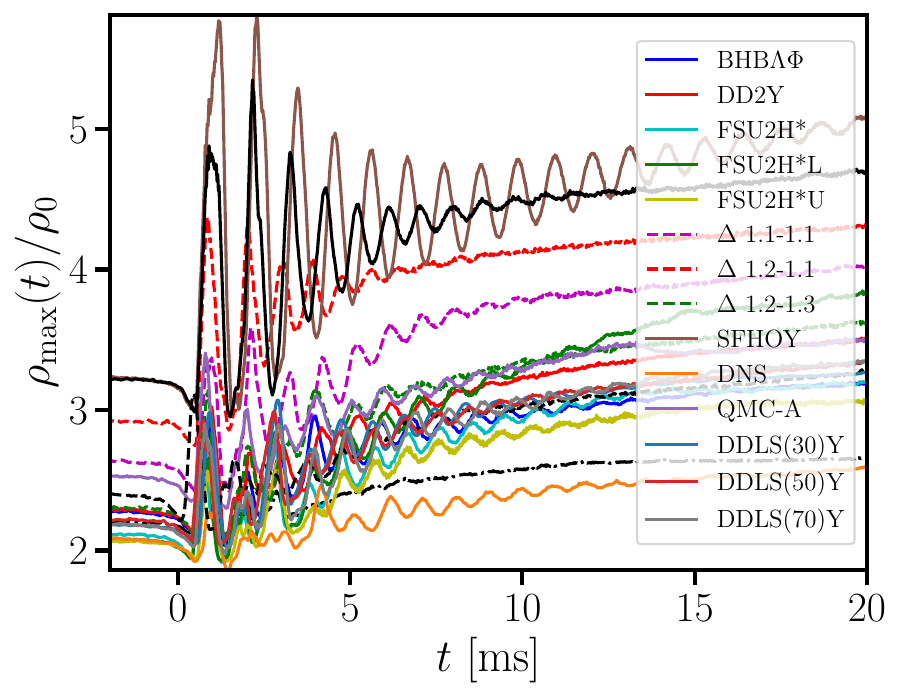} 
\caption{Maximum rest mass density in units of the nuclear saturation density, $\rho_{\mathrm{max}}(t)/\rho_0$, as a function of time.  
Black curves display results for exemplary purely nucleonic EoSs (solid line for SFHO, dashed line for FSU2R, dash-dotted line for DD2).}
\label{fig:8}
\end{figure}

In Fig.~\ref{fig:8} we show the maximum rest-mass density evolution $\rho_{\mathrm{max}}(t)$ of the different models, noting that $t=0$ corresponds to the time of merging. Our set of hyperonic models covers a wide range of densities before and after merging. Softer EoS models such as SFHOY and DD2Y$\Delta 1.1-1.1$, which can also be identified by higher density during the inspiral before $t=0$, feature stronger quasi-radial oscillations (cf.~\cite{PhysRevD.91.124056}). 
Comparing the purely nucleonic SFHO model (black solid line) and the corresponding hyperonic SFHOY one can clearly see how the occurrence of hyperons in the post-merger phase yields stronger and longer lasting radial oscillations. Note that the quasi-radial oscillations of the purely nucleonic SFHO are stronger than those of several hyperonic EoSs which are stiffer, and thus one would conclude that the strength of the quasi-radial oscillations is primarily determined by the softness of the EoS. 

\begin{figure}[h]
\centering
%\label{fig:5}
%\begin{subfigure}[t]{0.49\textwidth}
\centering \includegraphics[height = 2.2 in ]{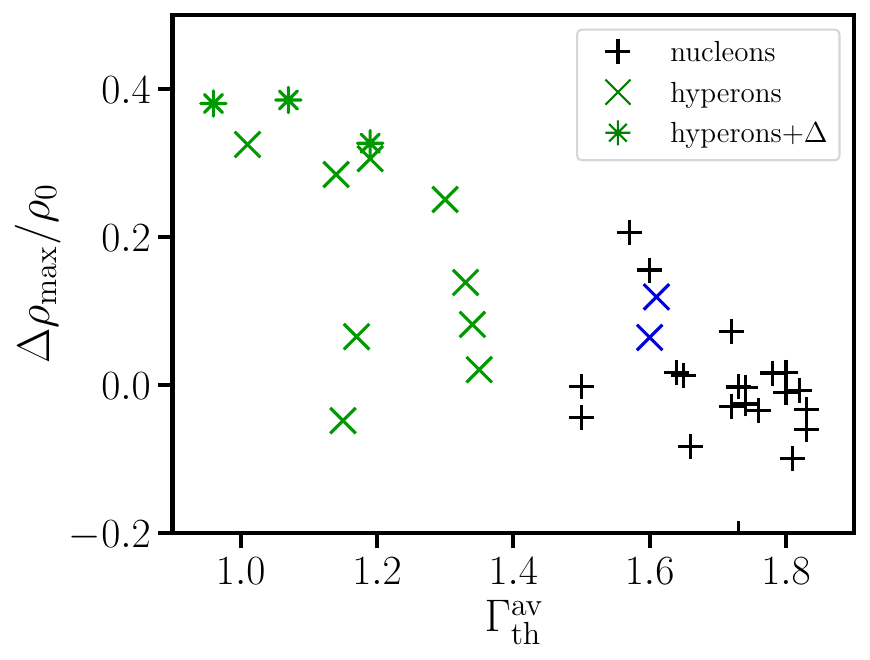}

\caption{Difference between maximum rest mass density in simulations with full EoS tables and with approximate thermal treatment as a function of the average thermal index inside the remnant at $t=10$ ms.  
The coloring indicates whenever heavy baryons represent more than 1\% of the baryon population (green points) or not (blue points).}
\label{fig:extra-1}
\end{figure}

As for the analysis of GWs, it is also instructive to compare the evolution of the maximum density in the remnant for the two different sets of simulations, i.e. $ \rho_{\mathrm{max}}(t)$ obtained with the fully temperature-dependent EoS tables, and $\rho_{\mathrm{max}}^{1.75}(t)$ obtained with the $\Gamma_{\mathrm{th}} = 1.75$ treatment. 
We evaluate $\Delta \rho_{\mathrm{max}}(t) = \rho_{\mathrm{\mathrm{max}}}(t)-\rho_{\mathrm{max}}^{1.75}(t)$ 10~ms after merging when the oscillations in the remnant have declined.
The results are shown in Fig.~\ref{fig:extra-1} for the hyperonic and the purely nucleonic EoSs, revealing a correlation between $\Delta \rho_{\mathrm{max}}(t = 10$ ms) and the mass-average thermal index extracted from the simulation with the fully temperature dependent EoS, also at $t = 10$~ms. The color of the data points indicates whether the fraction of hyperons and/or delta resonances is higher (green) or lower (blue) than 1\% at $t = 10$~ms.

As expected, simulations with $\Gamma_\mathrm{th} = 1.75$ underestimate the maximum density for models with heavier baryons because the reduction of the thermal index in the calculations with the complete temperature treatment implies an increase in the density. 
This is consistent with the GW frequency shift in these models, indicating a more compact merger remnant. Purely nucleonic models cluster around $\Delta \rho_{\mathrm{max}}/\rho_0 = 0$.

\begin{figure}[h]
\centering \includegraphics[height = 2.2 in ]{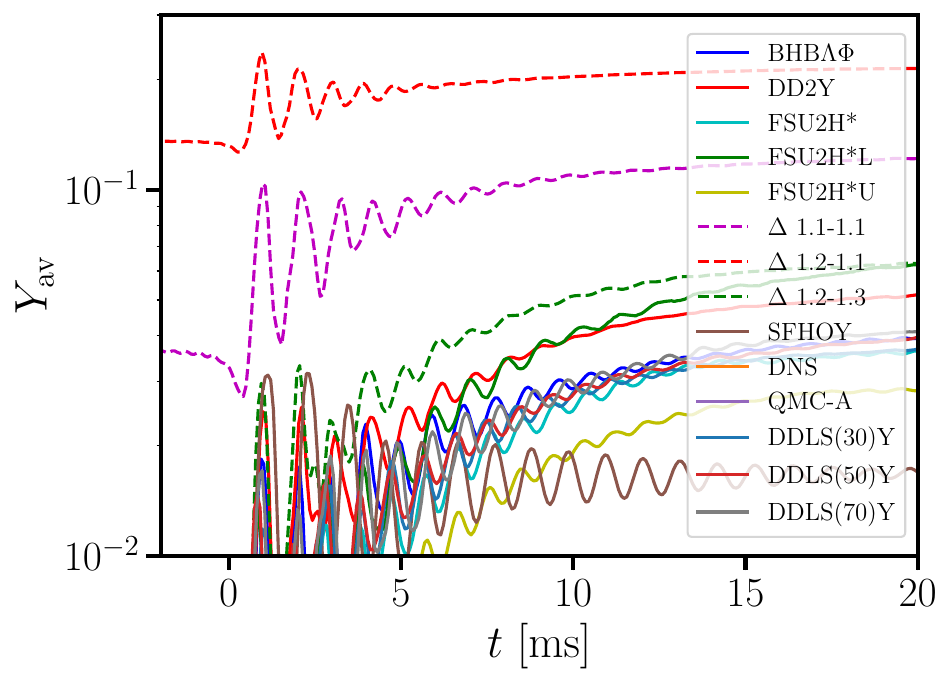} 
\caption{Hyperon fraction as a function of time. Note the logarithmic scale on the y-axis.}
\label{fig:9}
\end{figure}

For the hyperonic models, the hyperon abundance in the remnant is roughly correlated with the maximum rest mass density, as can be seen by comparing Fig.~\ref{fig:9}, where the time evolution of the average hyperon fraction is shown, with Fig.~\ref{fig:8}. This relates to the fact that the hyperon fraction increases with density for any EoS.

From Fig.~\ref{fig:9} it is clear that only in two of our models exotic species occur in significant quantities in the NSs before merging ($t<0$). Those models are the ones that also account for $\Delta$ baryons with the $\Delta^{-}$ being the dominant exotic component. The majority of models do not reach sufficiently high densities to produce (non-thermal) hyperons before merging. Consequently, the inspiral observables are insensitive to these exotic components. 
However, after merging, a significant amount of hyperons is produced by most of the models, except for DNS and QMC-A, which do not reach hyperon fractions within the range plotted in Fig.~\ref{fig:9}. Those two models are the ones that essentially behave as purely nucleonic EoSs, see e.g. Fig \ref{fig:5}. 
For these EoSs the nucleonic regime at lower densities is too stiff to compress matter to densities where hyperons occur in the post-merger phase for the binary masses considered here.

\begin{figure}[h]
\centering \includegraphics[height = 2.2 in ]{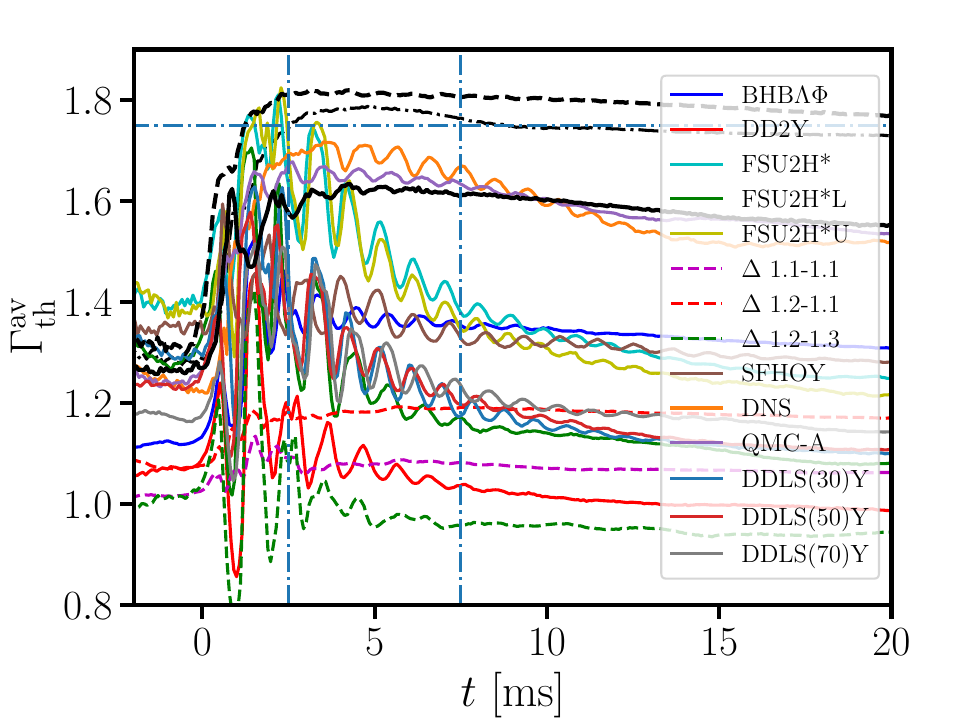} 
\caption{Average thermal index as a function of time for all hyperonic models. For comparison, with black lines, three nucleonic models (DD2 - dotted line, SFHO - solid line, FSU2R - dashed line) are also shown. The vertical blue lines show the characteristic time window that we use to compute $\bar{\Gamma}_{\mathrm{th}}$. The horizontal blue line shows the value of $\Gamma_{\mathrm{th}} = 1.75$}.
\label{fig:10}
\end{figure}

The evolution of the mass-averaged thermal index with time is shown in Fig.~\ref{fig:10}.
For most of the models, $\Gamma^\mathrm{av}_\mathrm{th}(t)$ exhibits strong oscillations, which follow the oscillations of the hyperon abundance. At later times $\Gamma^\mathrm{av}_\mathrm{th}(t)$ becomes roughly constant.
The oscillations of the thermal index of nucleonic models are either absent or much smaller than those of the hyperonic models. As expected, hyperonic models that produce a significant amount of hyperons have a much lower mass-averaged thermal index than nucleonic models, reaching values $\Gamma^{\mathrm{av}}_{\mathrm{th}} < 1.4$ after merging. One of the models with a large hyperon abundance, DD2Y$\Delta$1.2-1.3, even exhibits a mass-averaged thermal index smaller than one after a few milliseconds (see Fig.~\ref{fig:10}). 
Clearly, in such hyperonic models the thermal pressure support as a mechanism to stabilize the merger remnant in addition to the strong rotation is significantly reduced or even absent.

\begin{figure}[h]
\centering \includegraphics[height = 2.2 in ]{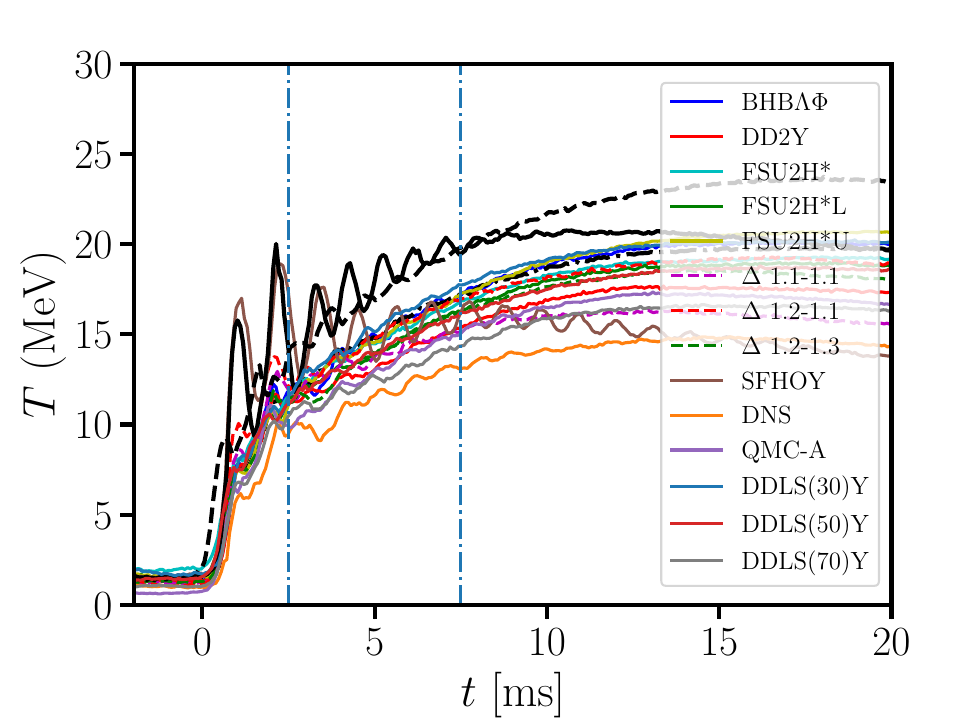} 
\caption{Average temperature of the remnant as a function of time for all hyperonic models. For comparison, three nucleonic models (DD2 - dotted line, SFHO - solid line, FSU2R - dashed line) are also shown with black lines. The vertical lines show the characteristic time window that we use to compute $\bar{\Gamma}_{\mathrm{th}}$ and $\bar{T}$}.
\label{fig:11}
\end{figure}

\begin{figure}[h]
\centering \includegraphics[height = 2.2 in ]{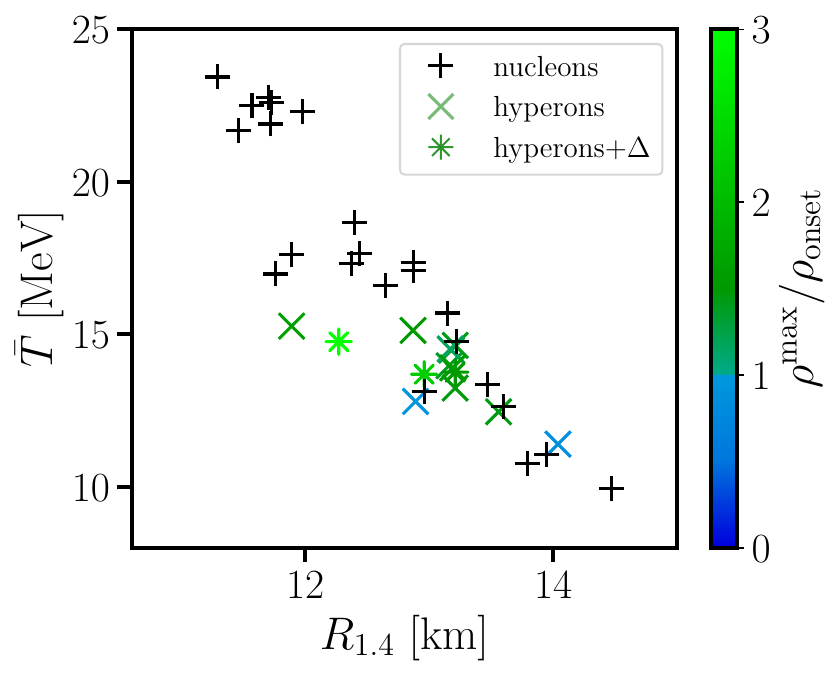} 
\caption{Mass and time averaged temperature inside of the remnant 5 ms after merging as a function of the radius of a $1.4M_{\odot}$ NS.}
\label{fig:avgT_R14}
\end{figure}

Also, we show the mass-averaged temperature evolution of the remnant in Fig.~\ref{fig:11}. After some initial oscillations, the temperature in the remnant becomes roughly constant. 
It is interesting to note that there is a strong model dependence of the predicted temperature inside the remnant.
Stiffer EoSs tend to yield colder remnants (see e.g.~\cite{PhysRevD.108.083029}), which also approximately holds for the set of hyperonic EoSs we are investigating. However, the presence of hyperons has a significant impact on the average temperature of the remnant, which is significantly reduced for hyperonic models as compared to nucleonic models (compare black curves of the nucleonic models in Fig.~\ref{fig:11} to the colored lines for hyperonic models). This becomes even more obvious upon comparing pairs of models with and without hyperons, which are based on the same nucleonic interaction, e.g. DD2Y family versus DD2, SFHOY versus SFHO, and FSU2H* family versus FSU2R. Although hyperonic EoSs are softer, the average temperatures in the simulations with these models are lower than those of the purely nucleonic models. As visible in Fig.~\ref{fig:4}, the specific heat is significantly larger for hyperonic matter than for purely nucleonic matter. Therefore, for a given amount of thermal energy, the temperature increase of matter is lower in the hyperonic  models compared to the nucleonic ones. See also Refs.~\cite{PhysRevD.108.083029,Fields:2023bhs} for an analysis of models considering only nucleonic degrees of freedom.

Finally, we quantitatively assess the relation between EoS stiffness and the average temperature in Fig.~\ref{fig:avgT_R14}. We extract the mass and time averaged temperature in the remnant 5 ms after merging in a time window of 2.5 ms. We plot the average temperature as function of the radius of a $1.4M_{\odot}$ star as proxy for the EoS stiffness. Fig.~\ref{fig:avgT_R14} illustrates that the average temperature roughly anticorrelates with the NS radius and that hyperonic models yield temperatures which are reduced compared to purely nucleonic models with the same radius, i.e.~EoS stiffness.

\subsection{Mass ejection}

\begin{figure*}[htbp]
\centering
\subfigure[]{\includegraphics[ width=0.45\textwidth ]{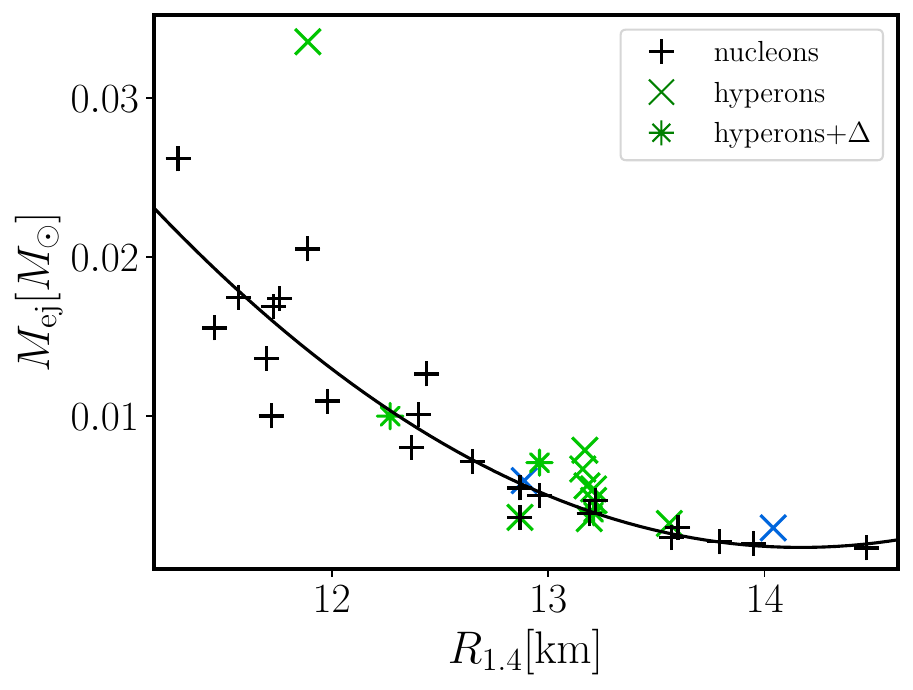}}
\subfigure[]{\includegraphics[  width=0.45\textwidth ]{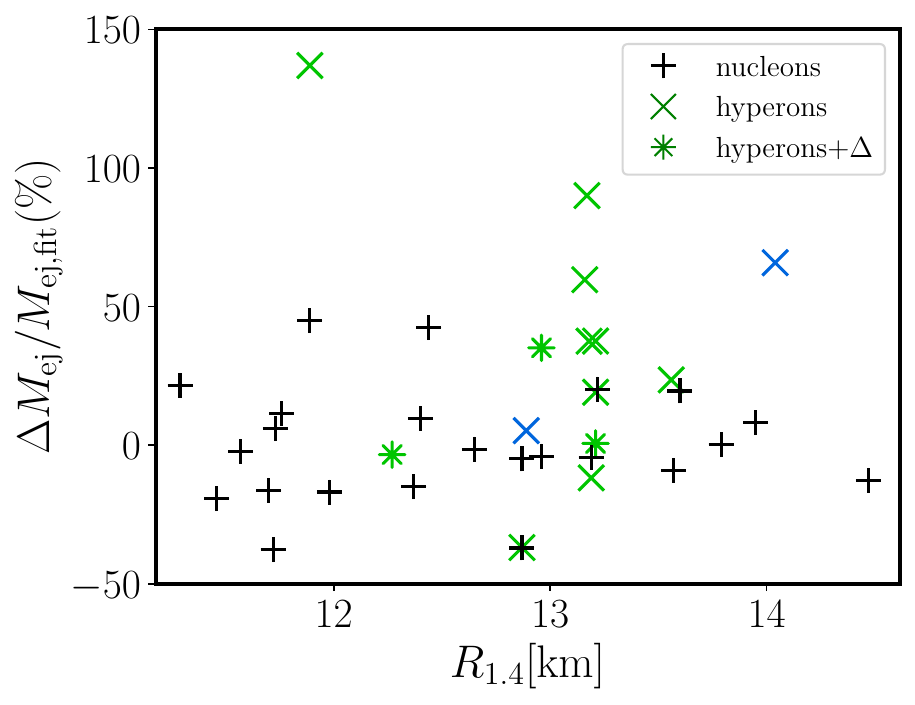}}

\caption{Ejecta mass $M_{\mathrm{ej}}$ as a function of the radius $R_{1.4}$ of 1.4\,$M_{\odot}$ NS (left panel). Crosses display results obtained for 
models with hyperons and asterisks mark models that additionally include $\Delta$ resonances. The coloring indicates  whether the heavy baryons represent more than 1\% of the baryon population (green points) or not (blue points). Nucleonic models are labeled with pluses. The black curve shows a parabolic fit to the purely nucleonic models. Right panel displays the relative difference between $M_{\mathrm{ej}}$ and a parabolic fit $M_{\mathrm{ej}}^\mathrm{fit}(R_{1.4})$ to the purely nucleonic models.} 
\label{fig:mass-ejecta}
\end{figure*}

We next analyze the mass ejection of mergers focusing on a potential difference between hyperonic and purely nucleonic models. Unbound matter from NS mergers undergoes the rapid-neutron capture process forming heavy elements and producing an electromagnetic transient powered by the radioactive decays~\cite{Fernandez2016,Baiotti2017-vm, ShibataMasura2019, Metzger2019, RadiceBernuzzi2020, NAKAR20201, RevModPhys.93.015002,Rosswog:2022tus, Janka:2022krt, Raffaella2021, Burns2020, Sarin2021}. The ejecta mass is of particular interest since it determines the total amount of synthesized elements and the brightness of the electromagnetic emission in the optical and infrared. In Fig.~\ref{fig:mass-ejecta} we report the ejecta mass of simulations of the 1.4-1.4~$M_\odot$ binaries. The amount of unbound material is determined 10~ms after merging noting that a significant amount of ejecta is produced on longer time scales of up to seconds, which we cannot capture by our relatively short simulations. Nonetheless, the early, so-called dynamical, ejecta may be most affected by the presence of hyperons. 

In Fig.~\ref{fig:mass-ejecta} { (left panel)} we show the ejecta mass as function of the radius $R_{1.4}$ of a 1.4~$M_\odot$ NS as an approximate measure for the EoS stiffness. We employ the same notation as in previous plots with the coloring indicating the number fraction of heavy baryons (green data points denoting models with more than one per cent; blue symbols for models with less than one per cent; black plus signs for purely nucleonic EoSs).
There is a clear trend for softer EoSs, i.e. EoSs with smaller radii, to produce significantly higher ejecta masses (cf.~\cite{Hotokezaka2013,Bauswein2013}). Most hyperonic models follow the behavior of purely nucleonic simulations. Some of the hyperonic models, however, yield ejecta masses which are up to a factor two increased compared to nucleonic models with the same $R_{1.4}$  (see right panel in Fig.~\ref{fig:mass-ejecta}). Such a strong increase represents a very significant effect and may be used to discriminate nucleonic and hyperonic models in future measurements. This requires to better understand mass ejection and kilonova emission in general, which are still subject to various uncertainties, and to determine the dependence of the ejecta increase on the hyperon content.   The most pronounced increase is found for the simulation with the SFHOY EoS. The likely reason for this strong increase is the fact that the total binary mass for this simulation is very close to the threshold binary for direct black-hole formation (cf. Tab.~\ref{tab:2}).

\section{Simulations of massive binaries}

\label{sec:moremassice}

In the previous section we investigated the correlation between the hyperon content in the remnant and the frequency shift of the dominant postmerger oscillation relative to a fiducial nucleonic model ($\Delta f = f_{\mathrm{peak}} - f_{\mathrm{peak}}^{1.75}$).
Hyperonic EoS models with a very low hyperon abundance in the remnant do not yield any sizable frequency shift; in fact, some simulations even feature a negative shift (see Fig.~\ref{fig:6}).

In simulations with more massive binaries the central density of the remnant will be increased, and hence, a significant amount of hyperons may occur.  
Here, we present the results from simulations of heavier binaries for five hyperonic EoS models, DNS, QMC, FSU2H*, FSU2H*L, FSU2H*U, including those that yield a negative shift for 1.4-1.4~$M_\odot$ mergers. 
We run simulations with total binary masses of either 2.9~$M_\odot$, 3.0~$M_\odot$ or 3.1~$M_\odot$, respectively, choosing for each EoS the configuration which is below but close to the threshold mass for prompt black hole formation.
In Table \ref{tab:1} we collect the results 
including the dominant peak frequency obtained with the 3D EoSs and the one resulting from an the approximate thermal treatment, as well as the mass and time averaged thermal index  $\bar{\Gamma}_{\mathrm{th}}$. 
For comparison, we also provide the results for $1.4 M_{\odot} - 1.4 M_{\odot}$ simulations.

\begin{figure*}[htbp]
    \centering
    \subfigure[]{
        \includegraphics[height = 2.5 in,  width=0.45\textwidth]{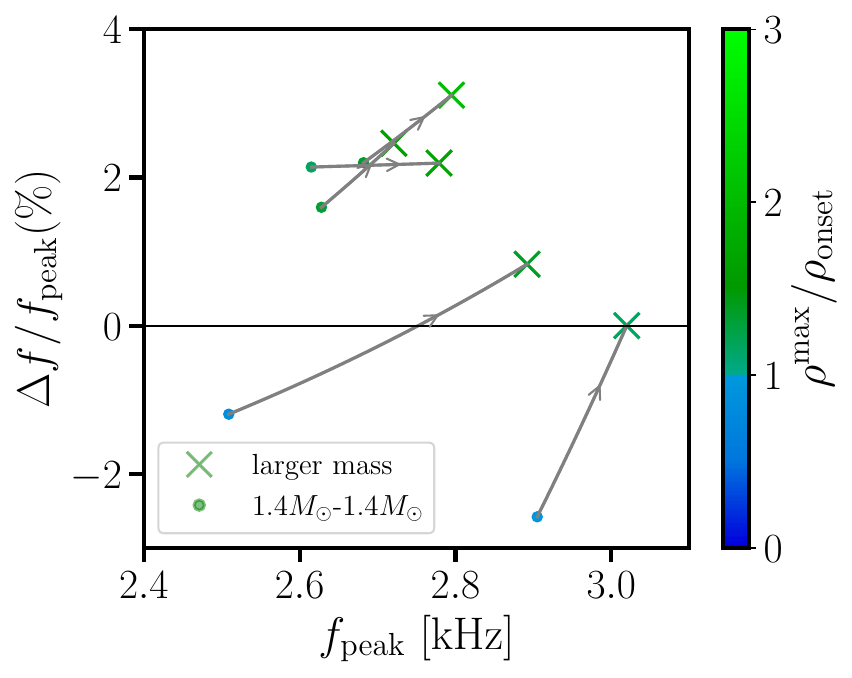}
    }
    \hspace{0.05\textwidth}
    \subfigure[]{
        \includegraphics[height = 2.5 in,  width=0.45\textwidth]{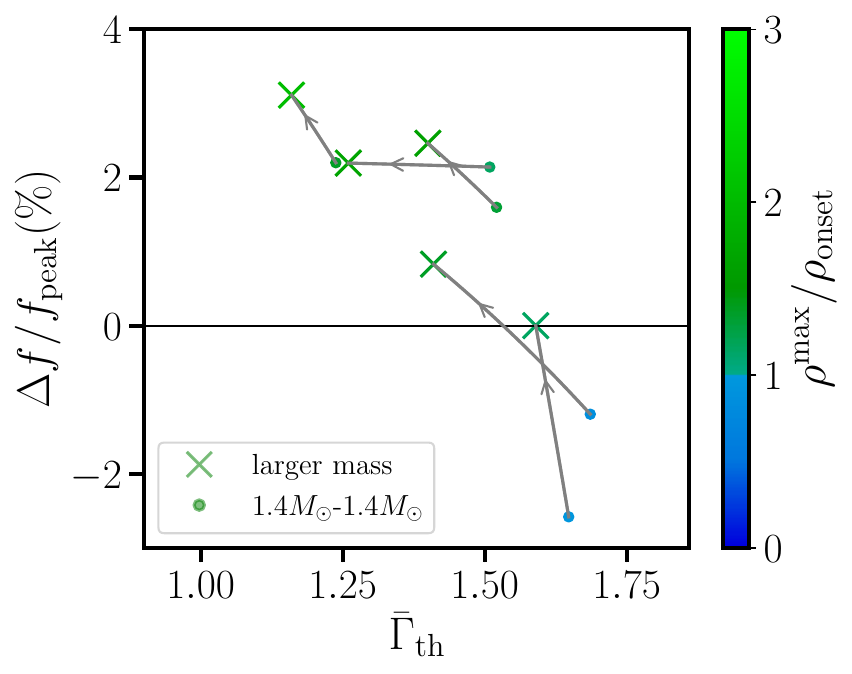}
    }
    \caption{ Relative frequency shift $\Delta f/f_{\mathrm{peak}}$ as a function of the peak frequency $f_{\mathrm{peak}}$ (left panel) and the average thermal index of the star $\bar{\Gamma}_\mathrm{th}$  (right panel). 
    For each EoS model, the gray line connects the results for the different total binary masses. The arrow indicates the direction towards larger masses.}
    \label{fig:extra-2}
\end{figure*}

The relative shift of the dominant postmerger frequency is shown in Fig.~\ref{fig:extra-2} as a function of frequency (left panel) and the averaged thermal index (right panel). We include the 1.4-1.4~$M_\odot$ data and the results from the more massive binaries. For each hyperonic EoS model, a gray line connects the data points for 1.4-1.4~$M_\odot$ mergers (dots) and more massive binaries (crosses). 
More massive binaries yield higher postmerger gravitational-wave frequencies.
As anticipated, the value of the shift $\Delta f = f_{\mathrm{peak}} - f_{\mathrm{peak}}^{1.75}$ increases for more massive binaries. This effect is most significant for the EoS models which yield a negative shift for 1.4-1.4~$M_\odot$ mergers, albeit the absolute values of the shift still do not become very large for these cases. These EoSs produce remnants that have a relatively low amount of hyperons. This directly reflects the average thermal index which is also not significantly lowered.  
For EoSs that feature a positive shift for 1.4-1.4~$M_\odot$ binaries, the change is smaller keeping in mind that the increase in the total binary mass differs among the models. In all cases the shift correlates with the average thermal index in the remnant (right panel in Fig.~\ref{fig:extra-2}).

\section{Simulations of asymmetric binaries}
\label{sec:asymmetric}

In this section, we extend our study to asymmetric binaries with a mass ratio $q=M_1/M_2=0.8$ and a total binary mass of $M_\mathrm{tot}=M_1+M_2=2.8~M_\odot$.
To this end, we performed an analogous simulation campaign as the one described in Section~\ref{sec:Simulations}, namely, a set of simulations with the full 3D EoS tables and another set employing the cold $\beta$ equilibrium slice supplemented with the $\Gamma_{\mathrm{th}} = 1.75$ treatment.

\begin{figure*}[htbp]
    \centering
    \subfigure[]{
        \includegraphics[height = 2.5 in,  width=0.45\textwidth]{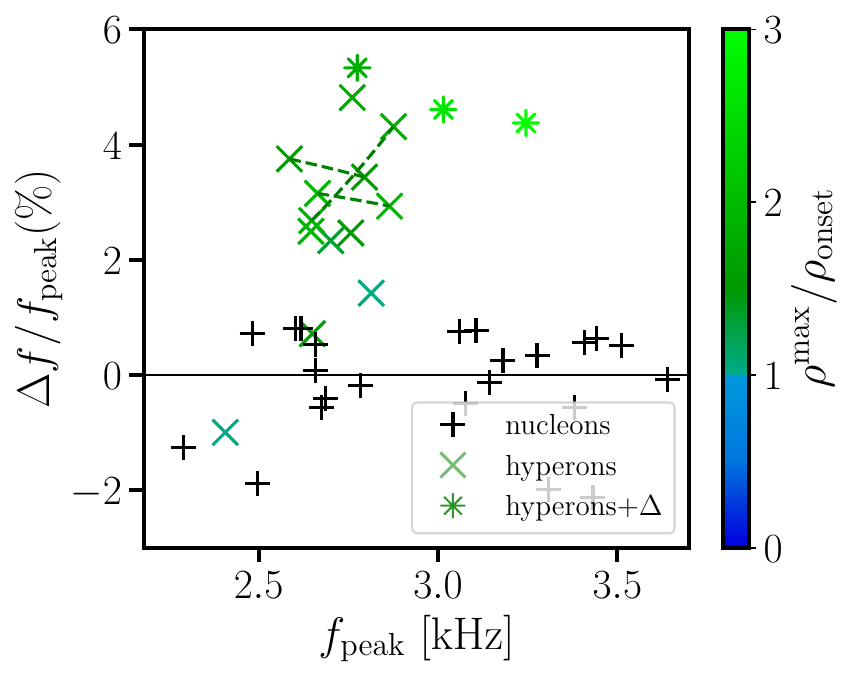}
    }
    \hspace{0.05\textwidth}
    \subfigure[]{
        \includegraphics[height = 2.5 in,  width=0.45\textwidth]{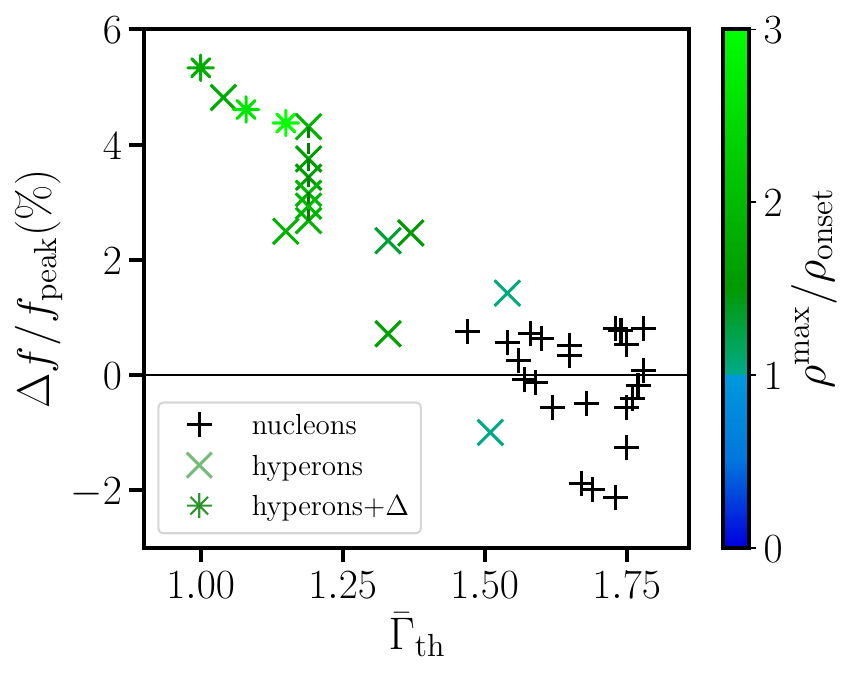}
    }
    \caption{Relative frequency shift $\Delta f/f_{\mathrm{peak}}$ as a function of the peak frequency $f_{\mathrm{peak}}$ (left panel) and the average thermal index of the star $\bar{\Gamma}_\mathrm{th}$ (right panel) for mergers with a total mass of $2.8 M_{\odot}$ and binary mass asymmetry of $q=0.8$. The symbols and the coloring have the same meaning as in Fig.~\ref{fig:extra-2}. For a small number of models two dominant peaks occur in the GW spectrum, which are roughly equally strong. In these cases the frequencies of both peaks are displayed and connected with a dashed line.}
    \label{fig:extra-8}
\end{figure*}

The results from the simulations are given in Table \ref{tab:2}, while Fig.~\ref{fig:extra-8} shows the relative frequency shift $\Delta f/f_{\mathrm{peak}}$ as a function of both the peak frequency $f_{\mathrm{peak}}$ (left panel) and the average thermal index of the remnant $\bar{\Gamma}_\mathrm{th}$ (right panel).  
The frequency shift, which is characteristic of EoSs with exotic baryons, is clearly visible also for asymmetric binaries. For some of the EoSs with hyperons and $\Delta$ baryons, the relative shift in asymmetric mergers is even more enhanced, reaching values up to 6\% as compared to at most 4\% for equal-mass systems.  Note that only 13 different EoS data points are displayed in Fig.~\ref{fig:extra-8}, although we consider 14 EoS models with heavy baryons in our calculations. The reason is that, for the softest EoS model (SFHOY), the remnant undergoes a prompt collapse in the simulation with the full 3D EoS table. In the corresponding simulation with $\Gamma_{\mathrm{th}} = 1.75$ the central object does not immediately form a black hole. This different qualitative behavior is another clear signature of the effects that thermal hyperons have on the remnant. The influence of the thermal behavior of EoSs with heavy baryons on the threshold mass will be discussed in Section \ref{sec:threshold}. In addition, some of the EoSs  (hyperonic models from~\cite{tsiopelas2024finitetemperature}) produce two  near-by dominant peaks with a comparable strength (those points are connected with a green dashed line; see Tab.~\ref{tab:2}). The exact mechanism that produces these peaks is still unknown.

In Fig.~\ref{fig:extra-8}, it is also noticeable that for asymmetric binaries nucleonic models show slightly smaller deviations from the reference value of $\Delta f = 0$ in the positive direction compared to the data for equal-mass systems. Models with heavy baryons are thus potentially easier to distinguish, which somewhat improves the prospects to identify exotic degrees of freedom in an observation of an asymmetric merger.

Some of the models with hyperons  do not stick out significantly with respect to the nucleonic models. These simulations feature a higher average thermal index in the remnant, as can be seen from panel b) in Fig.~\ref{fig:extra-8}. As for equal-mass binaries we find a correlation between the relative frequency shift and the average thermal index, and small frequency shifts occur for models where the density in the remnant does not significantly exceed the onset density of hyperon production at zero temperature. 

\begin{figure}[h]
\centering \includegraphics[height = 2.2 in ]{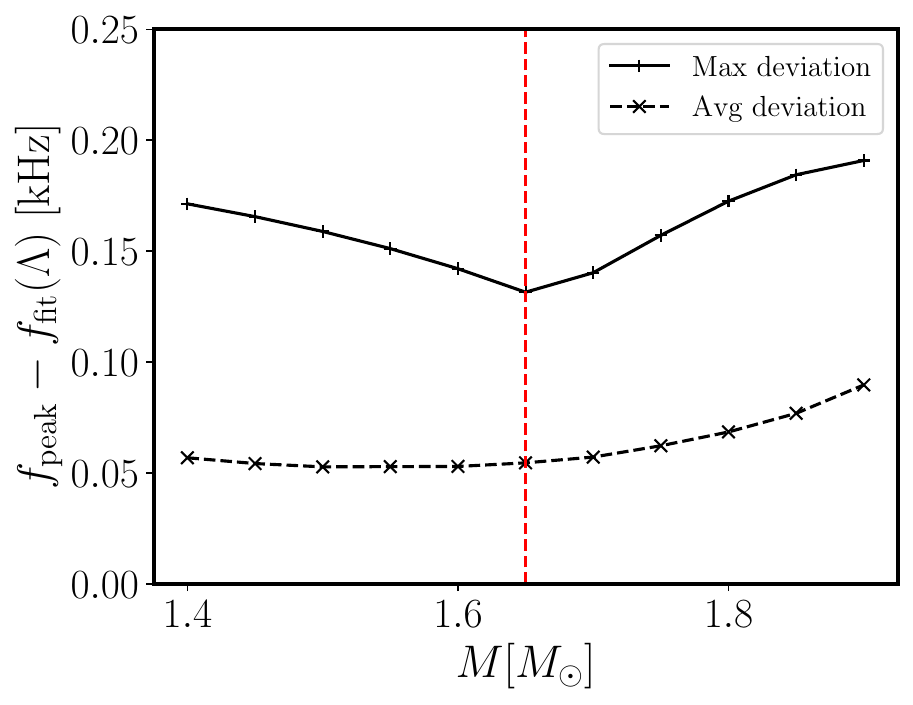} 
\caption{Average and maximum deviation of our data for purely nucleonic EoSs from quadratic least-squares fits $f_\mathrm{peak}-\Lambda_M$ for binary mass ratio $q=0.8$. The fits have been performed for different reference masses $M$ at which $\Lambda$ is evaluated ($\Lambda_M=\Lambda(M)$). The dashed vertical line indicates at which $M$ the maximum deviation reaches a minimum.}
\label{fig:extra-9}
\end{figure}

\begin{figure}[h]
\includegraphics[width=0.9\linewidth]{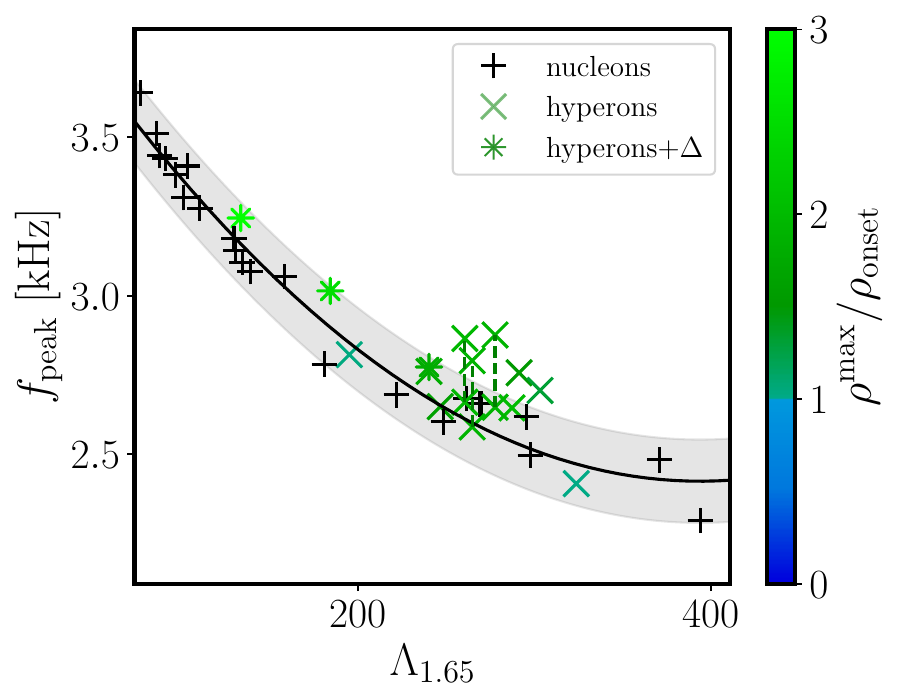}
\caption{Dominant postmerger GW frequency as function of tidal deformability of a 1.65~$M_\odot$ NS for asymmetric mergers with a binary mass ratio $q=0.8$. Crosses display results obtained for 
models with hyperons, where the coloring indicates the ratio between the
maximum rest-mass density in the postmerger remnant and
the onset rest-mass density of hyperon production at zero
temperature. Models that also include $\Delta$-baryons are labeled with asterisks,  whereas the nucleonic models are shown by plus signs. For a small number of models two dominant peaks occur in the GW spectrum, which are roughly equally strong. In these cases the frequencies of both peaks are displayed and connected with a dashed line. Black curve is a least-squares quadratic fit to purely nucleonic models. Gray band indicates maximum residual of purely nucleonic models from the fit.}
\label{fig:extra-10}
\end{figure}

As for the equal-mass binaries in subsect.~\ref{subsec:fpeak}, we explore direct relations between the tidal deformability and the dominant postmerger GW frequency from the simulations with 3D EoS tables for asymmetric mergers. From Fig.~\ref{fig:extra-9} we read off that the reference mass with the small maximum residual between a quadratic fit and the data occurs for $M=1.65~M_\odot$. We show the corresponding $f_\mathrm{peak}-\Lambda_\mathrm{M}$ relation in Fig.~\ref{fig:extra-10}, which exhibits a frequency increase of hyperonic models compared to purely nucleonic models also for asymmetric binaries similar to the results for equal-mass systems. Considering the findings for $q=1$ and $q=0.8$, we expect that a similar behavior will be found for binaries with other mass ratios as well. 

\section{Threshold mass for prompt collapse}
\label{sec:threshold}
All simulations discussed in the previous sections lead to the formation of a NS remnant. If one increases the initial mass of the system, the lifetime of the remnant, i.e. the time until black hole formation, gradually decreases.
Considering a set of simulations with different total binary masses for a given EoS model, one can identify the highest total mass $M_{\mathrm{tot}} = M_{\mathrm{stable}}$ for which the remnant does not collapse promptly to a black hole (on a timescale of about a millisecond after the merging). Similarly, one can find the lowest mass $M_{\mathrm{tot}} =M_{\mathrm{unstable}} $ for which it does directly form a black hole. This determines $M_{\mathrm{thres}} = (M_{\mathrm{stable}}+M_{\mathrm{unstable}})/2$ as the threshold binary mass for prompt black hole formation\footnote{Within this study we determine $M_\mathrm{thres}$ to within $\pm 0.01 M_{\odot}$ by the simulations, which, however, does not take into the account actual uncertainties of the calculations.}. $M_\mathrm{thres}$ is very sensitive to the EoS, and a determination of $M_\mathrm{thres}$ would put tight constraints on the NS properties such as tidal deformabilities, radii and maximum mass~\cite{PhysRevLett.111.131101,Bauswein2017,10.1093/mnras/stx1983, Köppel_2019,PhysRevD.103.123004,Agathos2020,Tootle_2021,Koelsch2022,Kashyap2022}. We refer the reader to Refs.~\cite{Bauswein2017,Sneppen2024} for applications employing the electromagnetic emission of GW170817 and its measured total binary mass to constrain $M_\mathrm{thres}$. 

The detailed investigation of the EoS dependence of $M_\mathrm{thres}$ in~\cite{PhysRevD.103.123004} also includes a few hyperonic models. The study has shown that a softening at higher densities reduces $M_\mathrm{thres}$ but increases $\Lambda_\mathrm{thres}$ being defined as $\Lambda(M_\mathrm{thres}/2)$ for equal-mass binaries. While a softening of the EoS at higher densities is typical for the appearance of hyperons, such an effect may not be unambiguously related to hyperons but could in principle also occur for purely nucleonic matter as discussed above. In fact, the small sample of hyperonic models considered in~\cite{PhysRevD.103.123004} does not prominently deviate from empirical relations connecting $M_\mathrm{thres}$ and stellar parameters of cold NSs such as $M_\mathrm{max}$, NS radii and the tidal deformability. One example is a bi-linear relation $M_{\mathrm{thres}}(M_\mathrm{max},R_{1.6}) = aM_{\mathrm{max}} + bR_{1.6}+c$ with the maximum mass of a cold NS $M_{\mathrm{max}}$, the radius $R_{1.6}$ of a $1.6M_{\odot}$ NS and fit parameters $a$, $b$ and $c$~\cite{PhysRevD.103.123004}.
We revisit this type of empirical relations using our larger set of hyperonic EoS models and investigate in particular whether the characteristic thermal behavior of hyperonic EoSs may lead to specific features that discriminate hyperonic models from nucleonic ones through $M_\mathrm{thres}$.

We therefore perform the two types of simulations that are already described in Section \ref{sec:Simulations}, namely using the full temperature-dependent EoS and the cold EoS supplemented with $\Gamma_{\mathrm{th}} = 1.75$. The latter is meant to mimic the thermal properties of purely nucleonic matter. Employing these approaches we determine $M_\mathrm{thres}$ for all nucleonic and hyperonic EoSs. For simplicity, we consider only equal-mass binaries ($q=1$).

\begin{figure}[h]
\centering \includegraphics[height = 2.2 in ]{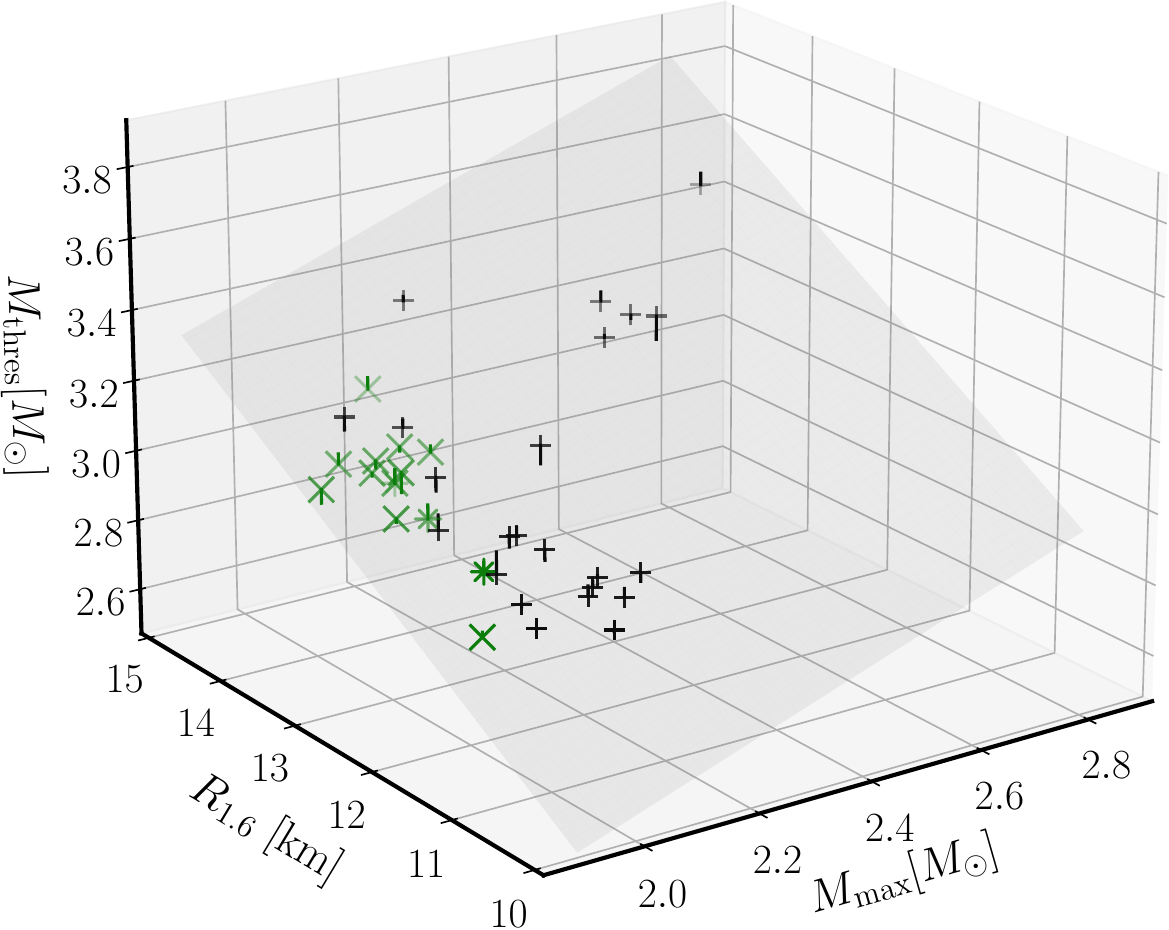} 
\caption{Threshold binary mass for prompt black hole formation, $M_{\mathrm{thres}}$, as function of $M_{\mathrm{max}}$ and $R_{1.6}$. Black plus signs show purely nucleonic models,  while models that also consider hyperons (and delta resonances) are labeled by green crosses (asterisks). The gray plane visualizes a bi-linear fit of the nucleonic models. The vertical lines at each data point indicate the deviation of the data point from the bi-linear fit (for most models those lines are so short that they are hardly visible).}
\label{fig:extra-3}
\end{figure}

In Fig. \ref{fig:extra-3} we show $M_{\mathrm{thres}}$ as a function of $M_{\mathrm{max}}$ and $R_{1.6}$ using the results from the calculations with the full 3D EoS tables. Green crosses display results from EoSs that consider hyperons, while black plus signs correspond to EoSs which only consider nucleons as degrees of freedom. The figure indicates that it is not possible to distinguish between the two classes of EoSs, confirming the conclusion that these empirical relations hold for all type of hadronic EoSs. In the figure, the gray plane shows a bi-linear fit $M_{\mathrm{thres}}(M_\mathrm{max},R_{1.6}) = aM_{\mathrm{max}} + bR_{1.6}+c$ considering only the nucleonic models. The average deviation of data points for nucleonic EoSs from the fit is $ 0.03 M_{\odot}$ and the maximum residual amounts to $0.07 M_{\odot}$. These numbers are rather similar for the hyperonic models, which were not used in the fitting procedure. For those EoSs, the average deviation is $0.03 M_{\odot}$, while the maximum deviation is only marginally lower ($0.06 M_{\odot}$).

\begin{figure}[h]
\centering \includegraphics[height = 2.2 in ]{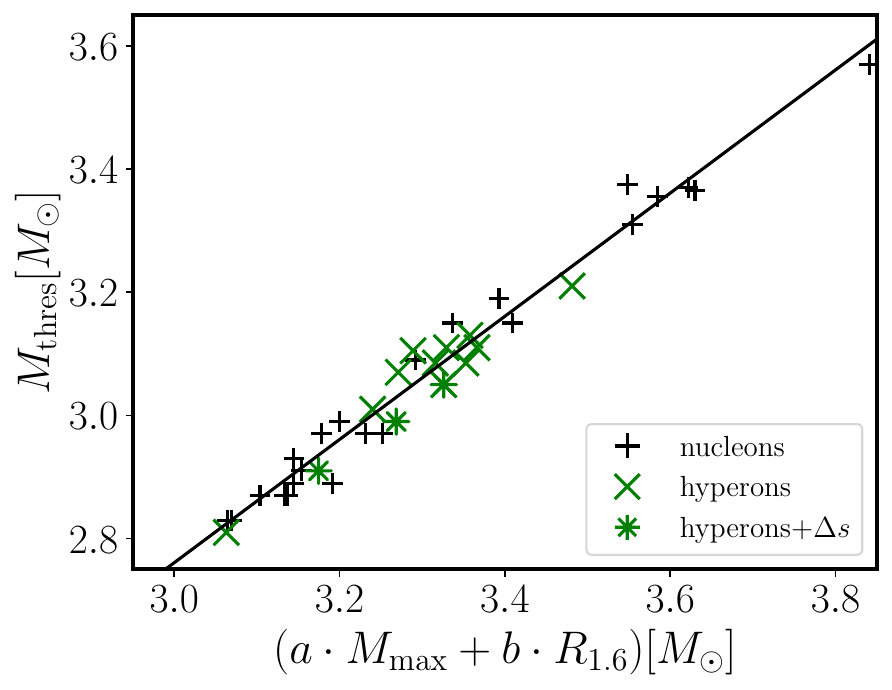} 
\caption{$M_{\mathrm{thres}}$ as function of $aM_{\mathrm{max}}+bR_{1.6}$, where coefficients $a$ and $b$ are obtained using the bi-linear fit shown in Fig. \ref{fig:extra-3}. Purely nucleonic models are shown by plus signs, while models that also consider hyperons (and delta resonances) are labeled by green crosses (asterisks). The straight line is the projection of the plane, i.e. the bi-linear fit, shown in \ref{fig:extra-3}.}
\label{fig:extra-4}
\end{figure}

The indistinguishability of the two classes of EoS models can be seen even better in Fig.~\ref{fig:extra-4} by a two dimensional plot showing $M_{\mathrm{thres}}$ as a function of the combination $ aM_{\mathrm{max}} + bR_{1.6}$, where $a$ and $b$ are the coefficients of the bi-linear fit. The data points for EoSs which include heavy baryons, scatter around the straight line. Similar results are obtained if other pairs of independent variables are used in a respective fit such as ($M_{\mathrm{max}}$, $R_{1.4}$) or ($M_{\mathrm{max}}$, $\Lambda_{1.4}$). 
Hence, we conclude that empirical bi-linear relations for $M_\mathrm{thres}$ hold to a high degree even for models which include heavy baryons, with the downside that these relations cannot be employed to infer the presence of hyperons in NSs.

\begin{figure}[h]
\centering \includegraphics[height = 2.2 in ]{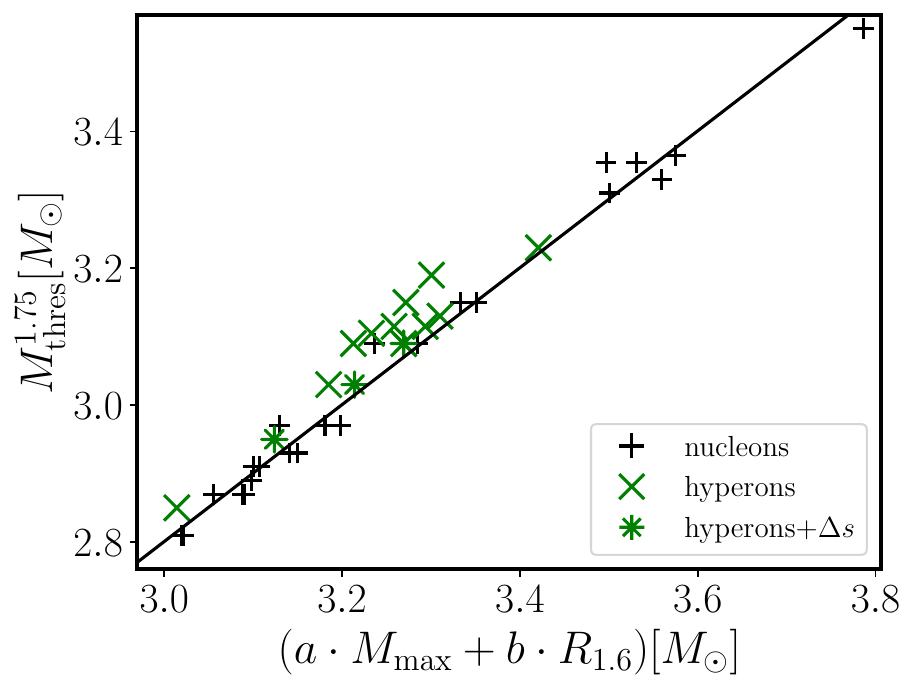} 
\caption{Same as Fig.~\ref{fig:extra-4} but for $M_{\mathrm{\mathrm{thres}}}^{1.75}$, i.e. data obtained from the simulations employing an approximate thermal treatment with $\Gamma_\mathrm{th}=1.75$ in combination with nucleonic and hyperonic models at $T=0$. Straight line visualizes the fit to the data considering only the nucleonic models with the $\Gamma_\mathrm{th}$ approach.}
\label{fig:extra-6}
\end{figure}

One may wonder why the models with heavy baryons do not stick out more clearly in these bi-linear relations considering that at higher densities their thermal index is strongly reduced in comparison to purely nucleonic EoSs (cf. the effects on $f_\mathrm{peak}$ discussed in the previous sections). In this regard, it is instructive to consider the set of simulations with the approximate thermal treatment adopting $\Gamma_\mathrm{th}=1.75$, i.e.~the same thermal properties. Figure~\ref{fig:extra-6} illustrates the deviations from a bi-linear fit to the nucleonic models similar to Fig.~\ref{fig:extra-4} but for the $\Gamma_\mathrm{th}=1.75$ data. In this figure in comparison to Fig.~\ref{fig:extra-4} some hyperonic models with the $\Gamma_\mathrm{th}$ treatment do show some small deviations from the bi-linear fit with a slightly higher threshold mass although they assume the same thermal behavior. 
We interpret this finding as follows. Inspecting the stellar properties of cold NSs within our sample of EoSs, we realize that the hyperonic models and nucleonic models do not exactly cover the same ranges in the parameter space, as seen in Fig.~\ref{fig:extra-3}. Hyperonic EoSs tend to have relatively large $R_{1.6}$ for a given $M_\mathrm{max}$\footnote{In this regard we remark that in order for models with heavy baryon models to be realistic and to reach sufficiently high maximum masses of cold NSs, a certain stiffness of the low and intermediate density region is needed. Therefore, these models tend to yield relatively large NS radii in the intermediate mass range although the maximum mass only slightly exceeds two solar masses.}. If the hyperonic models were purely nucleonic models with the same stellar parameters (as assumed in Fig.~\ref{fig:extra-6} by our $\Gamma_\mathrm{th}$ approach), we would conclude that the relation $M_\mathrm{thres}(M_\mathrm{max},R_{1.6})$ is not exactly bi-linear. Instead, in the area of the parameter range with small $M_\mathrm{max}$ and relatively large $R_{1.6}$, the threshold mass $M_\mathrm{thres}$ is slightly increased with respect to a bi-linear relation. Within our EoS sample, this area is not well covered by actual purely nucleonic models but mostly by hyperonic EoSs. These hyperonic models, however, yield a slightly lower $M_\mathrm{thres}$ because of their thermal properties, which effectively leads to the bi-linear behavior of $M_\mathrm{thres}(M_\mathrm{max},R_{1.6})$ if we use the full 3D EoS tables.

These considerations also suggest to revisit the exact behavior of $M_\mathrm{thres}(M_\mathrm{max},R_{1.6})$ for purely nucleonic EoSs, which is currently beyond the scope with the available sample of purely nucleonic models especially those with small $M_\mathrm{max}$ and relatively large $R_{1.6}$. Figure~\ref{fig:extra-6} indicates that bi-linear fits may not be the optimal relations to describe the data.

\begin{figure}[h]
\centering \includegraphics[height = 2.2 in ]{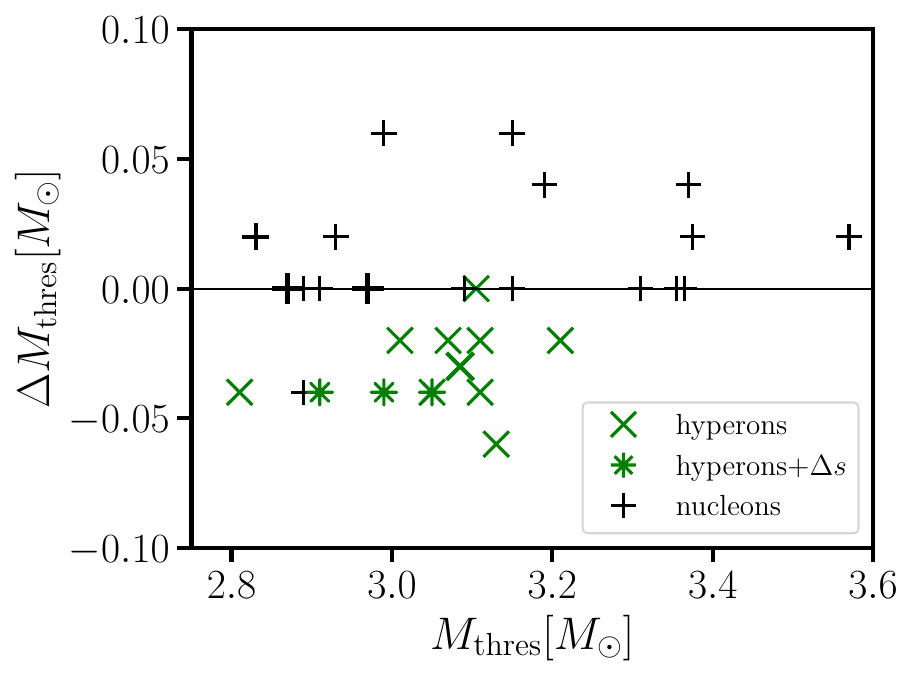} 
\caption{$\Delta M_{\mathrm{thres}}=M_\mathrm{thres}-M_{\mathrm{\mathrm{thres}}}^{1.75}$ as function of $M_{\mathrm{thres}}$. Black plus signs display nucleonic models, while models that also consider hyperons (and delta resonances) are labeled by green crosses (asterisks). If two or more nucleonic models produce equal $\Delta M_{\mathrm{thres}}$ and $M_{\mathrm{thres}}$, the corresponding data point is labeled with a thicker symbol. Hyperonic models that overlap are labeled with a thicker green cross.}
\label{fig:extra-7}
\end{figure}

The promising aspect of this intricacy is, however, the prospect that $M_\mathrm{thres}$ may reveal the presence of heavy baryons if measured well enough. In Fig.~\ref{fig:extra-7} we compare the threshold mass resulting from the simulations with the fully temperature dependent EoS tables and with the approximate thermal treatment. Hyperons can reduce $M_\mathrm{thres}$ by about 0.05~$M_\odot$ in comparison to the threshold mass one would expect for a purely nucleonic system with the same stellar parameters of cold NSs.\footnote{One may also conclude from Fig.~\ref{fig:extra-7} that choosing $\Gamma_\mathrm{1.75}$ for purely nucleonic EoSs to determine $M_\mathrm{thres}$ may yield a slight underestimation of the threshold mass.} Note that inferring the occurrence of hyperons would require a well measured $M_\mathrm{thres}$ as well as precise knowledge of $M_\mathrm{max}$ and $R_{1.6}$ to compare the measured threshold mass with the one expected for nucleonic EoSs.  Also, the exact numerical determination of $M_\mathrm{thres}$ depends on its precise definition as well as the physical model, numerical scheme and resolution~\cite{Kashyap2022,Koelsch2022}.

%% file: sections/section04.tex
\section{Conclusions} \label{sec:conclusions}
In this work we extend our previous investigation of the effects of hyperons in BNS mergers~\cite{Blacker:2023opp}. To this end, we consider a large sample of 14 temperature dependent hyperonic EoSs developed within different phenomenological approaches, which we systematically compare to a set of 23 purely nucleonic models. A subset of the hyperonic EoS models also includes Delta resonances.

Analyzing the EoS tables, we discuss the changes that the presence of hyperons at finite temperature induce in the thermal energy density and thermal pressure as well as the specific heat at constant pressure, for conditions that are relevant in BNSs. The appearance of hyperons strongly influences all thermodynamic quantities. In particular, the thermal pressure is significantly reduced, showing a characteristic drop. We find this pressure drop to be directly related to the excess amount of hyperons that are produced due to  thermal excitations. As a consequence, the thermal index of hyperonic EoSs shows a strong density dependence. On average, the thermal index of hyperonic models is much lower than that of nucleonic EoSs. The appearance of hyperons also modifies the thermal energy density and the specific heat, which both are systematically larger than the ones of nucleonic models.

For all EoS models we run relativistic hydrodynamics simulations of BNS mergers focusing first on symmetric binaries with a total mass of 2.8~$M_\odot$. 
Considering the characteristic impact of heavy baryons on the thermal behavior of the EoS, we analyze in great detail the dominant postmerger GW frequency, which, in contrast to GW inspiral parameters like the tidal deformability, is influenced by finite-temperature effects. As a general goal of this analysis we intend to discriminate purely nucleonic and hyperonic EoS models by means of their thermal properties noting that the stellar parameters of cold NSs may not easily allow a distinction since mass-radius relations resulting from purely nucleonic and hyperonic models, respectively, can look very similar. Following~\cite{Blacker:2023opp} we quantify the impact of the thermal EoS properties on the dominant GW frequency by running an additional simulation for each hyperonic EoS model where we supplement the $T=0$ EoS slice with an approximate thermal treatment which mimics the behavior of purely nucleonic matter. Comparing the resulting postmerger GW frequency to that of the original simulation with the fully temperature dependent EoS table we quantify the frequency shift due to the hyperonic thermal behavior compared to an idealized nucleonic behavior. For hyperonic models the GW postmerger frequency is characteristically increased by a few per cent because of the reduced thermal pressure, which can be in principle used to discern hyperonic EoSs from purely nucleonic models (see also~\cite{Blacker:2023opp}).

We corroborate this finding in several regards. (i) Compared to~\cite{Blacker:2023opp} we include a larger sample of hyperonic and purely nucleonic models, which all follow this behavior. (ii) Simulating BNS mergers with a binary mass ratio $q=0.8$ confirms that this effect also occurs in asymmetric binaries. (iii) In simulations with a higher total binary mass, we find an increased frequency shift. This is understandable because higher densities in the remnant lead to more hyperons and an overall reduced average thermal index. In line with this explanation, models with a high onset density of hyperon production which yield negative or zero shifts for binaries with $M_\mathrm{tot}=2.8~M_\odot$, lead to positive frequency shifts for higher binary masses. (iv) We run merger simulations with an independent relativistic hydrodynamics code based on a moving-mesh technique and observe the same effects of hyperonic models.  Future work should also consider the potential influence of neutrinos, magnetic fields and full general relativity.

We also confirm for asymmetric binaries with $q=0.8$ that relating the dominant postmerger GW frequency to the tidal deformability of massive stars (more massive than the inspiralling stars) yields particularly tight relations if one considers purely nucleonic EoSs. Hyperonic models tend to slightly stick out in these relations as a feature that potentially allows to identify the occurrence of heavy baryons through observations. For equal-mass binaries this has already been recognized in~\cite{Blacker:2023opp}.

Analyzing the detailed conditions in the merger remnant, we find that hyperonic models lead to increased maximum densities compared to nucleonic models. A tight relation between the dominant postmerger GW frequency and the maximum density during the early remnant evolution holds for purely nucleonic as well as hyperonic models. The thermal index averaged over the fluid elements of the remnant is significantly reduced if hyperons occur, which explains the characteristic GW frequency increase. Simulations with hyperonic EoSs feature systematically lower average temperatures in the remnant in line with the larger heat capacity of hyperonic matter. 

We investigate subdominant spectral features of the postmerger GW spectra~\cite{PhysRevD.91.124056} and define a frequency shift for secondary GW frequencies analogous to the shift of the main peak. As for the dominant peak, this shift quantifies the impact of only the thermal EoS properties but does not account for changes of the cold EoS by hyperonic degrees of freedom. Interestingly, the two secondary features which we extracted behave qualitatively different. The spectral feature $f_{2-0}$, which results from a coupling of the quasi-radial and the quadrupolar fluid oscillation mode of the remnant, shows a frequency shift of up to 8 per cent if hyperons are present. The relative frequency shift is even stronger than that of the dominant GW frequency. In contrast, the $f_\mathrm{spiral}$ feature, originating from orbiting tidal bulges during the early postmerger phase, does not exhibit any sizable differences between purely nucleonic and hyperonic models (except for one outlier). This might be explained by the specific origin of this feature and the fact that it is generated during the early evolution, when the average thermal index in hyperonic models is still more comparable to that of nucleonic systems and fewer hyperons are present (see Fig.~\ref{fig:3}). 

The fact that the various GW spectral features exhibit qualitatively different frequency shifts if hyperons occur could serve as a discriminator between nucleonic and hyperonic systems. However, directly relating GW frequencies, hyperonic models do not strongly stick out because the scatter in such frequency relations is too large. A more detailed understanding of these relations, i.e.~their scatter, could be a promising route to identify a feature that reveals the presence of heavy baryons.

Mergers with hyperonic EoSs yield tentatively more ejecta than nucleonic models with a similar stellar radius. Although not all hyperonic models follow this trend of a strongly increased ejecta mass and we only consider the early ejecta, the comparison between radius (or tidal deformability) and the amount of dynamical, relatively fast ejecta may provide a valuable signature for the occurrence of hyperons as both quantities are accessible in observations.

For our large set of EoSs, we determine the threshold binary mass for prompt black hole formation as another characteristic quantity, which can be measured or constrained in future observations (see~\cite{Sneppen2024} for a discussion of current limits). We find that hyperonic degrees of freedom can lead to a reduction of the threshold mass by about 0.05~$M_\odot$ in comparison to a purely nucleonic model with the same stellar parameters of cold NSs. 

We find that bi-linear relations of the type $M_\mathrm{thres}(M_\mathrm{max},R_{1.6})$ describe the full set of EoS models very well. However, in our set of models purely nucleonic and hyperonic EoSs are not exactly covering the same parameter range with regards to the stellar properties of isolated cold NSs. Considering the reduction of $M_\mathrm{thres}$ by the thermal effects induced by hyperons, this in turn implies that purely nucleonic models may not exactly follow bi-linear relations over the full range but instead yield slightly higher $M_\mathrm{thres}$ in the regime of $M_\mathrm{max}\approx2.0~M_\odot$ and relatively large $R_{1.6}$. In any case these deviations are relatively small and of the same order of the intrinsic scatter in these relations.

These considerations show that even larger sets of temperature-dependent hyperonic and purely nucleonic EoS tables are needed for future studies to fully understand the differences between those models. To our knowledge we employed all currently available temperature dependent hyperonic EoS models (broadly compatible with current NS constraints), but the EoSs considered here are not derived within a single, consistent framework. The slightly different coverage of the parameter space by hyperonic and nucleonic models reflects to a certain extend the principle effect of heavy baryons to soften the cold EoS in comparison to a purely nucleonic system. In this study we did not explicitly consider this effect as discriminator and instead assumed that the stellar properties of cold NSs would not be informative about the presence of heavy baryons. This is clearly not correct and it is an assumption, which we adopted to be conservative and to identify an unambiguous signature of heavy baryons. Thus, future work should also attempt to take into account the effects on the cold EoS slice. We also remark that some of the effects discussed in this study might be ambiguous in the sense that they may be similarly created by other degrees of freedom, such as, for instance, quark matter or even purely nucleonic matter (see e.g.~\cite{Blacker:2024tet,2024arXiv241100939T,PhysRevC.103.055806,Keller2023}).

%% file: sections/acknowledgements.tex
\section*{Acknowledgements} \label{sec:acknowledgements}
We would like to thank Vimal Vijayan and Arnau Rios for useful discussions. We acknowledge funds by the State of Hesse within the Cluster Project ELEMENTS supporting H.K.'s visit during which this project was initialized. 
This research has been supported from the projects CEX2019-000918-M, CEX2020-001058-M (Unidades de Excelencia ``Mar\'{\i}a de Maeztu"), PID2023-147112NB-C21 and PID2022-139427NB-I00 financed by MCIN/AEI/10.13039/501100011033/FEDER, UE, as well as by the EU STRONG-2020 project under the program H2020-INFRAIA-2018-1 grant agreement no. 824093. H.K. acknowledges support from the PRE2020-093558 Doctoral Grant of the spanish MCIN/ AEI/10.13039/501100011033/. 
S.B. and A.B. acknowledge support by Deutsche Forschungsgemeinschaft (DFG, German Research Foundation) through Project-ID 279384907 -- SFB 1245 (subproject B07). A.B. acknowledges support by the European Research Council (ERC) under the European Union’s Horizon 2020 research and innovation program under grant agreement No. 759253 and ERC Grant HEAVYMETAL No. 101071865 and support by the State of Hesse within the Cluster Project ELEMENTS.
L.T. also acknowledges support from the Generalitat Valenciana under contract CIPROM/2023/59, from the Generalitat de Catalunya under contract 2021 SGR 171, and from the CRC-TR 211 'Strong-interaction matter under extreme conditions'- project Nr. 315477589 - TRR 211. 
G.L acknowledges support by the Deutsche Forschungsgemeinschaft (DFG, German Research Foundation) - MA 4248/3-1 and support by the Klaus Tschira Foundation.

%% file: sections/appendix1.tex
\section{Results from the simulations}
\label{sec:app}
\onecolumngrid

In this appendix we provide tables, listing various results from our simulations that were discussed in this study along with characteristic properties of cold neutron stars for all hyperonic and nucleonic models used.

\begin{table*}[h]
    \centering
    \begin{tabular}{|c|c|c|c|c|c|c|c|c|c|c|c|c|c|}
    \hline
        EoS & $M_\mathrm{max}$ & $R_{1.4}$ & $f_\mathrm{peak}$  & $f_{\mathrm{peak}} ^{ 1.75}$  & $f_{\mathrm{2-0}}$ & $f^{1.75}_{\mathrm{2-0}}$  & $f_{\mathrm{spiral}}$  & $f^{ 1.75}_{\mathrm{spiral}}$  & $M_{\mathrm{ej}}$ & $\bar{\Gamma}_{\mathrm{th}}$ & $M_{\mathrm{thresh}}$  & $M_{\mathrm{thresh}}^{1.75}$ & Ref. \\ 
          &  [$M_{\odot}$] & [km] & [kHz] &  [kHz] & [kHz] & [kHz] &[kHz] & [kHz] & [$10^{-3}M_{\odot}$] & &  [$M_{\odot}$] &  [$M_{\odot}$] &  \\ \hline
        BHB$\Lambda\phi$ & 2.10 & 13.21 & 2.76 & 2.68 & 1.67 & 1.64 & 1.98 & 1.95 & 4.7 & 1.37 & 3.11 & 3.13 & \cite{Banik2014} \\ \hline
        DD2Y & 2.03 & 13.21 & 2.82 & 2.73 & 1.83 & 1.68 & 2.07 & 2.11 & 5.4 & 1.08 & 3.05 & 3.09 & \cite{Marques2017} \\ \hline
        DDLS(30)-Y & 2.02 & 12.87 & 2.69 & 2.63 & 1.62 & 1.57 & 1.97 & 2.03 & 3.7 & 1.28 & 3.11 & 3.11 & \cite{tsiopelas2024finitetemperature}\\ \hline
        DDLS(50)-Y & 2.00 & 13.19 & 2.74 & 2.67 & 1.74 & 1.60 & 2.02 & 2.00 & 3.6 & 1.29 & 3.09 & 3.12 & \cite{tsiopelas2024finitetemperature} \\ \hline
        DDLS(70)-Y & 1.98 & 13.56 & 2.71 & 2.64 & 1.73 & 1.61 & 1.97 & 1.93 & 3.3 & 1.29 & 3.09 & 3.12 & \cite{tsiopelas2024finitetemperature} \\ \hline
        DNS & 2.09 & 14.04 & 2.51 & 2.54 & 1.49 & 1.50 & 1.78 & 1.81 & 3.0 & 1.69 & 3.21 & 3.23 & \cite{Dexheimer2017} \\ \hline
        FSU2H* & 2.01 & 13.18 & 2.63 & 2.59 & 1.55 & 1.49 & 2.03 & 1.99 & 5.6 & 1.52 & 3.11 & 3.15 & \cite{Kochankovski2022} \\ \hline
        FSU2H*L & 1.91 & 13.16 & 2.68 & 2.62 & 1.63 & 1.51 & 1.93 & 1.89 & 6.7 & 1.24 & 3.07 & 3.09 & \cite{Kochankovski:2023trc} \\ \hline
        FSU2H*U & 2.06 & 13.17 & 2.62 & 2.56 & 1.52 & 1.44 & 1.91 & 1.89 & 7.9 & 1.51 & 3.13 & 3.19 & \cite{Kochankovski:2023trc} \\ \hline
        QMC-A & 1.99 & 12.89 & 2.91 & 2.98 & 1.85 & 1.85 & 2.17 & 2.22 & 6.0 & 1.65& 3.01 &3.03 & \cite{Stone2021} \\ \hline
        DD2Y$\Delta$1.1-1.1 & 2.04 & 12.96 & 3.03 & 2.93 & 2.07 & 1.95 & 2.26 & 2.24 & 7.1 & 1.08 & 2.99 & 3.03 & \cite{Raduta:2022elz}\\ \hline
        DD2Y$\Delta$1.2-1.1 & 2.05 & 12.27 & 3.26 & 3.14 & 2.09 & 1.99 & 2.48 & 2.49 & 10.0 & 1.18 & 2.91 & 2.95 & \cite{Raduta:2022elz} \\ \hline
        DD2Y$\Delta$1.2-1.3 & 2.03 & 13.21 & 2.82 & 2.72 & 1.81 & 1.69 & 2.04 & 2.08 & 4.0 & 0.99 & 3.05 & 3.09 & \cite{Raduta:2022elz}\\ \hline
        SFHOY & 1.99 & 11.89 & 3.60 & 3.46 & 2.65 & 2.52 & 2.94 & 2.69 & 33.5 & 1.38 & 2.81 & 2.85 & \cite{Fortin:2017dsj} \\ \hline
        
    \end{tabular}
    \caption{Hyperonic EoSs used to simulate equal-mass binary mergers a total mass of $2.8~M_{\odot}$. Columns give the acronyms of the EoS models, maximum mass of cold non-rotating NS, dominant postmerger frequency using a full 3D EoS, dominant postmerger frequency using constant thermal index $\Gamma = 1.75$,
    secondary peak $f_{2-0}$ using a full 3D EoS,
    secondary peak $f_{2-0}$ using a constant thermal index $\Gamma = 1.75$,
    secondary peak $f_{\mathrm{spiral}}$ using a full 3D EoS,
    secondary peak $f_{\mathrm{spiral}}$ using a constant thermal index $\Gamma = 1.75$,
    dynamical mass ejected in the first 10 ms after merging, 
    mass-averaged thermal index of the remnant's fluid in a time windows of 5 ms starting 2.5 ms after merging, threshold mass obtained using the full temperature dependent EoS, threshold mass obtained using constant thermal index $\Gamma_{\mathrm{th}} = 1.75$  and references of the original works where the models are presented. }
    \label{tab:app-1}
\end{table*}

\begin{table*}[t]
    \centering
    \begin{tabular}{|c|c|c|c|c|c|c|c|c|c|c|c|c|c|}
    \hline
        EoS & $M_{\mathrm{max}}$ & $R_{1.4}$ & $f_\mathrm{peak}$  & $f_{\mathrm{peak}} ^{1.75}$  & $f_{\mathrm{2-0}}$ & $f^{ 1.75}_{\mathrm{2-0}}$  & $f_{\mathrm{spiral}}$  & $f^{ 1.75}_{\mathrm{spiral}}$  & $M_{\mathrm{ej}}$ & $\bar{\Gamma}_{\mathrm{th}}$ & $M_{\mathrm{thresh}}$  & $M_{\mathrm{thresh}}^{1.75}$& Ref.\\ 
          &  [$M_{\odot}$] & [km] & [kHz] &  [kHz] & [kHz] & [kHz] &[kHz] & [kHz] & [$10^{-3}M_{\odot}$] &  & [$M_{\odot}$] & [$M_{\odot}$] & \\ \hline
        APR & 2.20 & 11.57 & 3.51 & 3.46 & 2.26 & 2.26 & 2.79 & 2.71 & 17.5 & 1.74 & 2.87 & 2.87&  \cite{Akmal:1998cf,Schneider:2019vdm} \\ \hline
        DD2 & 2.42 & 13.22 & 2.64 & 2.68 & 1.66 & 1.62 & 2.01 & 2.01 & 4.7 & 1.78 & 3.31 & 3.31 & \cite{Typel:2009sy,Hempel2010a}\\ \hline
        DD2F & 2.08 & 12.40 & 3.30 & 3.30 & 2.30 & 2.27 & 2.50 & 2.44 & 10.1 & 1.66 & 2.89 & 2.93 & \cite{Typel:2009sy,Alvarez-Castillo:2016oln} \\ \hline
        DDLS(30)-N & 2.48 & 12.87 & 2.62 & 2.62 & 1.48 & 1.45 & 1.95 & 1.94 & 3.6 & 1.83 & 3.38 & 3.36 & \cite{tsiopelas2024finitetemperature} \\ \hline
        DDLS(50)-N & 2.47 & 13.19 & 2.56 & 2.60 & 1.40 & 1.47 & 1.99 & 2.06 & 3.9 & 1.78 & 3.36 & 3.36 & \cite{tsiopelas2024finitetemperature} \\ \hline
        DDLS(70)-N & 2.46 & 13.56 & 2.62 & 2.60 & 1.58 & 1.52 & 1.91 & 1.89 & 2.4 & 1.79 & 3.37 & 3.37 & \cite{tsiopelas2024finitetemperature} \\ \hline
        DSH Fiducial & 2.17 & 11.73 & 3.44 & 3.40 & 2.26 & 2.26 & 2.55 & 2.67 & 16.9 & 1.77 & 2.89 & 2.89 & \cite{Du:2021rhq} \\ \hline
        DSH Large Mmax & 2.22 & 12.65 & 2.93 & 2.91 & 1.74 & 1.76 & 2.18 & 2.23 & 7.2 & 1.79 & 3.15 & 3.09 & \cite{Du:2021rhq} \\ \hline
        DSH Large SL & 2.16 & 11.76 & 3.51 & 3.46 & 2.41 & 2.30 & 2.72 & 2.72 & 17.4 & 1.52 & 2.87 & 2.87 & \cite{Du:2021rhq}\\ \hline
        DSH Large R & 2.13 & 12.44 & 3.16 & 3.18 & 2.10 & 2.10 & 2.46 & 2.41 & 12.7 & 1.72 & 2.97 & 2.97& \cite{Du:2021rhq} \\ \hline
        DSH Small SL & 2.18 & 11.70 & 3.31 & 3.33 & 2.07 & 2.16 & 2.52 & 2.54 & 13.6 & 1.76 & 2.91 & 2.91& \cite{Du:2021rhq} \\ \hline
        DSH Smaller R & 2.14 & 11.29 & 3.62 & 3.60 & 2.40 & 2.50 & 3.04 & 2.96 & 26.2 & 1.72 & 2.83 & 2.81& \cite{Du:2021rhq} \\ \hline
        FSU2R & 2.06 & 12.87 & 2.80 & 2.81 & 1.71 & 1.71 & 2.05 & 2.13 & 5.5 & 1.81 & 3.09 & 3.09 & \cite{Tolos:2017lgv}\\ \hline
        FTNS & 2.22 & 11.46 & 3.34 & 3.40 & 2.10 & 2.12 & 2.49 & 2.59 & 15.6 & 1.73 & 2.93 & 2.91  & \cite{Furusawa:2017auz,Togashi:2017mjp} \\ \hline
        GS2 & 2.09 & 13.60 & 2.73 & 2.70 & 1.68 & 1.71 & 2.01 & 1.98 & 3.0 & 1.76 & 3.15 & 3.15  & \cite{Shen:2011kr} \\ \hline
        LPB & 2.10 & 12.37 & 3.23 & 3.23 & 2.19 & 2.18 & 2.43 & 2.39 & 8.1 & 1.68 & 2.99 & 2.93 & \cite{Bombaci:2018ksa,Logoteta:2020yxf} \\ \hline
        LS220 & 2.04 & 12.96 & 3.09 & 3.06 & 2.08 & 2.10 & 2.27 & 2.25 & 5.0 & 1.54 & 2.97 & 2.97 & \cite{Lattimer:1991nc} \\ \hline
        LS375 & 2.71 & 13.95 & 2.44 & 2.44 & 1.39 & 1.40 & 1.77 & 1.74 & 2.0 & 1.63 & 3.57 & 3.55 & ~\cite{Lattimer:1991nc}\\ \hline
        SFHo & 2.06 & 11.89 & 3.43 & 3.45 & 2.50 & 2.38 & 2.77 & 2.71 & 20.5 & 1.62 & 2.87 & 2.87 & \cite{Hempel2010a,Steiner2013} \\ \hline
        SFHx & 2.13 & 11.98 & 3.16 & 3.18 & 2.03 & 2.01 & 2.39 & 2.37 & 11.0 & 1.82 & 2.97 & 2.97 &\cite{Hempel2010a,Steiner2013} \\ \hline
        SRO(SLy4) & 2.05 & 11.72 & 3.51 & 3.50 & 2.44 & 2.48 & 3.00 & 2.96 & 10.0 & 1.78 & 2.83 & 2.81 & \cite{Chabanat:1997un,Schneider:2017tfi}\\ \hline
        TM1 & 2.21 & 14.47 & 2.38 & 2.40 & 1.31 & 1.36 & 1.66 & 1.71 & 1.7 & 1.82 & 3.37 & 3.33 & \cite{Sugahara:1993wz,Hempel:2011mk} \\ \hline
        TMA & 2.01 & 13.79 & 2.58 & 2.57 & 1.54 & 1.57 & 1.78 & 1.73 & 2.1 & 1.74 & 3.19 & 3.15 & \cite{Toki:1995ya,Hempel:2011mk} \\
        \hline
    \end{tabular}
    \caption{Same as Table \ref{tab:app-1}, but for nucleonic EoSs.}
    \label{tab:app-nuc}
\end{table*}

\begin{table*}[t!]
\begin{tabular}{|c|c|c|c|c|c|c|c|c|}
\hline
EoS  & Simulated binaries & $f_{\mathrm{peak}}$ & 
$f_{\mathrm{peak}}^{1.75}$  & $\bar{\Gamma}_{\mathrm{th}}$ & $f_{\mathrm{peak}}^{1.4M_{\odot}}$ & $f_{\mathrm{peak}}^{1.75,1.4M_{\odot}}$ & $\bar{\Gamma}_{\mathrm{th}}^{1.4M_{\odot}}$ & Ref. \\
 &  & [kHz] & 
[kHz] &  & [kHz] & [kHz] &  & \\
\hline
DNS    & $1.55M_{\odot}-1.55 M_{\odot}$          & 2.89    & 2.87       & 1.41        & 2.51        & 2.54 &1.69 & \cite{Dexheimer2017}  \\
\hline
FSU2H$^{*}$  & $1.45 M_{\odot}-1.45 M_{\odot}$          & 2.71    & 2.65       & 1.40         & 2.63        & 2.59 &  1.52 & \cite{Kochankovski2022} \\
\hline
FSU2HL$^{*}$ & $1.45 M_{\odot}-1.45 M_{\odot}$          & 2.80    & 2.71       & 1.16        & 2.68        & 2.62 & 1.24 & \cite{Kochankovski:2023trc} \\
\hline

FSU2HU$^{*}$ & $1.50M_{\odot}-1.50 M_{\odot}$          & 2.78    & 2.72       & 1.26        & 2.62        & 2.56 & 1.51 & \cite{Kochankovski:2023trc}\\    
\hline
QMC-A    & $1.45 M_{\odot} -1.45 M_{\odot}$          & 3.02     & 3.02        & 1.59        & 2.91        & 2.98  &1.65 &  \cite{Stone2021} \\
\hline
\end{tabular}
\caption{Sample of hyperonic EoSs that we have used to simulate more massive equal-mass binaries. Third to fifth column provide the dominant frequency peak obtained using 3D EoS tables, dominant frequency peak obtained using the approximate thermal treatment and the mass and time averaged thermal index in the remnant. As a reference, sixth to eighth column provide the same quantities but for the less massive, $1.4 M_{\odot} - 1.4 M_{\odot}$ merger simulations. The ninth column gives references of the original works where the models are presented.}
\label{tab:1}
\end{table*}

\begin{table*}[!ht]
    \centering
    \begin{tabular}{|c|c|c|c|c|c|c|c|}
    \hline
        EoS & Species & $f_{\mathrm{peak}}$ & $f_{\mathrm{peak}}^{1.75}$ & $\rho_{\mathrm{max}}/\rho_{0}$ & $\bar{\Gamma}_{\mathrm{th}}$ & $\Lambda_{1.65}$ & Ref. \\ \hline
         BHB$\Lambda\phi$ & N-Y & 2.65 & 2.63 & 1.61 & 1.33 & 246.8 & \cite{Banik2014} \\ \hline
        DD2Y & N-Y & 2.76 & 2.63 & 1.78 & 1.04 & 240.4 & \cite{Marques2017}\\ \hline
        DDLS(30)-Y & N-Y & 2.59/2.80 & 2.49/2.70 & 1.58 & 1.19 & 264.9 & \cite{tsiopelas2024finitetemperature} \\ \hline
        DDLS(50)-Y & N-Y & 2.66/2.87 & 2.58/2.78 & 1.95 & 1.19 & 260.7 & \cite{tsiopelas2024finitetemperature} \\ \hline
        DDLS(70)-Y & N-Y & 2.65/2.88 & 2.58/2.75 & 1.89 & 1.19 & 277.8 & \cite{tsiopelas2024finitetemperature} \\ \hline 
        DNS & N-Y & 2.41 & 2.43 & 1.01 & 1.51 & 323.8 & \cite{Dexheimer2017} \\ \hline
        FSU2H$^*$ & N-Y & 2.76 & 2.69 & 1.50 & 1.37 & 291.5 & \cite{Kochankovski2022} \\ \hline
        FSU2HL$^*$ & N-Y & 2.65 & 2.58 & 1.90 & 1.15 & 287.5 & \cite{Kochankovski:2023trc} \\ \hline
        FSU2HU$^*$ & N-Y & 2.70 & 2.64 & 1.30 & 1.33 & 303.3 & \cite{Kochankovski:2023trc} \\ \hline
        QMC-A & N-Y & 2.81 & 2.77 & 1.04 & 1.54 & 195.3 & \cite{Stone2021} \\ \hline
        DD2Y$\Delta 1.1-1.1$ & N-Y-$\Delta$ & 3.02 & 2.88 & 2.52 & 1.08 & 184.5 & \cite{Raduta:2022elz} \\ \hline
        DD2Y$\Delta 1.1-1.2$ & N-Y-$\Delta$ & 3.25 & 3.10 & 3.25 & 1.15 & 133.7 & \cite{Raduta:2022elz}\\ \hline
        DD2Y$\Delta 1.2-1.3$ & N-Y-$\Delta$ & 2.78 & 2.63 & 1.79 & 1.00 & 240.4 & \cite{Raduta:2022elz} \\ \hline
        SFHOY & N-Y & / & 3.43 & / & / & 102.8 & \cite{Fortin:2017dsj} \\ \hline
        APR & N & 3.51 & 3.494 & / & 1.65 & 86.04 & \cite{Akmal:1998cf,Schneider:2019vdm} \\ \hline
        DD2 & N & 2.68 & 2.69 & / & 1.75 & 261.5 & \cite{Typel:2009sy,Hempel2010a} \\ \hline
        DD2F & N & 3.18 & 3.17 & / & 1.56 & 130.0 & \cite{Typel:2009sy,Alvarez-Castillo:2016oln}\\ \hline
        DDLS(30)-N & N & 2.60 & 2.66 & / & 1.78 & 268.8 & \cite{tsiopelas2024finitetemperature} \\ \hline
        DDLS(50)-N & N & 2.60 & 2.65 & / & 1.75 & 270.3 & \cite{tsiopelas2024finitetemperature} \\ \hline
        DDLS(70)-N & N & 2.62 & 2.60 & / & 1.73 & 295.6 & \cite{tsiopelas2024finitetemperature} \\ \hline 
        DSH Fiducial & N & 3.38 & 3.40 & / & 1.62 & 96.91 & \cite{Du:2021rhq} \\ \hline
        DSH Large Mmax & N & 2.78 & 2.79 & / & 1.77 & 181.1  & \cite{Du:2021rhq} \\ \hline
        DSH Large SL & N & 3.44 & 3.42 & / & 1.60 & 87.74  & \cite{Du:2021rhq}\\ \hline
        DSH Large R & N & 3.08 & 3.09 & / & 1.68 & 139.1  & \cite{Du:2021rhq} \\ \hline
        DSH Small SL & N & 3.28 & 3.27 & / & 1.65 & 110.6  & \cite{Du:2021rhq} \\ \hline
        DSH Smaller R & N & 3.64 & 3.64 & / & 1.57 & 76.73  & \cite{Du:2021rhq} \\ \hline
        FSU2R & N & 2.69 & 2.70 & / & 1.76 & 221.9 & \cite{Tolos:2017lgv}\\ \hline
        FTNS & N & 3.31 & 3.38 & / & 1.69 & 101.6 & \cite{Furusawa:2017auz,Togashi:2017mjp} \\ \hline
        GS2 & N & 2.60 & 2.58 & / & 1.78 & 248.5 & \cite{Shen:2011kr}\\ \hline
        LPB & N & 3.14 & 3.15 & / & 1.59 & 130.8 & \cite{Bombaci:2018ksa,Logoteta:2020yxf}\\ \hline
        LS220 & N & 3.06 & 3.04 & / & 1.47 & 158.4 & \cite{Lattimer:1991nc} \\ \hline
        LS375 & N & 2.48 & 2.46 & / & 1.58 & 371.1 & ~\cite{Lattimer:1991nc} \\ \hline
        SFHO & N & 3.41 & 3.39 & / & 1.54 & 103.4 & \cite{Hempel2010a,Steiner2013}\\ \hline
        SFHX & N & 3.11 & 3.08 & / & 1.74 & 134.6 &\cite{Hempel2010a,Steiner2013} \\ \hline
        SRO(Sly4) & N & 3.43 & 3.51 & / & 1.73 & 90.90 & \cite{Chabanat:1997un,Schneider:2017tfi} \\ \hline
        TM1 & N & 2.29 & 2.32 & / & 1.75 & 394.1 & \cite{Sugahara:1993wz,Hempel:2011mk} \\ \hline
        TMA & N & 2.50 & 2.54 & / & 1.67 & 297.8 & \cite{Toki:1995ya,Hempel:2011mk} \\ \hline
    \end{tabular}
    \caption{EoSs used to simulate the asymmetric binary mergers with a total mass of $2.8 M_{\odot}$ and a binary mass ratio of $q=0.8$. First to seventh column give the acronym of the models, hadronic species that the EoS is taking into account, dominant postmerger frequency using a full 3D EoS table, dominant postmerger frequency $f_{\mathrm{peak}}^{1.75}$ using a cold slice supplemented with a constant thermal index $\Gamma_{\mathrm{th}} = 1.75$, ratio $\rho_{\mathrm{max}}/\rho_{0}$ between the maximum density reached in the remnant 5~ms after the merger and the onset density of heavy baryons, the time-mass averaged thermal $\bar{\Gamma}_{\mathrm{th}}$, and tidal deformability $\Lambda_{1.65}$ of a star with mass of $1.65 M_{\odot}$. The last column provides the references for the EoS models. Some of the EoSs produce two dominant peaks with a comparable strength. 
    }
\label{tab:2}
\end{table*}

%% file: main.bbl
\begin{thebibliography}{173}%
\makeatletter
\providecommand \@ifxundefined [1]{%
 \@ifx{#1\undefined}
}%
\providecommand \@ifnum [1]{%
 \ifnum #1\expandafter \@firstoftwo
 \else \expandafter \@secondoftwo
 \fi
}%
\providecommand \@ifx [1]{%
 \ifx #1\expandafter \@firstoftwo
 \else \expandafter \@secondoftwo
 \fi
}%
\providecommand \natexlab [1]{#1}%
\providecommand \enquote  [1]{``#1''}%
\providecommand \bibnamefont  [1]{#1}%
\providecommand \bibfnamefont [1]{#1}%
\providecommand \citenamefont [1]{#1}%
\providecommand \href@noop [0]{\@secondoftwo}%
\providecommand \href [0]{\begingroup \@sanitize@url \@href}%
\providecommand \@href[1]{\@@startlink{#1}\@@href}%
\providecommand \@@href[1]{\endgroup#1\@@endlink}%
\providecommand \@sanitize@url [0]{\catcode `\\12\catcode `\$12\catcode `\&12\catcode `\#12\catcode `\^12\catcode `\_12\catcode `\%12\relax}%
\providecommand \@@startlink[1]{}%
\providecommand \@@endlink[0]{}%
\providecommand \url  [0]{\begingroup\@sanitize@url \@url }%
\providecommand \@url [1]{\endgroup\@href {#1}{\urlprefix }}%
\providecommand \urlprefix  [0]{URL }%
\providecommand \Eprint [0]{\href }%
\providecommand \doibase [0]{https://doi.org/}%
\providecommand \selectlanguage [0]{\@gobble}%
\providecommand \bibinfo  [0]{\@secondoftwo}%
\providecommand \bibfield  [0]{\@secondoftwo}%
\providecommand \translation [1]{[#1]}%
\providecommand \BibitemOpen [0]{}%
\providecommand \bibitemStop [0]{}%
\providecommand \bibitemNoStop [0]{.\EOS\space}%
\providecommand \EOS [0]{\spacefactor3000\relax}%
\providecommand \BibitemShut  [1]{\csname bibitem#1\endcsname}%
\let\auto@bib@innerbib\@empty
%</preamble>
\bibitem [{\citenamefont {{Ambartsumyan}}\ and\ \citenamefont {{Saakyan}}(1960)}]{1960SvA.....4..187A}%
  \BibitemOpen
  \bibfield  {author} {\bibinfo {author} {\bibfnamefont {V.~A.}\ \bibnamefont {{Ambartsumyan}}}\ and\ \bibinfo {author} {\bibfnamefont {G.~S.}\ \bibnamefont {{Saakyan}}},\ }\href@noop {} {\bibfield  {journal} {\bibinfo  {journal} {Soviet Ast.}\ }\textbf {\bibinfo {volume} {4}},\ \bibinfo {pages} {187} (\bibinfo {year} {1960})}\BibitemShut {NoStop}%
\bibitem [{\citenamefont {Glendenning}(1982)}]{Glendenning:1982nc}%
  \BibitemOpen
  \bibfield  {author} {\bibinfo {author} {\bibfnamefont {N.~K.}\ \bibnamefont {Glendenning}},\ }\href {https://doi.org/10.1016/0370-2693(82)90078-8} {\bibfield  {journal} {\bibinfo  {journal} {Phys. Lett. B}\ }\textbf {\bibinfo {volume} {114}},\ \bibinfo {pages} {392} (\bibinfo {year} {1982})}\BibitemShut {NoStop}%
\bibitem [{\citenamefont {{Glendenning}}(1985)}]{1985ApJ...293..470G}%
  \BibitemOpen
  \bibfield  {author} {\bibinfo {author} {\bibfnamefont {N.~K.}\ \bibnamefont {{Glendenning}}},\ }\href {https://doi.org/10.1086/163253} {\bibfield  {journal} {\bibinfo  {journal} {\apj}\ }\textbf {\bibinfo {volume} {293}},\ \bibinfo {pages} {470} (\bibinfo {year} {1985})}\BibitemShut {NoStop}%
\bibitem [{\citenamefont {Glendenning}(1987)}]{Glendenning:1987wb}%
  \BibitemOpen
  \bibfield  {author} {\bibinfo {author} {\bibfnamefont {N.~K.}\ \bibnamefont {Glendenning}},\ }\href {https://doi.org/10.1007/BF01294571} {\bibfield  {journal} {\bibinfo  {journal} {Z. Phys. A}\ }\textbf {\bibinfo {volume} {326}},\ \bibinfo {pages} {57} (\bibinfo {year} {1987})}\BibitemShut {NoStop}%
\bibitem [{\citenamefont {Weber}\ and\ \citenamefont {Weigel}(1989)}]{Weber:1989qf}%
  \BibitemOpen
  \bibfield  {author} {\bibinfo {author} {\bibfnamefont {F.}~\bibnamefont {Weber}}\ and\ \bibinfo {author} {\bibfnamefont {M.}~\bibnamefont {Weigel}},\ }\href {https://doi.org/https://doi.org/10.1016/0375-9474(89)90041-9} {\bibfield  {journal} {\bibinfo  {journal} {Nucl. Phys. A}\ }\textbf {\bibinfo {volume} {505}},\ \bibinfo {pages} {779} (\bibinfo {year} {1989})}\BibitemShut {NoStop}%
\bibitem [{\citenamefont {Glendenning}\ and\ \citenamefont {Moszkowski}(1991)}]{Glendenning:1991es}%
  \BibitemOpen
  \bibfield  {author} {\bibinfo {author} {\bibfnamefont {N.~K.}\ \bibnamefont {Glendenning}}\ and\ \bibinfo {author} {\bibfnamefont {S.~A.}\ \bibnamefont {Moszkowski}},\ }\href {https://doi.org/10.1103/PhysRevLett.67.2414} {\bibfield  {journal} {\bibinfo  {journal} {Phys. Rev. Lett.}\ }\textbf {\bibinfo {volume} {67}},\ \bibinfo {pages} {2414} (\bibinfo {year} {1991})}\BibitemShut {NoStop}%
\bibitem [{\citenamefont {Knrren}\ \emph {et~al.}(1995)\citenamefont {Knrren}, \citenamefont {Prakash},\ and\ \citenamefont {Ellis}}]{Knrren:1995rv}%
  \BibitemOpen
  \bibfield  {author} {\bibinfo {author} {\bibfnamefont {R.}~\bibnamefont {Knrren}}, \bibinfo {author} {\bibfnamefont {M.}~\bibnamefont {Prakash}},\ and\ \bibinfo {author} {\bibfnamefont {P.~J.}\ \bibnamefont {Ellis}},\ }\href {https://doi.org/10.1103/PhysRevC.52.3470} {\bibfield  {journal} {\bibinfo  {journal} {Phys. Rev. C}\ }\textbf {\bibinfo {volume} {52}},\ \bibinfo {pages} {3470} (\bibinfo {year} {1995})}\BibitemShut {NoStop}%
\bibitem [{\citenamefont {Schulze}\ \emph {et~al.}(1995)\citenamefont {Schulze}, \citenamefont {Lejeune}, \citenamefont {Cugnon}, \citenamefont {Baldo},\ and\ \citenamefont {Lombardo}}]{SCHULZE199521}%
  \BibitemOpen
  \bibfield  {author} {\bibinfo {author} {\bibfnamefont {H.-J.}\ \bibnamefont {Schulze}}, \bibinfo {author} {\bibfnamefont {A.}~\bibnamefont {Lejeune}}, \bibinfo {author} {\bibfnamefont {J.}~\bibnamefont {Cugnon}}, \bibinfo {author} {\bibfnamefont {M.}~\bibnamefont {Baldo}},\ and\ \bibinfo {author} {\bibfnamefont {U.}~\bibnamefont {Lombardo}},\ }\href {https://doi.org/https://doi.org/10.1016/0370-2693(95)00665-8} {\bibfield  {journal} {\bibinfo  {journal} {Phys. Lett. B}\ }\textbf {\bibinfo {volume} {355}},\ \bibinfo {pages} {21} (\bibinfo {year} {1995})}\BibitemShut {NoStop}%
\bibitem [{\citenamefont {Schaffner}\ and\ \citenamefont {Mishustin}(1996)}]{Schaffner:1995th}%
  \BibitemOpen
  \bibfield  {author} {\bibinfo {author} {\bibfnamefont {J.}~\bibnamefont {Schaffner}}\ and\ \bibinfo {author} {\bibfnamefont {I.~N.}\ \bibnamefont {Mishustin}},\ }\href {https://doi.org/10.1103/PhysRevC.53.1416} {\bibfield  {journal} {\bibinfo  {journal} {Phys. Rev. C}\ }\textbf {\bibinfo {volume} {53}},\ \bibinfo {pages} {1416} (\bibinfo {year} {1996})}\BibitemShut {NoStop}%
\bibitem [{\citenamefont {Balberg}\ and\ \citenamefont {Gal}(1997)}]{Balberg_1997}%
  \BibitemOpen
  \bibfield  {author} {\bibinfo {author} {\bibfnamefont {S.}~\bibnamefont {Balberg}}\ and\ \bibinfo {author} {\bibfnamefont {A.}~\bibnamefont {Gal}},\ }\href {https://doi.org/10.1016/s0375-9474(97)81465-0} {\bibfield  {journal} {\bibinfo  {journal} {Nucl. Phys. A}\ }\textbf {\bibinfo {volume} {625}},\ \bibinfo {pages} {435–472} (\bibinfo {year} {1997})}\BibitemShut {NoStop}%
\bibitem [{\citenamefont {Schulze}\ \emph {et~al.}(1998)\citenamefont {Schulze}, \citenamefont {Baldo}, \citenamefont {Lombardo}, \citenamefont {Cugnon},\ and\ \citenamefont {Lejeune}}]{PhysRevC.57.704}%
  \BibitemOpen
  \bibfield  {author} {\bibinfo {author} {\bibfnamefont {H.-J.}\ \bibnamefont {Schulze}}, \bibinfo {author} {\bibfnamefont {M.}~\bibnamefont {Baldo}}, \bibinfo {author} {\bibfnamefont {U.}~\bibnamefont {Lombardo}}, \bibinfo {author} {\bibfnamefont {J.}~\bibnamefont {Cugnon}},\ and\ \bibinfo {author} {\bibfnamefont {A.}~\bibnamefont {Lejeune}},\ }\href {https://doi.org/10.1103/PhysRevC.57.704} {\bibfield  {journal} {\bibinfo  {journal} {Phys. Rev. C}\ }\textbf {\bibinfo {volume} {57}},\ \bibinfo {pages} {704} (\bibinfo {year} {1998})}\BibitemShut {NoStop}%
\bibitem [{\citenamefont {Baldo}\ \emph {et~al.}(1998)\citenamefont {Baldo}, \citenamefont {Burgio},\ and\ \citenamefont {Schulze}}]{PhysRevC.58.3688}%
  \BibitemOpen
  \bibfield  {author} {\bibinfo {author} {\bibfnamefont {M.}~\bibnamefont {Baldo}}, \bibinfo {author} {\bibfnamefont {G.~F.}\ \bibnamefont {Burgio}},\ and\ \bibinfo {author} {\bibfnamefont {H.-J.}\ \bibnamefont {Schulze}},\ }\href {https://doi.org/10.1103/PhysRevC.58.3688} {\bibfield  {journal} {\bibinfo  {journal} {Phys. Rev. C}\ }\textbf {\bibinfo {volume} {58}},\ \bibinfo {pages} {3688} (\bibinfo {year} {1998})}\BibitemShut {NoStop}%
\bibitem [{\citenamefont {{Balberg}}\ \emph {et~al.}(1999)\citenamefont {{Balberg}}, \citenamefont {{Lichtenstadt}},\ and\ \citenamefont {{Cook}}}]{1999ApJS..121..515B}%
  \BibitemOpen
  \bibfield  {author} {\bibinfo {author} {\bibfnamefont {S.}~\bibnamefont {{Balberg}}}, \bibinfo {author} {\bibfnamefont {I.}~\bibnamefont {{Lichtenstadt}}},\ and\ \bibinfo {author} {\bibfnamefont {G.~B.}\ \bibnamefont {{Cook}}},\ }\href {https://doi.org/10.1086/313196} {\bibfield  {journal} {\bibinfo  {journal} {\apjs}\ }\textbf {\bibinfo {volume} {121}},\ \bibinfo {pages} {515} (\bibinfo {year} {1999})},\ \Eprint {https://arxiv.org/abs/astro-ph/9810361} {arXiv:astro-ph/9810361 [astro-ph]} \BibitemShut {NoStop}%
\bibitem [{\citenamefont {Baldo}\ \emph {et~al.}(2000)\citenamefont {Baldo}, \citenamefont {Burgio},\ and\ \citenamefont {Schulze}}]{PhysRevC.61.055801}%
  \BibitemOpen
  \bibfield  {author} {\bibinfo {author} {\bibfnamefont {M.}~\bibnamefont {Baldo}}, \bibinfo {author} {\bibfnamefont {G.~F.}\ \bibnamefont {Burgio}},\ and\ \bibinfo {author} {\bibfnamefont {H.-J.}\ \bibnamefont {Schulze}},\ }\href {https://doi.org/10.1103/PhysRevC.61.055801} {\bibfield  {journal} {\bibinfo  {journal} {Phys. Rev. C}\ }\textbf {\bibinfo {volume} {61}},\ \bibinfo {pages} {055801} (\bibinfo {year} {2000})}\BibitemShut {NoStop}%
\bibitem [{\citenamefont {Vida\~na}\ \emph {et~al.}(2000{\natexlab{a}})\citenamefont {Vida\~na}, \citenamefont {Polls}, \citenamefont {Ramos}, \citenamefont {Hjorth-Jensen},\ and\ \citenamefont {Stoks}}]{PhysRevC.61.025802}%
  \BibitemOpen
  \bibfield  {author} {\bibinfo {author} {\bibfnamefont {I.}~\bibnamefont {Vida\~na}}, \bibinfo {author} {\bibfnamefont {A.}~\bibnamefont {Polls}}, \bibinfo {author} {\bibfnamefont {A.}~\bibnamefont {Ramos}}, \bibinfo {author} {\bibfnamefont {M.}~\bibnamefont {Hjorth-Jensen}},\ and\ \bibinfo {author} {\bibfnamefont {V.~G.~J.}\ \bibnamefont {Stoks}},\ }\href {https://doi.org/10.1103/PhysRevC.61.025802} {\bibfield  {journal} {\bibinfo  {journal} {Phys. Rev. C}\ }\textbf {\bibinfo {volume} {61}},\ \bibinfo {pages} {025802} (\bibinfo {year} {2000}{\natexlab{a}})}\BibitemShut {NoStop}%
\bibitem [{\citenamefont {Vida\~na}\ \emph {et~al.}(2000{\natexlab{b}})\citenamefont {Vida\~na}, \citenamefont {Polls}, \citenamefont {Ramos}, \citenamefont {Engvik},\ and\ \citenamefont {Hjorth-Jensen}}]{PhysRevC.62.035801}%
  \BibitemOpen
  \bibfield  {author} {\bibinfo {author} {\bibfnamefont {I.}~\bibnamefont {Vida\~na}}, \bibinfo {author} {\bibfnamefont {A.}~\bibnamefont {Polls}}, \bibinfo {author} {\bibfnamefont {A.}~\bibnamefont {Ramos}}, \bibinfo {author} {\bibfnamefont {L.}~\bibnamefont {Engvik}},\ and\ \bibinfo {author} {\bibfnamefont {M.}~\bibnamefont {Hjorth-Jensen}},\ }\href {https://doi.org/10.1103/PhysRevC.62.035801} {\bibfield  {journal} {\bibinfo  {journal} {Phys. Rev. C}\ }\textbf {\bibinfo {volume} {62}},\ \bibinfo {pages} {035801} (\bibinfo {year} {2000}{\natexlab{b}})}\BibitemShut {NoStop}%
\bibitem [{\citenamefont {Schulze}\ \emph {et~al.}(2006)\citenamefont {Schulze}, \citenamefont {Polls}, \citenamefont {Ramos},\ and\ \citenamefont {Vida\~na}}]{PhysRevC.73.058801}%
  \BibitemOpen
  \bibfield  {author} {\bibinfo {author} {\bibfnamefont {H.-J.}\ \bibnamefont {Schulze}}, \bibinfo {author} {\bibfnamefont {A.}~\bibnamefont {Polls}}, \bibinfo {author} {\bibfnamefont {A.}~\bibnamefont {Ramos}},\ and\ \bibinfo {author} {\bibfnamefont {I.}~\bibnamefont {Vida\~na}},\ }\href {https://doi.org/10.1103/PhysRevC.73.058801} {\bibfield  {journal} {\bibinfo  {journal} {Phys. Rev. C}\ }\textbf {\bibinfo {volume} {73}},\ \bibinfo {pages} {058801} (\bibinfo {year} {2006})}\BibitemShut {NoStop}%
\bibitem [{\citenamefont {Sammarruca}(2009)}]{PhysRevC.79.034301}%
  \BibitemOpen
  \bibfield  {author} {\bibinfo {author} {\bibfnamefont {F.}~\bibnamefont {Sammarruca}},\ }\href {https://doi.org/10.1103/PhysRevC.79.034301} {\bibfield  {journal} {\bibinfo  {journal} {Phys. Rev. C}\ }\textbf {\bibinfo {volume} {79}},\ \bibinfo {pages} {034301} (\bibinfo {year} {2009})}\BibitemShut {NoStop}%
\bibitem [{\citenamefont {\DH{}apo}\ \emph {et~al.}(2010)\citenamefont {\DH{}apo}, \citenamefont {Schaefer},\ and\ \citenamefont {Wambach}}]{PhysRevC.81.035803}%
  \BibitemOpen
  \bibfield  {author} {\bibinfo {author} {\bibfnamefont {H.}~\bibnamefont {\DH{}apo}}, \bibinfo {author} {\bibfnamefont {B.-J.}\ \bibnamefont {Schaefer}},\ and\ \bibinfo {author} {\bibfnamefont {J.}~\bibnamefont {Wambach}},\ }\href {https://doi.org/10.1103/PhysRevC.81.035803} {\bibfield  {journal} {\bibinfo  {journal} {Phys. Rev. C}\ }\textbf {\bibinfo {volume} {81}},\ \bibinfo {pages} {035803} (\bibinfo {year} {2010})}\BibitemShut {NoStop}%
\bibitem [{\citenamefont {Schulze}\ and\ \citenamefont {Rijken}(2011)}]{PhysRevC.84.035801}%
  \BibitemOpen
  \bibfield  {author} {\bibinfo {author} {\bibfnamefont {H.-J.}\ \bibnamefont {Schulze}}\ and\ \bibinfo {author} {\bibfnamefont {T.}~\bibnamefont {Rijken}},\ }\href {https://doi.org/10.1103/PhysRevC.84.035801} {\bibfield  {journal} {\bibinfo  {journal} {Phys. Rev. C}\ }\textbf {\bibinfo {volume} {84}},\ \bibinfo {pages} {035801} (\bibinfo {year} {2011})}\BibitemShut {NoStop}%
\bibitem [{\citenamefont {Bonanno}\ and\ \citenamefont {Sedrakian}(2012)}]{Bonanno:2011ch}%
  \BibitemOpen
  \bibfield  {author} {\bibinfo {author} {\bibfnamefont {L.}~\bibnamefont {Bonanno}}\ and\ \bibinfo {author} {\bibfnamefont {A.}~\bibnamefont {Sedrakian}},\ }\href {https://doi.org/10.1051/0004-6361/201117832} {\bibfield  {journal} {\bibinfo  {journal} {Astron. Astrophys.}\ }\textbf {\bibinfo {volume} {539}},\ \bibinfo {pages} {A16} (\bibinfo {year} {2012})},\ \Eprint {https://arxiv.org/abs/1108.0559} {arXiv:1108.0559 [astro-ph.SR]} \BibitemShut {NoStop}%
\bibitem [{\citenamefont {Weissenborn}\ \emph {et~al.}(2012)\citenamefont {Weissenborn}, \citenamefont {Chatterjee},\ and\ \citenamefont {Schaffner-Bielich}}]{Weissenborn_2012}%
  \BibitemOpen
  \bibfield  {author} {\bibinfo {author} {\bibfnamefont {S.}~\bibnamefont {Weissenborn}}, \bibinfo {author} {\bibfnamefont {D.}~\bibnamefont {Chatterjee}},\ and\ \bibinfo {author} {\bibfnamefont {J.}~\bibnamefont {Schaffner-Bielich}},\ }\href {https://doi.org/10.1016/j.nuclphysa.2012.02.012} {\bibfield  {journal} {\bibinfo  {journal} {Nucl. Phys. A}\ }\textbf {\bibinfo {volume} {881}},\ \bibinfo {pages} {62–77} (\bibinfo {year} {2012})}\BibitemShut {NoStop}%
\bibitem [{\citenamefont {Bednarek}\ \emph {et~al.}(2012)\citenamefont {Bednarek}, \citenamefont {Haensel}, \citenamefont {Zdunik}, \citenamefont {Bejger},\ and\ \citenamefont {Manka}}]{Bednarek:2012zs}%
  \BibitemOpen
  \bibfield  {author} {\bibinfo {author} {\bibfnamefont {I.}~\bibnamefont {Bednarek}}, \bibinfo {author} {\bibfnamefont {P.}~\bibnamefont {Haensel}}, \bibinfo {author} {\bibfnamefont {J.~L.}\ \bibnamefont {Zdunik}}, \bibinfo {author} {\bibfnamefont {M.}~\bibnamefont {Bejger}},\ and\ \bibinfo {author} {\bibfnamefont {R.}~\bibnamefont {Manka}},\ }\href {https://doi.org/10.1051/0004-6361/201118560} {\bibfield  {journal} {\bibinfo  {journal} {Astron. Astrophys.}\ }\textbf {\bibinfo {volume} {543}},\ \bibinfo {pages} {A157} (\bibinfo {year} {2012})},\ \Eprint {https://arxiv.org/abs/1111.6942} {arXiv:1111.6942 [astro-ph.SR]} \BibitemShut {NoStop}%
\bibitem [{\citenamefont {{Maslov}}\ \emph {et~al.}(2015)\citenamefont {{Maslov}}, \citenamefont {{Kolomeitsev}},\ and\ \citenamefont {{Voskresensky}}}]{Maslov:2015gxa}%
  \BibitemOpen
  \bibfield  {author} {\bibinfo {author} {\bibfnamefont {K.~A.}\ \bibnamefont {{Maslov}}}, \bibinfo {author} {\bibfnamefont {E.~E.}\ \bibnamefont {{Kolomeitsev}}},\ and\ \bibinfo {author} {\bibfnamefont {D.~N.}\ \bibnamefont {{Voskresensky}}},\ }\href {https://doi.org/10.1016/j.physletb.2015.07.032} {\bibfield  {journal} {\bibinfo  {journal} {Phys. Lett. B}\ }\textbf {\bibinfo {volume} {748}},\ \bibinfo {pages} {369} (\bibinfo {year} {2015})},\ \Eprint {https://arxiv.org/abs/1504.02915} {arXiv:1504.02915 [astro-ph.HE]} \BibitemShut {NoStop}%
\bibitem [{\citenamefont {Tolos}\ \emph {et~al.}(2017{\natexlab{a}})\citenamefont {Tolos}, \citenamefont {Centelles},\ and\ \citenamefont {Ramos}}]{Tolos:2017kfa}%
  \BibitemOpen
  \bibfield  {author} {\bibinfo {author} {\bibfnamefont {L.}~\bibnamefont {Tolos}}, \bibinfo {author} {\bibfnamefont {M.}~\bibnamefont {Centelles}},\ and\ \bibinfo {author} {\bibfnamefont {A.}~\bibnamefont {Ramos}},\ }\href {https://doi.org/10.3847/1538-4357/834/1/3} {\bibfield  {journal} {\bibinfo  {journal} {Astrophys. J.}\ }\textbf {\bibinfo {volume} {834}},\ \bibinfo {pages} {3} (\bibinfo {year} {2017}{\natexlab{a}})}\BibitemShut {NoStop}%
\bibitem [{\citenamefont {Tolos}\ \emph {et~al.}(2017{\natexlab{b}})\citenamefont {Tolos}, \citenamefont {Centelles},\ and\ \citenamefont {Ramos}}]{Tolos:2017lgv}%
  \BibitemOpen
  \bibfield  {author} {\bibinfo {author} {\bibfnamefont {L.}~\bibnamefont {Tolos}}, \bibinfo {author} {\bibfnamefont {M.}~\bibnamefont {Centelles}},\ and\ \bibinfo {author} {\bibfnamefont {A.}~\bibnamefont {Ramos}},\ }\href {https://doi.org/10.1017/pasa.2017.60} {\bibfield  {journal} {\bibinfo  {journal} {Publ. Astron. Soc. Austral.}\ }\textbf {\bibinfo {volume} {34}},\ \bibinfo {pages} {e065} (\bibinfo {year} {2017}{\natexlab{b}})},\ \Eprint {https://arxiv.org/abs/1708.08681} {arXiv:1708.08681 [astro-ph.HE]} \BibitemShut {NoStop}%
\bibitem [{\citenamefont {Fortin}\ \emph {et~al.}(2017)\citenamefont {Fortin}, \citenamefont {Avancini}, \citenamefont {Provid\^encia},\ and\ \citenamefont {Vida\~na}}]{Fortin:2017zck}%
  \BibitemOpen
  \bibfield  {author} {\bibinfo {author} {\bibfnamefont {M.}~\bibnamefont {Fortin}}, \bibinfo {author} {\bibfnamefont {S.~S.}\ \bibnamefont {Avancini}}, \bibinfo {author} {\bibfnamefont {C.}~\bibnamefont {Provid\^encia}},\ and\ \bibinfo {author} {\bibfnamefont {I.}~\bibnamefont {Vida\~na}},\ }\href {https://doi.org/10.1103/PhysRevC.95.065803} {\bibfield  {journal} {\bibinfo  {journal} {Phys. Rev. C}\ }\textbf {\bibinfo {volume} {95}},\ \bibinfo {pages} {065803} (\bibinfo {year} {2017})}\BibitemShut {NoStop}%
\bibitem [{\citenamefont {Colucci}\ and\ \citenamefont {Sedrakian}(2013)}]{PhysRevC.87.055806}%
  \BibitemOpen
  \bibfield  {author} {\bibinfo {author} {\bibfnamefont {G.}~\bibnamefont {Colucci}}\ and\ \bibinfo {author} {\bibfnamefont {A.}~\bibnamefont {Sedrakian}},\ }\href {https://doi.org/10.1103/PhysRevC.87.055806} {\bibfield  {journal} {\bibinfo  {journal} {Phys. Rev. C}\ }\textbf {\bibinfo {volume} {87}},\ \bibinfo {pages} {055806} (\bibinfo {year} {2013})}\BibitemShut {NoStop}%
\bibitem [{\citenamefont {{van Dalen}}\ \emph {et~al.}(2014)\citenamefont {{van Dalen}}, \citenamefont {{Colucci}},\ and\ \citenamefont {{Sedrakian}}}]{Dalen:2014mda}%
  \BibitemOpen
  \bibfield  {author} {\bibinfo {author} {\bibfnamefont {E.~N.~E.}\ \bibnamefont {{van Dalen}}}, \bibinfo {author} {\bibfnamefont {G.}~\bibnamefont {{Colucci}}},\ and\ \bibinfo {author} {\bibfnamefont {A.}~\bibnamefont {{Sedrakian}}},\ }\href {https://doi.org/10.1016/j.physletb.2014.06.002} {\bibfield  {journal} {\bibinfo  {journal} {Phys. Lett. B}\ }\textbf {\bibinfo {volume} {734}},\ \bibinfo {pages} {383} (\bibinfo {year} {2014})},\ \Eprint {https://arxiv.org/abs/1406.0744} {arXiv:1406.0744 [nucl-th]} \BibitemShut {NoStop}%
\bibitem [{\citenamefont {Lonardoni}\ \emph {et~al.}(2014)\citenamefont {Lonardoni}, \citenamefont {Pederiva},\ and\ \citenamefont {Gandolfi}}]{PhysRevC.89.014314}%
  \BibitemOpen
  \bibfield  {author} {\bibinfo {author} {\bibfnamefont {D.}~\bibnamefont {Lonardoni}}, \bibinfo {author} {\bibfnamefont {F.}~\bibnamefont {Pederiva}},\ and\ \bibinfo {author} {\bibfnamefont {S.}~\bibnamefont {Gandolfi}},\ }\href {https://doi.org/10.1103/PhysRevC.89.014314} {\bibfield  {journal} {\bibinfo  {journal} {Phys. Rev. C}\ }\textbf {\bibinfo {volume} {89}},\ \bibinfo {pages} {014314} (\bibinfo {year} {2014})}\BibitemShut {NoStop}%
\bibitem [{\citenamefont {Drago}\ \emph {et~al.}(2016)\citenamefont {Drago}, \citenamefont {Lavagno}, \citenamefont {Pagliara},\ and\ \citenamefont {Pigato}}]{Drago:2016srv}%
  \BibitemOpen
  \bibfield  {author} {\bibinfo {author} {\bibfnamefont {A.}~\bibnamefont {Drago}}, \bibinfo {author} {\bibfnamefont {A.}~\bibnamefont {Lavagno}}, \bibinfo {author} {\bibfnamefont {G.}~\bibnamefont {Pagliara}},\ and\ \bibinfo {author} {\bibfnamefont {D.}~\bibnamefont {Pigato}},\ }\href {https://doi.org/10.1140/epja/i2016-16040-3} {\bibfield  {journal} {\bibinfo  {journal} {Eur. Phys. J. A}\ }\textbf {\bibinfo {volume} {52}},\ \bibinfo {pages} {40} (\bibinfo {year} {2016})}\BibitemShut {NoStop}%
\bibitem [{\citenamefont {Drago}\ and\ \citenamefont {Pagliara}(2016)}]{Drago:2016kpi}%
  \BibitemOpen
  \bibfield  {author} {\bibinfo {author} {\bibfnamefont {A.}~\bibnamefont {Drago}}\ and\ \bibinfo {author} {\bibfnamefont {G.}~\bibnamefont {Pagliara}},\ }\href {https://doi.org/10.1140/epja/i2016-16041-2} {\bibfield  {journal} {\bibinfo  {journal} {Eur Phys. J. A}\ }\textbf {\bibinfo {volume} {52}},\ \bibinfo {pages} {41} (\bibinfo {year} {2016})}\BibitemShut {NoStop}%
\bibitem [{\citenamefont {Fortin}\ \emph {et~al.}(2018{\natexlab{a}})\citenamefont {Fortin}, \citenamefont {Raduta}, \citenamefont {Avancini},\ and\ \citenamefont {Providência}}]{Fortin:2018aa}%
  \BibitemOpen
  \bibfield  {author} {\bibinfo {author} {\bibfnamefont {M.}~\bibnamefont {Fortin}}, \bibinfo {author} {\bibfnamefont {A.~R.}\ \bibnamefont {Raduta}}, \bibinfo {author} {\bibfnamefont {S.~S.}\ \bibnamefont {Avancini}},\ and\ \bibinfo {author} {\bibfnamefont {C.}~\bibnamefont {Providência}},\ }\href {https://doi.org/10.1103/PhysRevD.101.034017} {\bibfield  {journal} {\bibinfo  {journal} {Phys. Rev. D}\ }\textbf {\bibinfo {volume} {101}},\ \bibinfo {pages} {034017} (\bibinfo {year} {2018}{\natexlab{a}})}\BibitemShut {NoStop}%
\bibitem [{\citenamefont {Li}\ \emph {et~al.}(2018{\natexlab{a}})\citenamefont {Li}, \citenamefont {Sedrakian},\ and\ \citenamefont {Weber}}]{Li:2018kyx}%
  \BibitemOpen
  \bibfield  {author} {\bibinfo {author} {\bibfnamefont {J.~J.}\ \bibnamefont {Li}}, \bibinfo {author} {\bibfnamefont {A.}~\bibnamefont {Sedrakian}},\ and\ \bibinfo {author} {\bibfnamefont {F.}~\bibnamefont {Weber}},\ }\href {https://doi.org/https://doi.org/10.1016/j.physletb.2018.06.051} {\bibfield  {journal} {\bibinfo  {journal} {Phys. Lett. B}\ }\textbf {\bibinfo {volume} {783}},\ \bibinfo {pages} {234} (\bibinfo {year} {2018}{\natexlab{a}})}\BibitemShut {NoStop}%
\bibitem [{\citenamefont {Li}\ \emph {et~al.}(2018{\natexlab{b}})\citenamefont {Li}, \citenamefont {Long},\ and\ \citenamefont {Sedrakian}}]{Li:2018qap}%
  \BibitemOpen
  \bibfield  {author} {\bibinfo {author} {\bibfnamefont {J.~J.}\ \bibnamefont {Li}}, \bibinfo {author} {\bibfnamefont {W.~H.}\ \bibnamefont {Long}},\ and\ \bibinfo {author} {\bibfnamefont {A.}~\bibnamefont {Sedrakian}},\ }\href {https://doi.org/10.1140/epja/i2018-12566-6} {\bibfield  {journal} {\bibinfo  {journal} {Eur Phys. J. A}\ }\textbf {\bibinfo {volume} {54}},\ \bibinfo {pages} {133} (\bibinfo {year} {2018}{\natexlab{b}})}\BibitemShut {NoStop}%
\bibitem [{\citenamefont {Providência}\ \emph {et~al.}(2019)\citenamefont {Providência}, \citenamefont {Fortin}, \citenamefont {Pais},\ and\ \citenamefont {Rabhi}}]{Providencia:2019pny}%
  \BibitemOpen
  \bibfield  {author} {\bibinfo {author} {\bibfnamefont {C.}~\bibnamefont {Providência}}, \bibinfo {author} {\bibfnamefont {M.}~\bibnamefont {Fortin}}, \bibinfo {author} {\bibfnamefont {H.}~\bibnamefont {Pais}},\ and\ \bibinfo {author} {\bibfnamefont {A.}~\bibnamefont {Rabhi}},\ }\href {https://www.frontiersin.org/articles/10.3389/fspas.2019.00013} {\bibfield  {journal} {\bibinfo  {journal} {Front. Astron. Space Sci.}\ }\textbf {\bibinfo {volume} {6}} (\bibinfo {year} {2019})}\BibitemShut {NoStop}%
\bibitem [{\citenamefont {Sedrakian}\ \emph {et~al.}(2020)\citenamefont {Sedrakian}, \citenamefont {Weber},\ and\ \citenamefont {Li}}]{Sedrakian:2020qja}%
  \BibitemOpen
  \bibfield  {author} {\bibinfo {author} {\bibfnamefont {A.}~\bibnamefont {Sedrakian}}, \bibinfo {author} {\bibfnamefont {F.}~\bibnamefont {Weber}},\ and\ \bibinfo {author} {\bibfnamefont {J.~J.}\ \bibnamefont {Li}},\ }\href {https://doi.org/10.1103/PhysRevD.102.041301} {\bibfield  {journal} {\bibinfo  {journal} {Phys. Rev. D}\ }\textbf {\bibinfo {volume} {102}},\ \bibinfo {pages} {041301} (\bibinfo {year} {2020})}\BibitemShut {NoStop}%
\bibitem [{\citenamefont {Thapa}\ \emph {et~al.}(2021{\natexlab{a}})\citenamefont {Thapa}, \citenamefont {Sinha}, \citenamefont {Li},\ and\ \citenamefont {Sedrakian}}]{Thapa:2021ysx}%
  \BibitemOpen
  \bibfield  {author} {\bibinfo {author} {\bibfnamefont {V.~B.}\ \bibnamefont {Thapa}}, \bibinfo {author} {\bibfnamefont {M.}~\bibnamefont {Sinha}}, \bibinfo {author} {\bibfnamefont {J.-J.}\ \bibnamefont {Li}},\ and\ \bibinfo {author} {\bibfnamefont {A.}~\bibnamefont {Sedrakian}},\ }\href {https://doi.org/10.1103/PhysRevD.103.063004} {\bibfield  {journal} {\bibinfo  {journal} {Phys. Rev. D}\ }\textbf {\bibinfo {volume} {103}},\ \bibinfo {pages} {063004} (\bibinfo {year} {2021}{\natexlab{a}})}\BibitemShut {NoStop}%
\bibitem [{\citenamefont {Motta}\ and\ \citenamefont {Thomas}(2022)}]{Motta:2022nlj}%
  \BibitemOpen
  \bibfield  {author} {\bibinfo {author} {\bibfnamefont {T.~F.}\ \bibnamefont {Motta}}\ and\ \bibinfo {author} {\bibfnamefont {A.~W.}\ \bibnamefont {Thomas}},\ }\href {https://doi.org/10.1142/S0217732322300014} {\bibfield  {journal} {\bibinfo  {journal} {Mod. Phys. Lett. A}\ }\textbf {\bibinfo {volume} {37}},\ \bibinfo {pages} {2230001} (\bibinfo {year} {2022})},\ \Eprint {https://arxiv.org/abs/2201.11549} {arXiv:2201.11549 [nucl-th]} \BibitemShut {NoStop}%
\bibitem [{\citenamefont {Leong}\ \emph {et~al.}(2023)\citenamefont {Leong}, \citenamefont {Motta}, \citenamefont {Thomas},\ and\ \citenamefont {Guichon}}]{Leong:2023yma}%
  \BibitemOpen
  \bibfield  {author} {\bibinfo {author} {\bibfnamefont {J.}~\bibnamefont {Leong}}, \bibinfo {author} {\bibfnamefont {T.~F.}\ \bibnamefont {Motta}}, \bibinfo {author} {\bibfnamefont {A.~W.}\ \bibnamefont {Thomas}},\ and\ \bibinfo {author} {\bibfnamefont {P.~A.~M.}\ \bibnamefont {Guichon}},\ }\href {https://doi.org/10.1103/PhysRevC.108.015804} {\bibfield  {journal} {\bibinfo  {journal} {Phys. Rev. C}\ }\textbf {\bibinfo {volume} {108}},\ \bibinfo {pages} {015804} (\bibinfo {year} {2023})}\BibitemShut {NoStop}%
\bibitem [{\citenamefont {Demorest}\ \emph {et~al.}(2010)\citenamefont {Demorest}, \citenamefont {Pennucci}, \citenamefont {Ransom}, \citenamefont {Roberts},\ and\ \citenamefont {Hessels}}]{Demorest2010ShapiroStar}%
  \BibitemOpen
  \bibfield  {author} {\bibinfo {author} {\bibfnamefont {P.}~\bibnamefont {Demorest}}, \bibinfo {author} {\bibfnamefont {T.}~\bibnamefont {Pennucci}}, \bibinfo {author} {\bibfnamefont {S.}~\bibnamefont {Ransom}}, \bibinfo {author} {\bibfnamefont {M.}~\bibnamefont {Roberts}},\ and\ \bibinfo {author} {\bibfnamefont {J.}~\bibnamefont {Hessels}},\ }\href {https://doi.org/10.1038/nature09466} {\bibfield  {journal} {\bibinfo  {journal} {Nature}\ }\textbf {\bibinfo {volume} {467}},\ \bibinfo {pages} {1081} (\bibinfo {year} {2010})}\BibitemShut {NoStop}%
\bibitem [{\citenamefont {Antoniadis}\ \emph {et~al.}(2013)\citenamefont {Antoniadis} \emph {et~al.}}]{Antoniadis:2013pzd}%
  \BibitemOpen
  \bibfield  {author} {\bibinfo {author} {\bibfnamefont {J.}~\bibnamefont {Antoniadis}} \emph {et~al.},\ }\href {https://doi.org/10.1126/science.1233232} {\bibfield  {journal} {\bibinfo  {journal} {Science}\ }\textbf {\bibinfo {volume} {340}},\ \bibinfo {pages} {6131} (\bibinfo {year} {2013})},\ \Eprint {https://arxiv.org/abs/1304.6875} {arXiv:1304.6875 [astro-ph.HE]} \BibitemShut {NoStop}%
\bibitem [{\citenamefont {Fonseca}\ \emph {et~al.}(2016)\citenamefont {Fonseca} \emph {et~al.}}]{Fonseca2016}%
  \BibitemOpen
  \bibfield  {author} {\bibinfo {author} {\bibfnamefont {E.}~\bibnamefont {Fonseca}} \emph {et~al.},\ }\href {https://doi.org/10.3847/0004-637X/832/2/167} {\bibfield  {journal} {\bibinfo  {journal} {Astrophys. J.}\ }\textbf {\bibinfo {volume} {832}},\ \bibinfo {pages} {167} (\bibinfo {year} {2016})},\ \Eprint {https://arxiv.org/abs/1603.00545} {arXiv:1603.00545 [astro-ph.HE]} \BibitemShut {NoStop}%
\bibitem [{\citenamefont {Cromartie}\ \emph {et~al.}(2019)\citenamefont {Cromartie} \emph {et~al.}}]{Cromartie2020RelativisticPulsar}%
  \BibitemOpen
  \bibfield  {author} {\bibinfo {author} {\bibfnamefont {H.~T.}\ \bibnamefont {Cromartie}} \emph {et~al.} (\bibinfo {collaboration} {NANOGrav}),\ }\href {https://doi.org/10.1038/s41550-019-0880-2} {\bibfield  {journal} {\bibinfo  {journal} {Nature Astron.}\ }\textbf {\bibinfo {volume} {4}},\ \bibinfo {pages} {72} (\bibinfo {year} {2019})},\ \Eprint {https://arxiv.org/abs/1904.06759} {arXiv:1904.06759 [astro-ph.HE]} \BibitemShut {NoStop}%
\bibitem [{\citenamefont {Romani}\ \emph {et~al.}(2022)\citenamefont {Romani}, \citenamefont {Kandel}, \citenamefont {Filippenko}, \citenamefont {Brink},\ and\ \citenamefont {Zheng}}]{Romani:2022jhd}%
  \BibitemOpen
  \bibfield  {author} {\bibinfo {author} {\bibfnamefont {R.~W.}\ \bibnamefont {Romani}}, \bibinfo {author} {\bibfnamefont {D.}~\bibnamefont {Kandel}}, \bibinfo {author} {\bibfnamefont {A.~V.}\ \bibnamefont {Filippenko}}, \bibinfo {author} {\bibfnamefont {T.~G.}\ \bibnamefont {Brink}},\ and\ \bibinfo {author} {\bibfnamefont {W.}~\bibnamefont {Zheng}},\ }\href {https://doi.org/10.3847/2041-8213/ac8007} {\bibfield  {journal} {\bibinfo  {journal} {Astrophys. J. Lett.}\ }\textbf {\bibinfo {volume} {934}},\ \bibinfo {pages} {L17} (\bibinfo {year} {2022})},\ \Eprint {https://arxiv.org/abs/2207.05124} {arXiv:2207.05124 [astro-ph.HE]} \BibitemShut {NoStop}%
\bibitem [{\citenamefont {Haidenbauer}\ \emph {et~al.}(2017)\citenamefont {Haidenbauer}, \citenamefont {Mei{\ss}ner}, \citenamefont {Kaiser},\ and\ \citenamefont {Weise}}]{Haidenbauer:2016vfq}%
  \BibitemOpen
  \bibfield  {author} {\bibinfo {author} {\bibfnamefont {J.}~\bibnamefont {Haidenbauer}}, \bibinfo {author} {\bibfnamefont {U.~G.}\ \bibnamefont {Mei{\ss}ner}}, \bibinfo {author} {\bibfnamefont {N.}~\bibnamefont {Kaiser}},\ and\ \bibinfo {author} {\bibfnamefont {W.}~\bibnamefont {Weise}},\ }\href {https://doi.org/10.1140/epja/i2017-12316-4} {\bibfield  {journal} {\bibinfo  {journal} {Eur. Phys. J.}\ }\textbf {\bibinfo {volume} {A53}},\ \bibinfo {pages} {121} (\bibinfo {year} {2017})},\ \Eprint {https://arxiv.org/abs/1612.03758} {arXiv:1612.03758 [nucl-th]} \BibitemShut {NoStop}%
%%CITATION = ARXIV:1612.03758;%%
\bibitem [{\citenamefont {Logoteta}\ \emph {et~al.}(2019)\citenamefont {Logoteta}, \citenamefont {Vida\~na},\ and\ \citenamefont {Bombaci}}]{Logoteta:2019utx}%
  \BibitemOpen
  \bibfield  {author} {\bibinfo {author} {\bibfnamefont {D.}~\bibnamefont {Logoteta}}, \bibinfo {author} {\bibfnamefont {I.}~\bibnamefont {Vida\~na}},\ and\ \bibinfo {author} {\bibfnamefont {I.}~\bibnamefont {Bombaci}},\ }\href {https://doi.org/10.1140/epja/i2019-12909-9} {\bibfield  {journal} {\bibinfo  {journal} {Eur. Phys. J.}\ }\textbf {\bibinfo {volume} {A55}},\ \bibinfo {pages} {207} (\bibinfo {year} {2019})},\ \Eprint {https://arxiv.org/abs/1906.11722} {arXiv:1906.11722 [nucl-th]} \BibitemShut {NoStop}%
%%CITATION = ARXIV:1906.11722;%%
\bibitem [{\citenamefont {Gerstung}\ \emph {et~al.}(2020)\citenamefont {Gerstung}, \citenamefont {Kaiser},\ and\ \citenamefont {Weise}}]{Gerstung:2020ktv}%
  \BibitemOpen
  \bibfield  {author} {\bibinfo {author} {\bibfnamefont {D.}~\bibnamefont {Gerstung}}, \bibinfo {author} {\bibfnamefont {N.}~\bibnamefont {Kaiser}},\ and\ \bibinfo {author} {\bibfnamefont {W.}~\bibnamefont {Weise}},\ }\href {https://doi.org/10.1140/epja/s10050-020-00180-2} {\bibfield  {journal} {\bibinfo  {journal} {Eur. Phys. J. A}\ }\textbf {\bibinfo {volume} {56}},\ \bibinfo {pages} {175} (\bibinfo {year} {2020})},\ \Eprint {https://arxiv.org/abs/2001.10563} {arXiv:2001.10563 [nucl-th]} \BibitemShut {NoStop}%
\bibitem [{\citenamefont {Takatsuka}\ \emph {et~al.}(2002)\citenamefont {Takatsuka}, \citenamefont {Nishizaki},\ and\ \citenamefont {Yamamoto}}]{Takatsuka2002}%
  \BibitemOpen
  \bibfield  {author} {\bibinfo {author} {\bibfnamefont {T.}~\bibnamefont {Takatsuka}}, \bibinfo {author} {\bibfnamefont {S.}~\bibnamefont {Nishizaki}},\ and\ \bibinfo {author} {\bibfnamefont {Y.}~\bibnamefont {Yamamoto}},\ }\href {https://doi.org/10.1140/epja1339-35} {\bibfield  {journal} {\bibinfo  {journal} {Eur. Phys. J. A}\ }\textbf {\bibinfo {volume} {13}},\ \bibinfo {pages} {213} (\bibinfo {year} {2002})}\BibitemShut {NoStop}%
\bibitem [{\citenamefont {Takatsuka}(2004)}]{Takatsuka:2004ch}%
  \BibitemOpen
  \bibfield  {author} {\bibinfo {author} {\bibfnamefont {T.}~\bibnamefont {Takatsuka}},\ }\href {https://doi.org/10.1143/PTPS.156.84} {\bibfield  {journal} {\bibinfo  {journal} {Prog. Theor. Phys. Suppl.}\ }\textbf {\bibinfo {volume} {156}},\ \bibinfo {pages} {84} (\bibinfo {year} {2004})}\BibitemShut {NoStop}%
%%CITATION = PTPSA,156,84;%%
\bibitem [{\citenamefont {Vida\~na}\ \emph {et~al.}(2011)\citenamefont {Vida\~na}, \citenamefont {Logoteta}, \citenamefont {Providencia}, \citenamefont {Polls},\ and\ \citenamefont {Bombaci}}]{Vidana:2010ip}%
  \BibitemOpen
  \bibfield  {author} {\bibinfo {author} {\bibfnamefont {I.}~\bibnamefont {Vida\~na}}, \bibinfo {author} {\bibfnamefont {D.}~\bibnamefont {Logoteta}}, \bibinfo {author} {\bibfnamefont {C.}~\bibnamefont {Providencia}}, \bibinfo {author} {\bibfnamefont {A.}~\bibnamefont {Polls}},\ and\ \bibinfo {author} {\bibfnamefont {I.}~\bibnamefont {Bombaci}},\ }\href {https://doi.org/10.1209/0295-5075/94/11002} {\bibfield  {journal} {\bibinfo  {journal} {Europhys. Lett.}\ }\textbf {\bibinfo {volume} {94}},\ \bibinfo {pages} {11002} (\bibinfo {year} {2011})},\ \Eprint {https://arxiv.org/abs/1006.5660} {arXiv:1006.5660 [nucl-th]} \BibitemShut {NoStop}%
%%CITATION = ARXIV:1006.5660;%%
\bibitem [{\citenamefont {Yamamoto}\ \emph {et~al.}(2013)\citenamefont {Yamamoto}, \citenamefont {Furumoto}, \citenamefont {Yasutake},\ and\ \citenamefont {Rijken}}]{Yamamoto:2013ada}%
  \BibitemOpen
  \bibfield  {author} {\bibinfo {author} {\bibfnamefont {Y.}~\bibnamefont {Yamamoto}}, \bibinfo {author} {\bibfnamefont {T.}~\bibnamefont {Furumoto}}, \bibinfo {author} {\bibfnamefont {N.}~\bibnamefont {Yasutake}},\ and\ \bibinfo {author} {\bibfnamefont {T.~A.}\ \bibnamefont {Rijken}},\ }\href {https://doi.org/10.1103/PhysRevC.88.022801} {\bibfield  {journal} {\bibinfo  {journal} {Phys. Rev. C}\ }\textbf {\bibinfo {volume} {88}},\ \bibinfo {pages} {022801} (\bibinfo {year} {2013})},\ \Eprint {https://arxiv.org/abs/1308.2130} {arXiv:1308.2130 [nucl-th]} \BibitemShut {NoStop}%
%%CITATION = ARXIV:1308.2130;%%
\bibitem [{\citenamefont {Yamamoto}\ \emph {et~al.}(2014)\citenamefont {Yamamoto}, \citenamefont {Furumoto}, \citenamefont {Yasutake},\ and\ \citenamefont {Rijken}}]{Yamamoto:2014jga}%
  \BibitemOpen
  \bibfield  {author} {\bibinfo {author} {\bibfnamefont {Y.}~\bibnamefont {Yamamoto}}, \bibinfo {author} {\bibfnamefont {T.}~\bibnamefont {Furumoto}}, \bibinfo {author} {\bibfnamefont {N.}~\bibnamefont {Yasutake}},\ and\ \bibinfo {author} {\bibfnamefont {T.~A.}\ \bibnamefont {Rijken}},\ }\href {https://doi.org/10.1103/PhysRevC.90.045805} {\bibfield  {journal} {\bibinfo  {journal} {Phys. Rev. C}\ }\textbf {\bibinfo {volume} {90}},\ \bibinfo {pages} {045805} (\bibinfo {year} {2014})},\ \Eprint {https://arxiv.org/abs/1406.4332} {arXiv:1406.4332 [nucl-th]} \BibitemShut {NoStop}%
\bibitem [{\citenamefont {Lonardoni}\ \emph {et~al.}(2015)\citenamefont {Lonardoni}, \citenamefont {Lovato}, \citenamefont {Gandolfi},\ and\ \citenamefont {Pederiva}}]{PhysRevLett.114.092301}%
  \BibitemOpen
  \bibfield  {author} {\bibinfo {author} {\bibfnamefont {D.}~\bibnamefont {Lonardoni}}, \bibinfo {author} {\bibfnamefont {A.}~\bibnamefont {Lovato}}, \bibinfo {author} {\bibfnamefont {S.}~\bibnamefont {Gandolfi}},\ and\ \bibinfo {author} {\bibfnamefont {F.}~\bibnamefont {Pederiva}},\ }\href {https://doi.org/10.1103/PhysRevLett.114.092301} {\bibfield  {journal} {\bibinfo  {journal} {Phys. Rev. Lett.}\ }\textbf {\bibinfo {volume} {114}},\ \bibinfo {pages} {092301} (\bibinfo {year} {2015})}\BibitemShut {NoStop}%
\bibitem [{\citenamefont {Tong}\ \emph {et~al.}(2025)\citenamefont {Tong}, \citenamefont {Elhatisari},\ and\ \citenamefont {Meißner}}]{tong2024ab}%
  \BibitemOpen
  \bibfield  {author} {\bibinfo {author} {\bibfnamefont {H.}~\bibnamefont {Tong}}, \bibinfo {author} {\bibfnamefont {S.}~\bibnamefont {Elhatisari}},\ and\ \bibinfo {author} {\bibfnamefont {U.-G.}\ \bibnamefont {Meißner}},\ }\href {https://www.sciencedirect.com/science/article/pii/S2095927325000210} {\bibfield  {journal} {\bibinfo  {journal} {Sci. Bull.}\ } (\bibinfo {year} {2025})}\BibitemShut {NoStop}%
\bibitem [{\citenamefont {Burgio}\ \emph {et~al.}(2021)\citenamefont {Burgio}, \citenamefont {Schulze}, \citenamefont {Vidana},\ and\ \citenamefont {Wei}}]{Burgio:2021vgk}%
  \BibitemOpen
  \bibfield  {author} {\bibinfo {author} {\bibfnamefont {G.~F.}\ \bibnamefont {Burgio}}, \bibinfo {author} {\bibfnamefont {H.~J.}\ \bibnamefont {Schulze}}, \bibinfo {author} {\bibfnamefont {I.}~\bibnamefont {Vidana}},\ and\ \bibinfo {author} {\bibfnamefont {J.~B.}\ \bibnamefont {Wei}},\ }\href {https://doi.org/10.1016/j.ppnp.2021.103879} {\bibfield  {journal} {\bibinfo  {journal} {Prog. Part. Nucl. Phys.}\ }\textbf {\bibinfo {volume} {120}},\ \bibinfo {pages} {103879} (\bibinfo {year} {2021})},\ \Eprint {https://arxiv.org/abs/2105.03747} {arXiv:2105.03747 [nucl-th]} \BibitemShut {NoStop}%
\bibitem [{\citenamefont {{Sedrakian}}\ \emph {et~al.}(2022)\citenamefont {{Sedrakian}}, \citenamefont {{Li}},\ and\ \citenamefont {{Weber}}}]{Sedrakian2022}%
  \BibitemOpen
  \bibfield  {author} {\bibinfo {author} {\bibfnamefont {A.}~\bibnamefont {{Sedrakian}}}, \bibinfo {author} {\bibfnamefont {J.-J.}\ \bibnamefont {{Li}}},\ and\ \bibinfo {author} {\bibfnamefont {F.}~\bibnamefont {{Weber}}},\ }\href {https://doi.org/10.1142/9789811220944_0005} {\emph {\bibinfo {title} {Astrophysics in the XXI Century with Compact Stars.}}}\ (\bibinfo  {publisher} {World Scientific},\ \bibinfo {year} {2022})\ pp.\ \bibinfo {pages} {153--199}\BibitemShut {NoStop}%
\bibitem [{\citenamefont {Sedrakian}\ \emph {et~al.}(2023)\citenamefont {Sedrakian}, \citenamefont {Li},\ and\ \citenamefont {Weber}}]{Sedrakian:2022ata}%
  \BibitemOpen
  \bibfield  {author} {\bibinfo {author} {\bibfnamefont {A.}~\bibnamefont {Sedrakian}}, \bibinfo {author} {\bibfnamefont {J.-J.}\ \bibnamefont {Li}},\ and\ \bibinfo {author} {\bibfnamefont {F.}~\bibnamefont {Weber}},\ }\href {https://doi.org/10.1016/j.ppnp.2023.104041} {\bibfield  {journal} {\bibinfo  {journal} {Prog. Part. Nucl. Phys.}\ }\textbf {\bibinfo {volume} {131}},\ \bibinfo {pages} {104041} (\bibinfo {year} {2023})},\ \Eprint {https://arxiv.org/abs/2212.01086} {arXiv:2212.01086 [nucl-th]} \BibitemShut {NoStop}%
\bibitem [{\citenamefont {Ribes}\ \emph {et~al.}(2019)\citenamefont {Ribes}, \citenamefont {Ramos}, \citenamefont {Tolos}, \citenamefont {Gonzalez-Boquera},\ and\ \citenamefont {Centelles}}]{Ribes:2019kno}%
  \BibitemOpen
  \bibfield  {author} {\bibinfo {author} {\bibfnamefont {P.}~\bibnamefont {Ribes}}, \bibinfo {author} {\bibfnamefont {A.}~\bibnamefont {Ramos}}, \bibinfo {author} {\bibfnamefont {L.}~\bibnamefont {Tolos}}, \bibinfo {author} {\bibfnamefont {C.}~\bibnamefont {Gonzalez-Boquera}},\ and\ \bibinfo {author} {\bibfnamefont {M.}~\bibnamefont {Centelles}},\ }\href {https://doi.org/10.3847/1538-4357/ab3a93} {\bibfield  {journal} {\bibinfo  {journal} {Astrophys. J.}\ }\textbf {\bibinfo {volume} {883}},\ \bibinfo {pages} {168} (\bibinfo {year} {2019})},\ \Eprint {https://arxiv.org/abs/1907.08583} {arXiv:1907.08583 [astro-ph.HE]} \BibitemShut {NoStop}%
\bibitem [{\citenamefont {Muto}\ \emph {et~al.}(2021)\citenamefont {Muto}, \citenamefont {Maruyama},\ and\ \citenamefont {Tatsumi}}]{Muto:2021jms}%
  \BibitemOpen
  \bibfield  {author} {\bibinfo {author} {\bibfnamefont {T.}~\bibnamefont {Muto}}, \bibinfo {author} {\bibfnamefont {T.}~\bibnamefont {Maruyama}},\ and\ \bibinfo {author} {\bibfnamefont {T.}~\bibnamefont {Tatsumi}},\ }\href {https://doi.org/10.1016/j.physletb.2021.136587} {\bibfield  {journal} {\bibinfo  {journal} {Phys. Lett. B}\ }\textbf {\bibinfo {volume} {820}},\ \bibinfo {pages} {136587} (\bibinfo {year} {2021})},\ \Eprint {https://arxiv.org/abs/2106.03449} {arXiv:2106.03449 [nucl-th]} \BibitemShut {NoStop}%
\bibitem [{\citenamefont {Thapa}\ \emph {et~al.}(2021{\natexlab{b}})\citenamefont {Thapa}, \citenamefont {Sinha}, \citenamefont {Li},\ and\ \citenamefont {Sedrakian}}]{Thapa:2021kfo}%
  \BibitemOpen
  \bibfield  {author} {\bibinfo {author} {\bibfnamefont {V.~B.}\ \bibnamefont {Thapa}}, \bibinfo {author} {\bibfnamefont {M.}~\bibnamefont {Sinha}}, \bibinfo {author} {\bibfnamefont {J.~J.}\ \bibnamefont {Li}},\ and\ \bibinfo {author} {\bibfnamefont {A.}~\bibnamefont {Sedrakian}},\ }\href {https://doi.org/10.1103/PhysRevD.103.063004} {\bibfield  {journal} {\bibinfo  {journal} {Phys. Rev. D}\ }\textbf {\bibinfo {volume} {103}},\ \bibinfo {pages} {063004} (\bibinfo {year} {2021}{\natexlab{b}})},\ \Eprint {https://arxiv.org/abs/2102.08787} {arXiv:2102.08787 [astro-ph.HE]} \BibitemShut {NoStop}%
\bibitem [{\citenamefont {Alford}\ \emph {et~al.}(2005)\citenamefont {Alford}, \citenamefont {Braby}, \citenamefont {Paris},\ and\ \citenamefont {Reddy}}]{Alford:2004pf}%
  \BibitemOpen
  \bibfield  {author} {\bibinfo {author} {\bibfnamefont {M.}~\bibnamefont {Alford}}, \bibinfo {author} {\bibfnamefont {M.}~\bibnamefont {Braby}}, \bibinfo {author} {\bibfnamefont {M.~W.}\ \bibnamefont {Paris}},\ and\ \bibinfo {author} {\bibfnamefont {S.}~\bibnamefont {Reddy}},\ }\href {https://doi.org/10.1086/430902} {\bibfield  {journal} {\bibinfo  {journal} {Astrophys. J.}\ }\textbf {\bibinfo {volume} {629}},\ \bibinfo {pages} {969} (\bibinfo {year} {2005})},\ \Eprint {https://arxiv.org/abs/nucl-th/0411016} {arXiv:nucl-th/0411016} \BibitemShut {NoStop}%
\bibitem [{\citenamefont {Oertel}\ \emph {et~al.}(2016)\citenamefont {Oertel}, \citenamefont {Gulminelli}, \citenamefont {Provid\^encia},\ and\ \citenamefont {Raduta}}]{Oertel:2016xsn}%
  \BibitemOpen
  \bibfield  {author} {\bibinfo {author} {\bibfnamefont {M.}~\bibnamefont {Oertel}}, \bibinfo {author} {\bibfnamefont {F.}~\bibnamefont {Gulminelli}}, \bibinfo {author} {\bibfnamefont {C.}~\bibnamefont {Provid\^encia}},\ and\ \bibinfo {author} {\bibfnamefont {A.~R.}\ \bibnamefont {Raduta}},\ }\href {https://doi.org/10.1140/epja/i2016-16050-1} {\bibfield  {journal} {\bibinfo  {journal} {Eur. Phys. J. A}\ }\textbf {\bibinfo {volume} {52}},\ \bibinfo {pages} {50} (\bibinfo {year} {2016})},\ \Eprint {https://arxiv.org/abs/1601.00435} {arXiv:1601.00435 [nucl-th]} \BibitemShut {NoStop}%
\bibitem [{\citenamefont {Fortin}\ \emph {et~al.}(2018{\natexlab{b}})\citenamefont {Fortin}, \citenamefont {Oertel},\ and\ \citenamefont {Provid\^encia}}]{Fortin:2017dsj}%
  \BibitemOpen
  \bibfield  {author} {\bibinfo {author} {\bibfnamefont {M.}~\bibnamefont {Fortin}}, \bibinfo {author} {\bibfnamefont {M.}~\bibnamefont {Oertel}},\ and\ \bibinfo {author} {\bibfnamefont {C.}~\bibnamefont {Provid\^encia}},\ }\href {https://doi.org/10.1017/pasa.2018.32} {\bibfield  {journal} {\bibinfo  {journal} {Publ. Astron. Soc. Austral.}\ }\textbf {\bibinfo {volume} {35}},\ \bibinfo {pages} {44} (\bibinfo {year} {2018}{\natexlab{b}})},\ \Eprint {https://arxiv.org/abs/1711.09427} {arXiv:1711.09427 [astro-ph.HE]} \BibitemShut {NoStop}%
\bibitem [{\citenamefont {Lenka}\ \emph {et~al.}(2019)\citenamefont {Lenka}, \citenamefont {Char},\ and\ \citenamefont {Banik}}]{lenka2019properties}%
  \BibitemOpen
  \bibfield  {author} {\bibinfo {author} {\bibfnamefont {S.~S.}\ \bibnamefont {Lenka}}, \bibinfo {author} {\bibfnamefont {P.}~\bibnamefont {Char}},\ and\ \bibinfo {author} {\bibfnamefont {S.}~\bibnamefont {Banik}},\ }\href {https://doi.org/10.1088/1361-6471/ab36a2} {\bibfield  {journal} {\bibinfo  {journal} {J. Phys. G}\ }\textbf {\bibinfo {volume} {46}},\ \bibinfo {pages} {105201} (\bibinfo {year} {2019})},\ \Eprint {https://arxiv.org/abs/1805.09492} {arXiv:1805.09492 [astro-ph.HE]} \BibitemShut {NoStop}%
\bibitem [{\citenamefont {Peres}\ \emph {et~al.}(2013)\citenamefont {Peres}, \citenamefont {Oertel},\ and\ \citenamefont {Novak}}]{PhysRevD.87.043006}%
  \BibitemOpen
  \bibfield  {author} {\bibinfo {author} {\bibfnamefont {B.}~\bibnamefont {Peres}}, \bibinfo {author} {\bibfnamefont {M.}~\bibnamefont {Oertel}},\ and\ \bibinfo {author} {\bibfnamefont {J.}~\bibnamefont {Novak}},\ }\href {https://doi.org/10.1103/PhysRevD.87.043006} {\bibfield  {journal} {\bibinfo  {journal} {Phys. Rev. D}\ }\textbf {\bibinfo {volume} {87}},\ \bibinfo {pages} {043006} (\bibinfo {year} {2013})}\BibitemShut {NoStop}%
\bibitem [{\citenamefont {Raduta}\ \emph {et~al.}(2021{\natexlab{a}})\citenamefont {Raduta}, \citenamefont {Nacu},\ and\ \citenamefont {Oertel}}]{Raduta:2021coc}%
  \BibitemOpen
  \bibfield  {author} {\bibinfo {author} {\bibfnamefont {A.~R.}\ \bibnamefont {Raduta}}, \bibinfo {author} {\bibfnamefont {F.}~\bibnamefont {Nacu}},\ and\ \bibinfo {author} {\bibfnamefont {M.}~\bibnamefont {Oertel}},\ }\href {https://doi.org/10.1140/epja/s10050-021-00628-z} {\bibfield  {journal} {\bibinfo  {journal} {Eur. Phys. J. A}\ }\textbf {\bibinfo {volume} {57}},\ \bibinfo {pages} {329} (\bibinfo {year} {2021}{\natexlab{a}})},\ \Eprint {https://arxiv.org/abs/2109.00251} {arXiv:2109.00251 [nucl-th]} \BibitemShut {NoStop}%
\bibitem [{\citenamefont {Raduta}(2022)}]{Raduta:2022elz}%
  \BibitemOpen
  \bibfield  {author} {\bibinfo {author} {\bibfnamefont {A.~R.}\ \bibnamefont {Raduta}},\ }\href {https://doi.org/10.1140/epja/s10050-022-00772-0} {\bibfield  {journal} {\bibinfo  {journal} {Eur. Phys. J. A}\ }\textbf {\bibinfo {volume} {58}},\ \bibinfo {pages} {115} (\bibinfo {year} {2022})}\BibitemShut {NoStop}%
\bibitem [{\citenamefont {Sedrakian}\ and\ \citenamefont {Harutyunyan}(2021)}]{Sedrakian_2021}%
  \BibitemOpen
  \bibfield  {author} {\bibinfo {author} {\bibfnamefont {A.}~\bibnamefont {Sedrakian}}\ and\ \bibinfo {author} {\bibfnamefont {A.}~\bibnamefont {Harutyunyan}},\ }\href {https://doi.org/10.3390/universe7100382} {\bibfield  {journal} {\bibinfo  {journal} {Universe}\ }\textbf {\bibinfo {volume} {7}},\ \bibinfo {pages} {382} (\bibinfo {year} {2021})}\BibitemShut {NoStop}%
\bibitem [{\citenamefont {Malik}\ \emph {et~al.}(2021)\citenamefont {Malik}, \citenamefont {Banik},\ and\ \citenamefont {Bandyopadhyay}}]{Malik:2021nas}%
  \BibitemOpen
  \bibfield  {author} {\bibinfo {author} {\bibfnamefont {T.}~\bibnamefont {Malik}}, \bibinfo {author} {\bibfnamefont {S.}~\bibnamefont {Banik}},\ and\ \bibinfo {author} {\bibfnamefont {D.}~\bibnamefont {Bandyopadhyay}},\ }\href {https://doi.org/10.3847/1538-4357/abe860} {\bibfield  {journal} {\bibinfo  {journal} {Astrophys. J.}\ }\textbf {\bibinfo {volume} {910}},\ \bibinfo {pages} {96} (\bibinfo {year} {2021})},\ \Eprint {https://arxiv.org/abs/2104.00775} {arXiv:2104.00775 [astro-ph.HE]} \BibitemShut {NoStop}%
\bibitem [{\citenamefont {Sedrakian}\ and\ \citenamefont {Harutyunyan}(2022)}]{Sedrakian_2022}%
  \BibitemOpen
  \bibfield  {author} {\bibinfo {author} {\bibfnamefont {A.}~\bibnamefont {Sedrakian}}\ and\ \bibinfo {author} {\bibfnamefont {A.}~\bibnamefont {Harutyunyan}},\ }\href {https://doi.org/10.1140/epja/s10050-022-00792-w} {\bibfield  {journal} {\bibinfo  {journal} {Eur. Phys. J. A}\ }\textbf {\bibinfo {volume} {58}},\ \bibinfo {pages} {137} (\bibinfo {year} {2022})},\ \Eprint {https://arxiv.org/abs/2202.12083} {arXiv:2202.12083 [nucl-th]} \BibitemShut {NoStop}%
\bibitem [{\citenamefont {Guichon}\ \emph {et~al.}(2024)\citenamefont {Guichon}, \citenamefont {Stone},\ and\ \citenamefont {Thomas}}]{Guichon:2023iev}%
  \BibitemOpen
  \bibfield  {author} {\bibinfo {author} {\bibfnamefont {P.~A.~M.}\ \bibnamefont {Guichon}}, \bibinfo {author} {\bibfnamefont {J.~R.}\ \bibnamefont {Stone}},\ and\ \bibinfo {author} {\bibfnamefont {A.~W.}\ \bibnamefont {Thomas}},\ }\href {https://doi.org/10.1103/PhysRevD.109.083035} {\bibfield  {journal} {\bibinfo  {journal} {Phys. Rev. D}\ }\textbf {\bibinfo {volume} {109}},\ \bibinfo {pages} {083035} (\bibinfo {year} {2024})}\BibitemShut {NoStop}%
\bibitem [{\citenamefont {Stone}\ \emph {et~al.}(2021)\citenamefont {Stone}, \citenamefont {Dexheimer}, \citenamefont {Guichon}, \citenamefont {Thomas},\ and\ \citenamefont {Typel}}]{Stone2021}%
  \BibitemOpen
  \bibfield  {author} {\bibinfo {author} {\bibfnamefont {J.~R.}\ \bibnamefont {Stone}}, \bibinfo {author} {\bibfnamefont {V.}~\bibnamefont {Dexheimer}}, \bibinfo {author} {\bibfnamefont {P.~A.~M.}\ \bibnamefont {Guichon}}, \bibinfo {author} {\bibfnamefont {A.~W.}\ \bibnamefont {Thomas}},\ and\ \bibinfo {author} {\bibfnamefont {S.}~\bibnamefont {Typel}},\ }\href {https://doi.org/10.1093/mnras/staa4006} {\bibfield  {journal} {\bibinfo  {journal} {Mon. Not. Roy. Astron. Soc.}\ }\textbf {\bibinfo {volume} {502}},\ \bibinfo {pages} {3476} (\bibinfo {year} {2021})}\BibitemShut {NoStop}%
\bibitem [{\citenamefont {Thapa}\ \emph {et~al.}(2023)\citenamefont {Thapa}, \citenamefont {Beznogov}, \citenamefont {Raduta},\ and\ \citenamefont {Thakur}}]{PhysRevD.107.103054}%
  \BibitemOpen
  \bibfield  {author} {\bibinfo {author} {\bibfnamefont {V.~B.}\ \bibnamefont {Thapa}}, \bibinfo {author} {\bibfnamefont {M.~V.}\ \bibnamefont {Beznogov}}, \bibinfo {author} {\bibfnamefont {A.~R.}\ \bibnamefont {Raduta}},\ and\ \bibinfo {author} {\bibfnamefont {P.}~\bibnamefont {Thakur}},\ }\href {https://doi.org/10.1103/PhysRevD.107.103054} {\bibfield  {journal} {\bibinfo  {journal} {Phys. Rev. D}\ }\textbf {\bibinfo {volume} {107}},\ \bibinfo {pages} {103054} (\bibinfo {year} {2023})}\BibitemShut {NoStop}%
\bibitem [{\citenamefont {Tsiopelas}\ \emph {et~al.}(2024)\citenamefont {Tsiopelas}, \citenamefont {Sedrakian},\ and\ \citenamefont {Oertel}}]{tsiopelas2024finitetemperature}%
  \BibitemOpen
  \bibfield  {author} {\bibinfo {author} {\bibfnamefont {S.}~\bibnamefont {Tsiopelas}}, \bibinfo {author} {\bibfnamefont {A.}~\bibnamefont {Sedrakian}},\ and\ \bibinfo {author} {\bibfnamefont {M.}~\bibnamefont {Oertel}},\ }\href {https://doi.org/10.1140/epja/s10050-024-01351-1} {\bibfield  {journal} {\bibinfo  {journal} {Eur. Phys. J. A}\ }\textbf {\bibinfo {volume} {60}},\ \bibinfo {pages} {127} (\bibinfo {year} {2024})}\BibitemShut {NoStop}%
\bibitem [{\citenamefont {Chatterjee}\ and\ \citenamefont {Vida\~na}(2016)}]{Chatterjee:2015pua}%
  \BibitemOpen
  \bibfield  {author} {\bibinfo {author} {\bibfnamefont {D.}~\bibnamefont {Chatterjee}}\ and\ \bibinfo {author} {\bibfnamefont {I.}~\bibnamefont {Vida\~na}},\ }\href {https://doi.org/10.1140/epja/i2016-16029-x} {\bibfield  {journal} {\bibinfo  {journal} {Eur. Phys. J. A}\ }\textbf {\bibinfo {volume} {52}},\ \bibinfo {pages} {29} (\bibinfo {year} {2016})},\ \Eprint {https://arxiv.org/abs/1510.06306} {arXiv:1510.06306 [nucl-th]} \BibitemShut {NoStop}%
\bibitem [{\citenamefont {Tolos}\ and\ \citenamefont {Fabbietti}(2020)}]{Tolos:2020aln}%
  \BibitemOpen
  \bibfield  {author} {\bibinfo {author} {\bibfnamefont {L.}~\bibnamefont {Tolos}}\ and\ \bibinfo {author} {\bibfnamefont {L.}~\bibnamefont {Fabbietti}},\ }\href {https://doi.org/10.1016/j.ppnp.2020.103770} {\bibfield  {journal} {\bibinfo  {journal} {Prog. Part. Nucl. Phys.}\ }\textbf {\bibinfo {volume} {112}},\ \bibinfo {pages} {103770} (\bibinfo {year} {2020})},\ \Eprint {https://arxiv.org/abs/2002.09223} {arXiv:2002.09223 [nucl-ex]} \BibitemShut {NoStop}%
\bibitem [{\citenamefont {Schaffner-Bielich}(2020)}]{Schaffner-Bielich:2020psc}%
  \BibitemOpen
  \bibfield  {author} {\bibinfo {author} {\bibfnamefont {J.}~\bibnamefont {Schaffner-Bielich}},\ }\href {https://doi.org/10.1017/9781316848357} {\emph {\bibinfo {title} {{Compact Star Physics}}}}\ (\bibinfo  {publisher} {Cambridge University Press},\ \bibinfo {year} {2020})\BibitemShut {NoStop}%
\bibitem [{\citenamefont {Logoteta}(2021)}]{Logoteta:2021iuy}%
  \BibitemOpen
  \bibfield  {author} {\bibinfo {author} {\bibfnamefont {D.}~\bibnamefont {Logoteta}},\ }\href {https://doi.org/10.3390/universe7110408} {\bibfield  {journal} {\bibinfo  {journal} {Universe}\ }\textbf {\bibinfo {volume} {7}},\ \bibinfo {pages} {408} (\bibinfo {year} {2021})}\BibitemShut {NoStop}%
\bibitem [{\citenamefont {Abbott}\ \emph {et~al.}(2017{\natexlab{a}})\citenamefont {Abbott} \emph {et~al.}}]{LIGOScientific:2017vwq}%
  \BibitemOpen
  \bibfield  {author} {\bibinfo {author} {\bibfnamefont {B.~P.}\ \bibnamefont {Abbott}} \emph {et~al.} (\bibinfo {collaboration} {LIGO Scientific, Virgo}),\ }\href {https://doi.org/10.1103/PhysRevLett.119.161101} {\bibfield  {journal} {\bibinfo  {journal} {Phys. Rev. Lett.}\ }\textbf {\bibinfo {volume} {119}},\ \bibinfo {pages} {161101} (\bibinfo {year} {2017}{\natexlab{a}})},\ \Eprint {https://arxiv.org/abs/1710.05832} {arXiv:1710.05832 [gr-qc]} \BibitemShut {NoStop}%
\bibitem [{\citenamefont {Abbott}\ \emph {et~al.}(2017{\natexlab{b}})\citenamefont {Abbott} \emph {et~al.}}]{Abbott2017em}%
  \BibitemOpen
  \bibfield  {author} {\bibinfo {author} {\bibfnamefont {B.~P.}\ \bibnamefont {Abbott}} \emph {et~al.} (\bibinfo {collaboration} {LIGO Scientific, Virgo}),\ }\href {https://doi.org/10.3847/2041-8213/aa91c9} {\bibfield  {journal} {\bibinfo  {journal} {\apjl}\ }\textbf {\bibinfo {volume} {848}},\ \bibinfo {eid} {L12} (\bibinfo {year} {2017}{\natexlab{b}})},\ \Eprint {https://arxiv.org/abs/1710.05833} {arXiv:1710.05833 [astro-ph.HE]} \BibitemShut {NoStop}%
\bibitem [{\citenamefont {Abbott}\ \emph {et~al.}(2018)\citenamefont {Abbott} \emph {et~al.}}]{Abbott2018GW170817:State}%
  \BibitemOpen
  \bibfield  {author} {\bibinfo {author} {\bibfnamefont {B.~P.}\ \bibnamefont {Abbott}} \emph {et~al.} (\bibinfo {collaboration} {LIGO Scientific, Virgo}),\ }\href {https://doi.org/10.1103/PhysRevLett.121.161101} {\bibfield  {journal} {\bibinfo  {journal} {Phys. Rev. Lett.}\ }\textbf {\bibinfo {volume} {121}},\ \bibinfo {pages} {161101} (\bibinfo {year} {2018})},\ \Eprint {https://arxiv.org/abs/1805.11581} {arXiv:1805.11581 [gr-qc]} \BibitemShut {NoStop}%
\bibitem [{\citenamefont {Abbott}\ \emph {et~al.}(2019)\citenamefont {Abbott} \emph {et~al.}}]{Abbott2019}%
  \BibitemOpen
  \bibfield  {author} {\bibinfo {author} {\bibfnamefont {B.~P.}\ \bibnamefont {Abbott}} \emph {et~al.} (\bibinfo {collaboration} {LIGO Scientific Collaboration and Virgo Collaboration}),\ }\href {https://doi.org/10.1103/PhysRevX.9.011001} {\bibfield  {journal} {\bibinfo  {journal} {Phys. Rev. X}\ }\textbf {\bibinfo {volume} {9}},\ \bibinfo {pages} {011001} (\bibinfo {year} {2019})}\BibitemShut {NoStop}%
\bibitem [{\citenamefont {Punturo}\ \emph {et~al.}(2010)\citenamefont {Punturo} \emph {et~al.}}]{Punturo:2010zz}%
  \BibitemOpen
  \bibfield  {author} {\bibinfo {author} {\bibfnamefont {M.}~\bibnamefont {Punturo}} \emph {et~al.},\ }\href {https://doi.org/10.1088/0264-9381/27/19/194002} {\bibfield  {journal} {\bibinfo  {journal} {Class. Quant. Grav.}\ }\textbf {\bibinfo {volume} {27}},\ \bibinfo {pages} {194002} (\bibinfo {year} {2010})}\BibitemShut {NoStop}%
\bibitem [{\citenamefont {Abbott}\ \emph {et~al.}(2017{\natexlab{c}})\citenamefont {Abbott} \emph {et~al.}}]{LIGOScientific:2016wof}%
  \BibitemOpen
  \bibfield  {author} {\bibinfo {author} {\bibfnamefont {B.~P.}\ \bibnamefont {Abbott}} \emph {et~al.} (\bibinfo {collaboration} {LIGO Scientific}),\ }\href {https://doi.org/10.1088/1361-6382/aa51f4} {\bibfield  {journal} {\bibinfo  {journal} {Class. Quant. Grav.}\ }\textbf {\bibinfo {volume} {34}},\ \bibinfo {pages} {044001} (\bibinfo {year} {2017}{\natexlab{c}})},\ \Eprint {https://arxiv.org/abs/1607.08697} {arXiv:1607.08697 [astro-ph.IM]} \BibitemShut {NoStop}%
\bibitem [{\citenamefont {Torres-Rivas}\ \emph {et~al.}(2019)\citenamefont {Torres-Rivas}, \citenamefont {Chatziioannou}, \citenamefont {Bauswein},\ and\ \citenamefont {Clark}}]{Torres-Rivas:2018svp}%
  \BibitemOpen
  \bibfield  {author} {\bibinfo {author} {\bibfnamefont {A.}~\bibnamefont {Torres-Rivas}}, \bibinfo {author} {\bibfnamefont {K.}~\bibnamefont {Chatziioannou}}, \bibinfo {author} {\bibfnamefont {A.}~\bibnamefont {Bauswein}},\ and\ \bibinfo {author} {\bibfnamefont {J.~A.}\ \bibnamefont {Clark}},\ }\href {https://doi.org/10.1103/PhysRevD.99.044014} {\bibfield  {journal} {\bibinfo  {journal} {Phys. Rev. D}\ }\textbf {\bibinfo {volume} {99}},\ \bibinfo {pages} {044014} (\bibinfo {year} {2019})},\ \Eprint {https://arxiv.org/abs/1811.08931} {arXiv:1811.08931 [gr-qc]} \BibitemShut {NoStop}%
\bibitem [{\citenamefont {Sekiguchi}\ \emph {et~al.}(2011)\citenamefont {Sekiguchi}, \citenamefont {Kiuchi}, \citenamefont {Kyutoku},\ and\ \citenamefont {Shibata}}]{Sekiguchi:2011mc}%
  \BibitemOpen
  \bibfield  {author} {\bibinfo {author} {\bibfnamefont {Y.}~\bibnamefont {Sekiguchi}}, \bibinfo {author} {\bibfnamefont {K.}~\bibnamefont {Kiuchi}}, \bibinfo {author} {\bibfnamefont {K.}~\bibnamefont {Kyutoku}},\ and\ \bibinfo {author} {\bibfnamefont {M.}~\bibnamefont {Shibata}},\ }\href {https://doi.org/10.1103/PhysRevLett.107.211101} {\bibfield  {journal} {\bibinfo  {journal} {Phys. Rev. Lett.}\ }\textbf {\bibinfo {volume} {107}},\ \bibinfo {pages} {211101} (\bibinfo {year} {2011})},\ \Eprint {https://arxiv.org/abs/1110.4442} {arXiv:1110.4442 [astro-ph.HE]} \BibitemShut {NoStop}%
\bibitem [{\citenamefont {Radice}\ \emph {et~al.}(2017)\citenamefont {Radice}, \citenamefont {Bernuzzi}, \citenamefont {Del~Pozzo}, \citenamefont {Roberts},\ and\ \citenamefont {Ott}}]{Radice:2016rys}%
  \BibitemOpen
  \bibfield  {author} {\bibinfo {author} {\bibfnamefont {D.}~\bibnamefont {Radice}}, \bibinfo {author} {\bibfnamefont {S.}~\bibnamefont {Bernuzzi}}, \bibinfo {author} {\bibfnamefont {W.}~\bibnamefont {Del~Pozzo}}, \bibinfo {author} {\bibfnamefont {L.~F.}\ \bibnamefont {Roberts}},\ and\ \bibinfo {author} {\bibfnamefont {C.~D.}\ \bibnamefont {Ott}},\ }\href {https://doi.org/10.3847/2041-8213/aa775f} {\bibfield  {journal} {\bibinfo  {journal} {Astrophys. J. Lett.}\ }\textbf {\bibinfo {volume} {842}},\ \bibinfo {pages} {L10} (\bibinfo {year} {2017})},\ \Eprint {https://arxiv.org/abs/1612.06429} {arXiv:1612.06429 [astro-ph.HE]} \BibitemShut {NoStop}%
\bibitem [{\citenamefont {Blacker}\ \emph {et~al.}(2024)\citenamefont {Blacker}, \citenamefont {Kochankovski}, \citenamefont {Bauswein}, \citenamefont {Ramos},\ and\ \citenamefont {Tolos}}]{Blacker:2023opp}%
  \BibitemOpen
  \bibfield  {author} {\bibinfo {author} {\bibfnamefont {S.}~\bibnamefont {Blacker}}, \bibinfo {author} {\bibfnamefont {H.}~\bibnamefont {Kochankovski}}, \bibinfo {author} {\bibfnamefont {A.}~\bibnamefont {Bauswein}}, \bibinfo {author} {\bibfnamefont {A.}~\bibnamefont {Ramos}},\ and\ \bibinfo {author} {\bibfnamefont {L.}~\bibnamefont {Tolos}},\ }\href {https://doi.org/10.1103/PhysRevD.109.043015} {\bibfield  {journal} {\bibinfo  {journal} {Phys. Rev. D}\ }\textbf {\bibinfo {volume} {109}},\ \bibinfo {pages} {043015} (\bibinfo {year} {2024})},\ \Eprint {https://arxiv.org/abs/2307.03710} {arXiv:2307.03710 [astro-ph.HE]} \BibitemShut {NoStop}%
\bibitem [{\citenamefont {Lattimer}\ and\ \citenamefont {Prakash}(2007)}]{LATTIMER2007}%
  \BibitemOpen
  \bibfield  {author} {\bibinfo {author} {\bibfnamefont {J.~M.}\ \bibnamefont {Lattimer}}\ and\ \bibinfo {author} {\bibfnamefont {M.}~\bibnamefont {Prakash}},\ }\href {https://doi.org/https://doi.org/10.1016/j.physrep.2007.02.003} {\bibfield  {journal} {\bibinfo  {journal} {Phys. Rep.}\ }\textbf {\bibinfo {volume} {442}},\ \bibinfo {pages} {109} (\bibinfo {year} {2007})}\BibitemShut {NoStop}%
\bibitem [{\citenamefont {Lattimer}(2010)}]{LATTIMER2010101}%
  \BibitemOpen
  \bibfield  {author} {\bibinfo {author} {\bibfnamefont {J.~M.}\ \bibnamefont {Lattimer}},\ }\href {https://doi.org/https://doi.org/10.1016/j.newar.2010.09.013} {\bibfield  {journal} {\bibinfo  {journal} {New Astron. Rev.}\ }\textbf {\bibinfo {volume} {54}},\ \bibinfo {pages} {101} (\bibinfo {year} {2010})},\ \bibinfo {note} {proceedings: A Life With Stars}\BibitemShut {NoStop}%
\bibitem [{\citenamefont {Lattimer}(2012)}]{Lattimer2012}%
  \BibitemOpen
  \bibfield  {author} {\bibinfo {author} {\bibfnamefont {J.~M.}\ \bibnamefont {Lattimer}},\ }\href {https://doi.org/https://doi.org/10.1146/annurev-nucl-102711-095018} {\bibfield  {journal} {\bibinfo  {journal} {Annu. Rev. Nucl. Part. Sci.}\ }\textbf {\bibinfo {volume} {62}},\ \bibinfo {pages} {485} (\bibinfo {year} {2012})}\BibitemShut {NoStop}%
\bibitem [{\citenamefont {Özel}\ and\ \citenamefont {Freire}(2016)}]{Ozel2016}%
  \BibitemOpen
  \bibfield  {author} {\bibinfo {author} {\bibfnamefont {F.}~\bibnamefont {Özel}}\ and\ \bibinfo {author} {\bibfnamefont {P.}~\bibnamefont {Freire}},\ }\href {https://doi.org/https://doi.org/10.1146/annurev-astro-081915-023322} {\bibfield  {journal} {\bibinfo  {journal} {Annu. Rev. Astron. Astrophys.}\ }\textbf {\bibinfo {volume} {54}},\ \bibinfo {pages} {401} (\bibinfo {year} {2016})}\BibitemShut {NoStop}%
\bibitem [{\citenamefont {Lattimer}\ and\ \citenamefont {Prakash}(2016)}]{LATTIMER2016}%
  \BibitemOpen
  \bibfield  {author} {\bibinfo {author} {\bibfnamefont {J.~M.}\ \bibnamefont {Lattimer}}\ and\ \bibinfo {author} {\bibfnamefont {M.}~\bibnamefont {Prakash}},\ }\href {https://doi.org/https://doi.org/10.1016/j.physrep.2015.12.005} {\bibfield  {journal} {\bibinfo  {journal} {Phys. Rep.}\ }\textbf {\bibinfo {volume} {621}},\ \bibinfo {pages} {127} (\bibinfo {year} {2016})}\BibitemShut {NoStop}%
\bibitem [{\citenamefont {Oertel}\ \emph {et~al.}(2017)\citenamefont {Oertel}, \citenamefont {Hempel}, \citenamefont {Kl\"ahn},\ and\ \citenamefont {Typel}}]{RevModPhys.89.015007}%
  \BibitemOpen
  \bibfield  {author} {\bibinfo {author} {\bibfnamefont {M.}~\bibnamefont {Oertel}}, \bibinfo {author} {\bibfnamefont {M.}~\bibnamefont {Hempel}}, \bibinfo {author} {\bibfnamefont {T.}~\bibnamefont {Kl\"ahn}},\ and\ \bibinfo {author} {\bibfnamefont {S.}~\bibnamefont {Typel}},\ }\href {https://doi.org/10.1103/RevModPhys.89.015007} {\bibfield  {journal} {\bibinfo  {journal} {Rev. Mod. Phys.}\ }\textbf {\bibinfo {volume} {89}},\ \bibinfo {pages} {015007} (\bibinfo {year} {2017})}\BibitemShut {NoStop}%
\bibitem [{\citenamefont {Raduta}\ \emph {et~al.}(2021{\natexlab{b}})\citenamefont {Raduta}, \citenamefont {Nacu},\ and\ \citenamefont {Oertel}}]{Raduta2021}%
  \BibitemOpen
  \bibfield  {author} {\bibinfo {author} {\bibfnamefont {A.~R.}\ \bibnamefont {Raduta}}, \bibinfo {author} {\bibfnamefont {F.}~\bibnamefont {Nacu}},\ and\ \bibinfo {author} {\bibfnamefont {M.}~\bibnamefont {Oertel}},\ }\href {https://doi.org/10.1140/epja/s10050-021-00628-z} {\bibfield  {journal} {\bibinfo  {journal} {Eur. Phys. J. A}\ }\textbf {\bibinfo {volume} {57}},\ \bibinfo {pages} {329} (\bibinfo {year} {2021}{\natexlab{b}})}\BibitemShut {NoStop}%
\bibitem [{\citenamefont {Burgio}\ and\ \citenamefont {Vidana}(2020)}]{Burgio:2020fom}%
  \BibitemOpen
  \bibfield  {author} {\bibinfo {author} {\bibfnamefont {G.~F.}\ \bibnamefont {Burgio}}\ and\ \bibinfo {author} {\bibfnamefont {I.}~\bibnamefont {Vidana}},\ }\href {https://doi.org/10.3390/universe6080119} {\bibfield  {journal} {\bibinfo  {journal} {Universe}\ }\textbf {\bibinfo {volume} {6}},\ \bibinfo {pages} {119} (\bibinfo {year} {2020})},\ \Eprint {https://arxiv.org/abs/2007.04427} {arXiv:2007.04427 [nucl-th]} \BibitemShut {NoStop}%
\bibitem [{\citenamefont {{Kochankovski}}\ \emph {et~al.}(2022)\citenamefont {{Kochankovski}}, \citenamefont {{Ramos}},\ and\ \citenamefont {{Tolos}}}]{Kochankovski2022}%
  \BibitemOpen
  \bibfield  {author} {\bibinfo {author} {\bibfnamefont {H.}~\bibnamefont {{Kochankovski}}}, \bibinfo {author} {\bibfnamefont {A.}~\bibnamefont {{Ramos}}},\ and\ \bibinfo {author} {\bibfnamefont {L.}~\bibnamefont {{Tolos}}},\ }\href {https://doi.org/10.1093/mnras/stac2671} {\bibfield  {journal} {\bibinfo  {journal} {Mon. Not. Roy. Astron. Soc.}\ }\textbf {\bibinfo {volume} {517}},\ \bibinfo {pages} {507} (\bibinfo {year} {2022})},\ \Eprint {https://arxiv.org/abs/2206.11266} {arXiv:2206.11266 [astro-ph.HE]} \BibitemShut {NoStop}%
\bibitem [{\citenamefont {Kochankovski}\ \emph {et~al.}(2024)\citenamefont {Kochankovski}, \citenamefont {Ramos},\ and\ \citenamefont {Tolos}}]{Kochankovski:2023trc}%
  \BibitemOpen
  \bibfield  {author} {\bibinfo {author} {\bibfnamefont {H.}~\bibnamefont {Kochankovski}}, \bibinfo {author} {\bibfnamefont {A.}~\bibnamefont {Ramos}},\ and\ \bibinfo {author} {\bibfnamefont {L.}~\bibnamefont {Tolos}},\ }\href {https://doi.org/10.1093/mnras/stae231} {\bibfield  {journal} {\bibinfo  {journal} {Mon. Not. Roy. Astron. Soc.}\ }\textbf {\bibinfo {volume} {528}},\ \bibinfo {pages} {2629} (\bibinfo {year} {2024})},\ \Eprint {https://arxiv.org/abs/2309.14879} {arXiv:2309.14879 [astro-ph.HE]} \BibitemShut {NoStop}%
\bibitem [{\citenamefont {Marques}\ \emph {et~al.}(2017)\citenamefont {Marques}, \citenamefont {Oertel}, \citenamefont {Hempel},\ and\ \citenamefont {Novak}}]{Marques2017}%
  \BibitemOpen
  \bibfield  {author} {\bibinfo {author} {\bibfnamefont {M.}~\bibnamefont {Marques}}, \bibinfo {author} {\bibfnamefont {M.}~\bibnamefont {Oertel}}, \bibinfo {author} {\bibfnamefont {M.}~\bibnamefont {Hempel}},\ and\ \bibinfo {author} {\bibfnamefont {J.}~\bibnamefont {Novak}},\ }\href {https://doi.org/10.1103/PhysRevC.96.045806} {\bibfield  {journal} {\bibinfo  {journal} {Phys. Rev. C}\ }\textbf {\bibinfo {volume} {96}},\ \bibinfo {pages} {045806} (\bibinfo {year} {2017})},\ \Eprint {https://arxiv.org/abs/1706.02913} {arXiv:1706.02913 [nucl-th]} \BibitemShut {NoStop}%
%%CITATION = ARXIV:1706.02913;%%
\bibitem [{\citenamefont {Banik}\ \emph {et~al.}(2014)\citenamefont {Banik}, \citenamefont {Hempel},\ and\ \citenamefont {Bandyopadhyay}}]{Banik2014}%
  \BibitemOpen
  \bibfield  {author} {\bibinfo {author} {\bibfnamefont {S.}~\bibnamefont {Banik}}, \bibinfo {author} {\bibfnamefont {M.}~\bibnamefont {Hempel}},\ and\ \bibinfo {author} {\bibfnamefont {D.}~\bibnamefont {Bandyopadhyay}},\ }\href {https://doi.org/10.1088/0067-0049/214/2/22} {\bibfield  {journal} {\bibinfo  {journal} {\apjs}\ }\textbf {\bibinfo {volume} {214}},\ \bibinfo {eid} {22} (\bibinfo {year} {2014})},\ \Eprint {https://arxiv.org/abs/1404.6173} {arXiv:1404.6173 [astro-ph.HE]} \BibitemShut {NoStop}%
\bibitem [{\citenamefont {{Dexheimer}}(2017)}]{Dexheimer2017}%
  \BibitemOpen
  \bibfield  {author} {\bibinfo {author} {\bibfnamefont {V.}~\bibnamefont {{Dexheimer}}},\ }\href {https://doi.org/10.1017/pasa.2017.61} {\bibfield  {journal} {\bibinfo  {journal} {Publ. Astron. Soc. Austral.}\ }\textbf {\bibinfo {volume} {34}},\ \bibinfo {eid} {e066} (\bibinfo {year} {2017})},\ \Eprint {https://arxiv.org/abs/1708.08342} {arXiv:1708.08342 [astro-ph.HE]} \BibitemShut {NoStop}%
\bibitem [{\citenamefont {Hempel}\ and\ \citenamefont {Schaffner-Bielich}(2010)}]{Hempel2010a}%
  \BibitemOpen
  \bibfield  {author} {\bibinfo {author} {\bibfnamefont {M.}~\bibnamefont {Hempel}}\ and\ \bibinfo {author} {\bibfnamefont {J.}~\bibnamefont {Schaffner-Bielich}},\ }\href {https://doi.org/10.1016/j.nuclphysa.2010.02.010} {\bibfield  {journal} {\bibinfo  {journal} {Nucl. Phys. A}\ }\textbf {\bibinfo {volume} {A837}},\ \bibinfo {pages} {210} (\bibinfo {year} {2010})},\ \Eprint {https://arxiv.org/abs/0911.4073} {arXiv:0911.4073 [nucl-th]} \BibitemShut {NoStop}%
\bibitem [{\citenamefont {Steiner}\ \emph {et~al.}(2013)\citenamefont {Steiner}, \citenamefont {Hempel},\ and\ \citenamefont {Fischer}}]{Steiner2013}%
  \BibitemOpen
  \bibfield  {author} {\bibinfo {author} {\bibfnamefont {A.~W.}\ \bibnamefont {Steiner}}, \bibinfo {author} {\bibfnamefont {M.}~\bibnamefont {Hempel}},\ and\ \bibinfo {author} {\bibfnamefont {T.}~\bibnamefont {Fischer}},\ }\href {https://doi.org/10.1088/0004-637X/774/1/17} {\bibfield  {journal} {\bibinfo  {journal} {\apj}\ }\textbf {\bibinfo {volume} {774}},\ \bibinfo {eid} {17} (\bibinfo {year} {2013})},\ \Eprint {https://arxiv.org/abs/1207.2184} {arXiv:1207.2184 [astro-ph.SR]} \BibitemShut {NoStop}%
\bibitem [{\citenamefont {Typel}\ \emph {et~al.}(2010)\citenamefont {Typel}, \citenamefont {Ropke}, \citenamefont {Klahn}, \citenamefont {Blaschke},\ and\ \citenamefont {Wolter}}]{Typel:2009sy}%
  \BibitemOpen
  \bibfield  {author} {\bibinfo {author} {\bibfnamefont {S.}~\bibnamefont {Typel}}, \bibinfo {author} {\bibfnamefont {G.}~\bibnamefont {Ropke}}, \bibinfo {author} {\bibfnamefont {T.}~\bibnamefont {Klahn}}, \bibinfo {author} {\bibfnamefont {D.}~\bibnamefont {Blaschke}},\ and\ \bibinfo {author} {\bibfnamefont {H.~H.}\ \bibnamefont {Wolter}},\ }\href {https://doi.org/10.1103/PhysRevC.81.015803} {\bibfield  {journal} {\bibinfo  {journal} {Phys. Rev. C}\ }\textbf {\bibinfo {volume} {81}},\ \bibinfo {pages} {015803} (\bibinfo {year} {2010})},\ \Eprint {https://arxiv.org/abs/0908.2344} {arXiv:0908.2344 [nucl-th]} \BibitemShut {NoStop}%
\bibitem [{\citenamefont {{Isenberg}}\ and\ \citenamefont {{Nester}}(1980)}]{1980grg1.conf...23I}%
  \BibitemOpen
  \bibfield  {author} {\bibinfo {author} {\bibfnamefont {J.}~\bibnamefont {{Isenberg}}}\ and\ \bibinfo {author} {\bibfnamefont {J.}~\bibnamefont {{Nester}}},\ }in\ \href@noop {} {\emph {\bibinfo {booktitle} {General Relativity and Gravitation. Vol. 1. One hundred years after the birth of Albert Einstein. Edited by A. Held. New York}}},\ Vol.~\bibinfo {volume} {1}\ (\bibinfo {year} {1980})\ p.~\bibinfo {pages} {23}\BibitemShut {NoStop}%
\bibitem [{\citenamefont {{Wilson}}\ \emph {et~al.}(1996)\citenamefont {{Wilson}}, \citenamefont {{Mathews}},\ and\ \citenamefont {{Marronetti}}}]{1996PhRvD..54.1317W}%
  \BibitemOpen
  \bibfield  {author} {\bibinfo {author} {\bibfnamefont {J.~R.}\ \bibnamefont {{Wilson}}}, \bibinfo {author} {\bibfnamefont {G.~J.}\ \bibnamefont {{Mathews}}},\ and\ \bibinfo {author} {\bibfnamefont {P.}~\bibnamefont {{Marronetti}}},\ }\href {https://doi.org/10.1103/PhysRevD.54.1317} {\bibfield  {journal} {\bibinfo  {journal} {\prd}\ }\textbf {\bibinfo {volume} {54}},\ \bibinfo {pages} {1317} (\bibinfo {year} {1996})},\ \Eprint {https://arxiv.org/abs/gr-qc/9601017} {arXiv:gr-qc/9601017 [gr-qc]} \BibitemShut {NoStop}%
\bibitem [{\citenamefont {Oechslin}\ \emph {et~al.}(2002)\citenamefont {Oechslin}, \citenamefont {Rosswog},\ and\ \citenamefont {Thielemann}}]{Oechslin:2001km}%
  \BibitemOpen
  \bibfield  {author} {\bibinfo {author} {\bibfnamefont {R.}~\bibnamefont {Oechslin}}, \bibinfo {author} {\bibfnamefont {S.}~\bibnamefont {Rosswog}},\ and\ \bibinfo {author} {\bibfnamefont {F.~K.}\ \bibnamefont {Thielemann}},\ }\href {https://doi.org/10.1103/PhysRevD.65.103005} {\bibfield  {journal} {\bibinfo  {journal} {Phys. Rev. D}\ }\textbf {\bibinfo {volume} {65}},\ \bibinfo {pages} {103005} (\bibinfo {year} {2002})},\ \Eprint {https://arxiv.org/abs/gr-qc/0111005} {arXiv:gr-qc/0111005} \BibitemShut {NoStop}%
\bibitem [{\citenamefont {{Oechslin}}\ \emph {et~al.}(2007)\citenamefont {{Oechslin}}, \citenamefont {{Janka}},\ and\ \citenamefont {{Marek}}}]{2007A&A...467..395O}%
  \BibitemOpen
  \bibfield  {author} {\bibinfo {author} {\bibfnamefont {R.}~\bibnamefont {{Oechslin}}}, \bibinfo {author} {\bibfnamefont {H.~T.}\ \bibnamefont {{Janka}}},\ and\ \bibinfo {author} {\bibfnamefont {A.}~\bibnamefont {{Marek}}},\ }\href {https://doi.org/10.1051/0004-6361:20066682} {\bibfield  {journal} {\bibinfo  {journal} {Astron. Astrophys.}\ }\textbf {\bibinfo {volume} {467}},\ \bibinfo {pages} {395} (\bibinfo {year} {2007})},\ \Eprint {https://arxiv.org/abs/astro-ph/0611047} {arXiv:astro-ph/0611047 [astro-ph]} \BibitemShut {NoStop}%
\bibitem [{\citenamefont {Bauswein}\ \emph {et~al.}(2010{\natexlab{a}})\citenamefont {Bauswein}, \citenamefont {Oechslin},\ and\ \citenamefont {Janka}}]{Bauswein:2009im}%
  \BibitemOpen
  \bibfield  {author} {\bibinfo {author} {\bibfnamefont {A.}~\bibnamefont {Bauswein}}, \bibinfo {author} {\bibfnamefont {R.}~\bibnamefont {Oechslin}},\ and\ \bibinfo {author} {\bibfnamefont {H.~T.}\ \bibnamefont {Janka}},\ }\href {https://doi.org/10.1103/PhysRevD.81.024012} {\bibfield  {journal} {\bibinfo  {journal} {Phys. Rev. D}\ }\textbf {\bibinfo {volume} {81}},\ \bibinfo {pages} {024012} (\bibinfo {year} {2010}{\natexlab{a}})},\ \Eprint {https://arxiv.org/abs/0910.5169} {arXiv:0910.5169 [astro-ph.SR]} \BibitemShut {NoStop}%
\bibitem [{\citenamefont {Akmal}\ \emph {et~al.}(1998)\citenamefont {Akmal}, \citenamefont {Pandharipande},\ and\ \citenamefont {Ravenhall}}]{Akmal:1998cf}%
  \BibitemOpen
  \bibfield  {author} {\bibinfo {author} {\bibfnamefont {A.}~\bibnamefont {Akmal}}, \bibinfo {author} {\bibfnamefont {V.~R.}\ \bibnamefont {Pandharipande}},\ and\ \bibinfo {author} {\bibfnamefont {D.~G.}\ \bibnamefont {Ravenhall}},\ }\href {https://doi.org/10.1103/PhysRevC.58.1804} {\bibfield  {journal} {\bibinfo  {journal} {Phys. Rev. C}\ }\textbf {\bibinfo {volume} {58}},\ \bibinfo {pages} {1804} (\bibinfo {year} {1998})},\ \Eprint {https://arxiv.org/abs/nucl-th/9804027} {arXiv:nucl-th/9804027} \BibitemShut {NoStop}%
\bibitem [{\citenamefont {Schneider}\ \emph {et~al.}(2019)\citenamefont {Schneider}, \citenamefont {Constantinou}, \citenamefont {Muccioli},\ and\ \citenamefont {Prakash}}]{Schneider:2019vdm}%
  \BibitemOpen
  \bibfield  {author} {\bibinfo {author} {\bibfnamefont {A.~S.}\ \bibnamefont {Schneider}}, \bibinfo {author} {\bibfnamefont {C.}~\bibnamefont {Constantinou}}, \bibinfo {author} {\bibfnamefont {B.}~\bibnamefont {Muccioli}},\ and\ \bibinfo {author} {\bibfnamefont {M.}~\bibnamefont {Prakash}},\ }\href {https://doi.org/10.1103/PhysRevC.100.025803} {\bibfield  {journal} {\bibinfo  {journal} {Phys. Rev. C}\ }\textbf {\bibinfo {volume} {100}},\ \bibinfo {pages} {025803} (\bibinfo {year} {2019})},\ \Eprint {https://arxiv.org/abs/1901.09652} {arXiv:1901.09652 [nucl-th]} \BibitemShut {NoStop}%
\bibitem [{\citenamefont {Alvarez-Castillo}\ \emph {et~al.}(2016)\citenamefont {Alvarez-Castillo}, \citenamefont {Ayriyan}, \citenamefont {Benic}, \citenamefont {Blaschke}, \citenamefont {Grigorian},\ and\ \citenamefont {Typel}}]{Alvarez-Castillo:2016oln}%
  \BibitemOpen
  \bibfield  {author} {\bibinfo {author} {\bibfnamefont {D.}~\bibnamefont {Alvarez-Castillo}}, \bibinfo {author} {\bibfnamefont {A.}~\bibnamefont {Ayriyan}}, \bibinfo {author} {\bibfnamefont {S.}~\bibnamefont {Benic}}, \bibinfo {author} {\bibfnamefont {D.}~\bibnamefont {Blaschke}}, \bibinfo {author} {\bibfnamefont {H.}~\bibnamefont {Grigorian}},\ and\ \bibinfo {author} {\bibfnamefont {S.}~\bibnamefont {Typel}},\ }\href {https://doi.org/10.1140/epja/i2016-16069-2} {\bibfield  {journal} {\bibinfo  {journal} {Eur. Phys. J. A}\ }\textbf {\bibinfo {volume} {52}},\ \bibinfo {pages} {69} (\bibinfo {year} {2016})},\ \Eprint {https://arxiv.org/abs/1603.03457} {arXiv:1603.03457 [nucl-th]} \BibitemShut {NoStop}%
\bibitem [{\citenamefont {Furusawa}\ \emph {et~al.}(2017)\citenamefont {Furusawa}, \citenamefont {Togashi}, \citenamefont {Nagakura}, \citenamefont {Sumiyoshi}, \citenamefont {Yamada}, \citenamefont {Suzuki},\ and\ \citenamefont {Takano}}]{Furusawa:2017auz}%
  \BibitemOpen
  \bibfield  {author} {\bibinfo {author} {\bibfnamefont {S.}~\bibnamefont {Furusawa}}, \bibinfo {author} {\bibfnamefont {H.}~\bibnamefont {Togashi}}, \bibinfo {author} {\bibfnamefont {H.}~\bibnamefont {Nagakura}}, \bibinfo {author} {\bibfnamefont {K.}~\bibnamefont {Sumiyoshi}}, \bibinfo {author} {\bibfnamefont {S.}~\bibnamefont {Yamada}}, \bibinfo {author} {\bibfnamefont {H.}~\bibnamefont {Suzuki}},\ and\ \bibinfo {author} {\bibfnamefont {M.}~\bibnamefont {Takano}},\ }\href {https://doi.org/10.1088/1361-6471/aa7f35} {\bibfield  {journal} {\bibinfo  {journal} {J. Phys. G}\ }\textbf {\bibinfo {volume} {44}},\ \bibinfo {pages} {094001} (\bibinfo {year} {2017})},\ \Eprint {https://arxiv.org/abs/1707.06410} {arXiv:1707.06410 [astro-ph.HE]} \BibitemShut {NoStop}%
\bibitem [{\citenamefont {Togashi}\ \emph {et~al.}(2017)\citenamefont {Togashi}, \citenamefont {Nakazato}, \citenamefont {Takehara}, \citenamefont {Yamamuro}, \citenamefont {Suzuki},\ and\ \citenamefont {Takano}}]{Togashi:2017mjp}%
  \BibitemOpen
  \bibfield  {author} {\bibinfo {author} {\bibfnamefont {H.}~\bibnamefont {Togashi}}, \bibinfo {author} {\bibfnamefont {K.}~\bibnamefont {Nakazato}}, \bibinfo {author} {\bibfnamefont {Y.}~\bibnamefont {Takehara}}, \bibinfo {author} {\bibfnamefont {S.}~\bibnamefont {Yamamuro}}, \bibinfo {author} {\bibfnamefont {H.}~\bibnamefont {Suzuki}},\ and\ \bibinfo {author} {\bibfnamefont {M.}~\bibnamefont {Takano}},\ }\href {https://doi.org/10.1016/j.nuclphysa.2017.02.010} {\bibfield  {journal} {\bibinfo  {journal} {Nucl. Phys. A}\ }\textbf {\bibinfo {volume} {961}},\ \bibinfo {pages} {78} (\bibinfo {year} {2017})},\ \Eprint {https://arxiv.org/abs/1702.05324} {arXiv:1702.05324 [nucl-th]} \BibitemShut {NoStop}%
\bibitem [{\citenamefont {Shen}\ \emph {et~al.}(2011)\citenamefont {Shen}, \citenamefont {Horowitz},\ and\ \citenamefont {Teige}}]{Shen:2011kr}%
  \BibitemOpen
  \bibfield  {author} {\bibinfo {author} {\bibfnamefont {G.}~\bibnamefont {Shen}}, \bibinfo {author} {\bibfnamefont {C.~J.}\ \bibnamefont {Horowitz}},\ and\ \bibinfo {author} {\bibfnamefont {S.}~\bibnamefont {Teige}},\ }\href {https://doi.org/10.1103/PhysRevC.83.035802} {\bibfield  {journal} {\bibinfo  {journal} {Phys. Rev. C}\ }\textbf {\bibinfo {volume} {83}},\ \bibinfo {pages} {035802} (\bibinfo {year} {2011})},\ \Eprint {https://arxiv.org/abs/1101.3715} {arXiv:1101.3715 [astro-ph.SR]} \BibitemShut {NoStop}%
\bibitem [{\citenamefont {Bombaci}\ and\ \citenamefont {Logoteta}(2018)}]{Bombaci:2018ksa}%
  \BibitemOpen
  \bibfield  {author} {\bibinfo {author} {\bibfnamefont {I.}~\bibnamefont {Bombaci}}\ and\ \bibinfo {author} {\bibfnamefont {D.}~\bibnamefont {Logoteta}},\ }\href {https://doi.org/10.1051/0004-6361/201731604} {\bibfield  {journal} {\bibinfo  {journal} {Astron. Astrophys.}\ }\textbf {\bibinfo {volume} {609}},\ \bibinfo {pages} {A128} (\bibinfo {year} {2018})},\ \Eprint {https://arxiv.org/abs/1805.11846} {arXiv:1805.11846 [astro-ph.HE]} \BibitemShut {NoStop}%
\bibitem [{\citenamefont {Logoteta}\ \emph {et~al.}(2021)\citenamefont {Logoteta}, \citenamefont {Perego},\ and\ \citenamefont {Bombaci}}]{Logoteta:2020yxf}%
  \BibitemOpen
  \bibfield  {author} {\bibinfo {author} {\bibfnamefont {D.}~\bibnamefont {Logoteta}}, \bibinfo {author} {\bibfnamefont {A.}~\bibnamefont {Perego}},\ and\ \bibinfo {author} {\bibfnamefont {I.}~\bibnamefont {Bombaci}},\ }\href {https://doi.org/10.1051/0004-6361/202039457} {\bibfield  {journal} {\bibinfo  {journal} {Astron. Astrophys.}\ }\textbf {\bibinfo {volume} {646}},\ \bibinfo {pages} {A55} (\bibinfo {year} {2021})},\ \Eprint {https://arxiv.org/abs/2012.03599} {arXiv:2012.03599 [nucl-th]} \BibitemShut {NoStop}%
\bibitem [{\citenamefont {Lattimer}\ and\ \citenamefont {Swesty}(1991)}]{Lattimer:1991nc}%
  \BibitemOpen
  \bibfield  {author} {\bibinfo {author} {\bibfnamefont {J.~M.}\ \bibnamefont {Lattimer}}\ and\ \bibinfo {author} {\bibfnamefont {F.~D.}\ \bibnamefont {Swesty}},\ }\href {https://doi.org/10.1016/0375-9474(91)90452-C} {\bibfield  {journal} {\bibinfo  {journal} {Nucl. Phys. A}\ }\textbf {\bibinfo {volume} {535}},\ \bibinfo {pages} {331} (\bibinfo {year} {1991})}\BibitemShut {NoStop}%
\bibitem [{\citenamefont {Chabanat}\ \emph {et~al.}(1998)\citenamefont {Chabanat}, \citenamefont {Bonche}, \citenamefont {Haensel}, \citenamefont {Meyer},\ and\ \citenamefont {Schaeffer}}]{Chabanat:1997un}%
  \BibitemOpen
  \bibfield  {author} {\bibinfo {author} {\bibfnamefont {E.}~\bibnamefont {Chabanat}}, \bibinfo {author} {\bibfnamefont {P.}~\bibnamefont {Bonche}}, \bibinfo {author} {\bibfnamefont {P.}~\bibnamefont {Haensel}}, \bibinfo {author} {\bibfnamefont {J.}~\bibnamefont {Meyer}},\ and\ \bibinfo {author} {\bibfnamefont {R.}~\bibnamefont {Schaeffer}},\ }\href {https://doi.org/10.1016/S0375-9474(98)00180-8} {\bibfield  {journal} {\bibinfo  {journal} {Nucl. Phys. A}\ }\textbf {\bibinfo {volume} {635}},\ \bibinfo {pages} {231} (\bibinfo {year} {1998})},\ \bibinfo {note} {[Erratum: Nucl.Phys.A 643, 441--441 (1998)]}\BibitemShut {NoStop}%
\bibitem [{\citenamefont {Schneider}\ \emph {et~al.}(2017)\citenamefont {Schneider}, \citenamefont {Roberts},\ and\ \citenamefont {Ott}}]{Schneider:2017tfi}%
  \BibitemOpen
  \bibfield  {author} {\bibinfo {author} {\bibfnamefont {A.~S.}\ \bibnamefont {Schneider}}, \bibinfo {author} {\bibfnamefont {L.~F.}\ \bibnamefont {Roberts}},\ and\ \bibinfo {author} {\bibfnamefont {C.~D.}\ \bibnamefont {Ott}},\ }\href {https://doi.org/10.1103/PhysRevC.96.065802} {\bibfield  {journal} {\bibinfo  {journal} {Phys. Rev. C}\ }\textbf {\bibinfo {volume} {96}},\ \bibinfo {pages} {065802} (\bibinfo {year} {2017})},\ \Eprint {https://arxiv.org/abs/1707.01527} {arXiv:1707.01527 [astro-ph.HE]} \BibitemShut {NoStop}%
\bibitem [{\citenamefont {Sugahara}\ and\ \citenamefont {Toki}(1994)}]{Sugahara:1993wz}%
  \BibitemOpen
  \bibfield  {author} {\bibinfo {author} {\bibfnamefont {Y.}~\bibnamefont {Sugahara}}\ and\ \bibinfo {author} {\bibfnamefont {H.}~\bibnamefont {Toki}},\ }\href {https://doi.org/10.1016/0375-9474(94)90923-7} {\bibfield  {journal} {\bibinfo  {journal} {Nucl. Phys. A}\ }\textbf {\bibinfo {volume} {579}},\ \bibinfo {pages} {557} (\bibinfo {year} {1994})}\BibitemShut {NoStop}%
\bibitem [{\citenamefont {Hempel}\ \emph {et~al.}(2012)\citenamefont {Hempel}, \citenamefont {Fischer}, \citenamefont {Schaffner-Bielich},\ and\ \citenamefont {Liebendorfer}}]{Hempel:2011mk}%
  \BibitemOpen
  \bibfield  {author} {\bibinfo {author} {\bibfnamefont {M.}~\bibnamefont {Hempel}}, \bibinfo {author} {\bibfnamefont {T.}~\bibnamefont {Fischer}}, \bibinfo {author} {\bibfnamefont {J.}~\bibnamefont {Schaffner-Bielich}},\ and\ \bibinfo {author} {\bibfnamefont {M.}~\bibnamefont {Liebendorfer}},\ }\href {https://doi.org/10.1088/0004-637X/748/1/70} {\bibfield  {journal} {\bibinfo  {journal} {Astrophys. J.}\ }\textbf {\bibinfo {volume} {748}},\ \bibinfo {pages} {70} (\bibinfo {year} {2012})},\ \Eprint {https://arxiv.org/abs/1108.0848} {arXiv:1108.0848 [astro-ph.HE]} \BibitemShut {NoStop}%
\bibitem [{\citenamefont {Toki}\ \emph {et~al.}(1995)\citenamefont {Toki}, \citenamefont {Hirata}, \citenamefont {Sugahara}, \citenamefont {Sumiyoshi},\ and\ \citenamefont {Tanihata}}]{Toki:1995ya}%
  \BibitemOpen
  \bibfield  {author} {\bibinfo {author} {\bibfnamefont {H.}~\bibnamefont {Toki}}, \bibinfo {author} {\bibfnamefont {D.}~\bibnamefont {Hirata}}, \bibinfo {author} {\bibfnamefont {Y.}~\bibnamefont {Sugahara}}, \bibinfo {author} {\bibfnamefont {K.}~\bibnamefont {Sumiyoshi}},\ and\ \bibinfo {author} {\bibfnamefont {I.}~\bibnamefont {Tanihata}},\ }\href {https://doi.org/10.1016/0375-9474(95)00161-S} {\bibfield  {journal} {\bibinfo  {journal} {Nucl. Phys. A}\ }\textbf {\bibinfo {volume} {588}},\ \bibinfo {pages} {c357} (\bibinfo {year} {1995})}\BibitemShut {NoStop}%
\bibitem [{\citenamefont {Du}\ \emph {et~al.}(2022)\citenamefont {Du}, \citenamefont {Steiner},\ and\ \citenamefont {Holt}}]{Du:2021rhq}%
  \BibitemOpen
  \bibfield  {author} {\bibinfo {author} {\bibfnamefont {X.}~\bibnamefont {Du}}, \bibinfo {author} {\bibfnamefont {A.~W.}\ \bibnamefont {Steiner}},\ and\ \bibinfo {author} {\bibfnamefont {J.~W.}\ \bibnamefont {Holt}},\ }\href {https://doi.org/10.1103/PhysRevC.105.035803} {\bibfield  {journal} {\bibinfo  {journal} {Phys. Rev. C}\ }\textbf {\bibinfo {volume} {105}},\ \bibinfo {pages} {035803} (\bibinfo {year} {2022})},\ \Eprint {https://arxiv.org/abs/2107.06697} {arXiv:2107.06697 [nucl-th]} \BibitemShut {NoStop}%
\bibitem [{\citenamefont {Typel}\ \emph {et~al.}(2015)\citenamefont {Typel}, \citenamefont {Oertel},\ and\ \citenamefont {Kl\"ahn}}]{Typel:2013rza}%
  \BibitemOpen
  \bibfield  {author} {\bibinfo {author} {\bibfnamefont {S.}~\bibnamefont {Typel}}, \bibinfo {author} {\bibfnamefont {M.}~\bibnamefont {Oertel}},\ and\ \bibinfo {author} {\bibfnamefont {T.}~\bibnamefont {Kl\"ahn}},\ }\href {https://doi.org/10.1134/S1063779615040061} {\bibfield  {journal} {\bibinfo  {journal} {Phys. Part. Nucl.}\ }\textbf {\bibinfo {volume} {46}},\ \bibinfo {pages} {633} (\bibinfo {year} {2015})},\ \Eprint {https://arxiv.org/abs/1307.5715} {arXiv:1307.5715 [astro-ph.SR]} \BibitemShut {NoStop}%
\bibitem [{\citenamefont {Typel}\ \emph {et~al.}(2022)\citenamefont {Typel} \emph {et~al.}}]{CompOSECoreTeam:2022ddl}%
  \BibitemOpen
  \bibfield  {author} {\bibinfo {author} {\bibfnamefont {S.}~\bibnamefont {Typel}} \emph {et~al.} (\bibinfo {collaboration} {CompOSE Core Team}),\ }\href {https://doi.org/10.1140/epja/s10050-022-00847-y} {\bibfield  {journal} {\bibinfo  {journal} {Eur. Phys. J. A}\ }\textbf {\bibinfo {volume} {58}},\ \bibinfo {pages} {221} (\bibinfo {year} {2022})},\ \Eprint {https://arxiv.org/abs/2203.03209} {arXiv:2203.03209 [astro-ph.HE]} \BibitemShut {NoStop}%
\bibitem [{\citenamefont {Dexheimer}\ \emph {et~al.}(2022)\citenamefont {Dexheimer}, \citenamefont {Mancini}, \citenamefont {Oertel}, \citenamefont {Providência}, \citenamefont {Tolos},\ and\ \citenamefont {Typel}}]{particles5030028}%
  \BibitemOpen
  \bibfield  {author} {\bibinfo {author} {\bibfnamefont {V.}~\bibnamefont {Dexheimer}}, \bibinfo {author} {\bibfnamefont {M.}~\bibnamefont {Mancini}}, \bibinfo {author} {\bibfnamefont {M.}~\bibnamefont {Oertel}}, \bibinfo {author} {\bibfnamefont {C.}~\bibnamefont {Providência}}, \bibinfo {author} {\bibfnamefont {L.}~\bibnamefont {Tolos}},\ and\ \bibinfo {author} {\bibfnamefont {S.}~\bibnamefont {Typel}},\ }\href {https://doi.org/10.3390/particles5030028} {\bibfield  {journal} {\bibinfo  {journal} {Particles}\ }\textbf {\bibinfo {volume} {5}},\ \bibinfo {pages} {346} (\bibinfo {year} {2022})}\BibitemShut {NoStop}%
\bibitem [{\citenamefont {{Janka}}\ \emph {et~al.}(1993)\citenamefont {{Janka}}, \citenamefont {{Zwerger}},\ and\ \citenamefont {{Moenchmeyer}}}]{Janka1993}%
  \BibitemOpen
  \bibfield  {author} {\bibinfo {author} {\bibfnamefont {H.~T.}\ \bibnamefont {{Janka}}}, \bibinfo {author} {\bibfnamefont {T.}~\bibnamefont {{Zwerger}}},\ and\ \bibinfo {author} {\bibfnamefont {R.}~\bibnamefont {{Moenchmeyer}}},\ }\href {https://ui.adsabs.harvard.edu/abs/1993A&A...268..360J} {\bibfield  {journal} {\bibinfo  {journal} {\aap}\ }\textbf {\bibinfo {volume} {268}},\ \bibinfo {pages} {360} (\bibinfo {year} {1993})}\BibitemShut {NoStop}%
\bibitem [{\citenamefont {Bauswein}\ \emph {et~al.}(2010{\natexlab{b}})\citenamefont {Bauswein}, \citenamefont {Janka},\ and\ \citenamefont {Oechslin}}]{Bauswein:2010dn}%
  \BibitemOpen
  \bibfield  {author} {\bibinfo {author} {\bibfnamefont {A.}~\bibnamefont {Bauswein}}, \bibinfo {author} {\bibfnamefont {H.~T.}\ \bibnamefont {Janka}},\ and\ \bibinfo {author} {\bibfnamefont {R.}~\bibnamefont {Oechslin}},\ }\href {https://doi.org/10.1103/PhysRevD.82.084043} {\bibfield  {journal} {\bibinfo  {journal} {Phys. Rev. D}\ }\textbf {\bibinfo {volume} {82}},\ \bibinfo {pages} {084043} (\bibinfo {year} {2010}{\natexlab{b}})},\ \Eprint {https://arxiv.org/abs/1006.3315} {arXiv:1006.3315 [astro-ph.SR]} \BibitemShut {NoStop}%
\bibitem [{\citenamefont {Lioutas}\ \emph {et~al.}(2021)\citenamefont {Lioutas}, \citenamefont {Bauswein},\ and\ \citenamefont {Stergioulas}}]{PhysRevD.104.043011}%
  \BibitemOpen
  \bibfield  {author} {\bibinfo {author} {\bibfnamefont {G.}~\bibnamefont {Lioutas}}, \bibinfo {author} {\bibfnamefont {A.}~\bibnamefont {Bauswein}},\ and\ \bibinfo {author} {\bibfnamefont {N.}~\bibnamefont {Stergioulas}},\ }\href {https://doi.org/10.1103/PhysRevD.104.043011} {\bibfield  {journal} {\bibinfo  {journal} {Phys. Rev. D}\ }\textbf {\bibinfo {volume} {104}},\ \bibinfo {pages} {043011} (\bibinfo {year} {2021})}\BibitemShut {NoStop}%
\bibitem [{\citenamefont {Bauswein}\ \emph {et~al.}(2019)\citenamefont {Bauswein}, \citenamefont {Bastian}, \citenamefont {Blaschke}, \citenamefont {Chatziioannou}, \citenamefont {Clark}, \citenamefont {Fischer},\ and\ \citenamefont {Oertel}}]{PhysRevLett.122.061102}%
  \BibitemOpen
  \bibfield  {author} {\bibinfo {author} {\bibfnamefont {A.}~\bibnamefont {Bauswein}}, \bibinfo {author} {\bibfnamefont {N.-U.~F.}\ \bibnamefont {Bastian}}, \bibinfo {author} {\bibfnamefont {D.~B.}\ \bibnamefont {Blaschke}}, \bibinfo {author} {\bibfnamefont {K.}~\bibnamefont {Chatziioannou}}, \bibinfo {author} {\bibfnamefont {J.~A.}\ \bibnamefont {Clark}}, \bibinfo {author} {\bibfnamefont {T.}~\bibnamefont {Fischer}},\ and\ \bibinfo {author} {\bibfnamefont {M.}~\bibnamefont {Oertel}},\ }\href {https://doi.org/10.1103/PhysRevLett.122.061102} {\bibfield  {journal} {\bibinfo  {journal} {Phys. Rev. Lett.}\ }\textbf {\bibinfo {volume} {122}},\ \bibinfo {pages} {061102} (\bibinfo {year} {2019})}\BibitemShut {NoStop}%
\bibitem [{\citenamefont {Blacker}\ \emph {et~al.}(2020)\citenamefont {Blacker}, \citenamefont {Bastian}, \citenamefont {Bauswein}, \citenamefont {Blaschke}, \citenamefont {Fischer}, \citenamefont {Oertel}, \citenamefont {Soultanis},\ and\ \citenamefont {Typel}}]{Blacker:2020nlq}%
  \BibitemOpen
  \bibfield  {author} {\bibinfo {author} {\bibfnamefont {S.}~\bibnamefont {Blacker}}, \bibinfo {author} {\bibfnamefont {N.-U.~F.}\ \bibnamefont {Bastian}}, \bibinfo {author} {\bibfnamefont {A.}~\bibnamefont {Bauswein}}, \bibinfo {author} {\bibfnamefont {D.~B.}\ \bibnamefont {Blaschke}}, \bibinfo {author} {\bibfnamefont {T.}~\bibnamefont {Fischer}}, \bibinfo {author} {\bibfnamefont {M.}~\bibnamefont {Oertel}}, \bibinfo {author} {\bibfnamefont {T.}~\bibnamefont {Soultanis}},\ and\ \bibinfo {author} {\bibfnamefont {S.}~\bibnamefont {Typel}},\ }\href {https://doi.org/10.1103/PhysRevD.102.123023} {\bibfield  {journal} {\bibinfo  {journal} {Phys. Rev. D}\ }\textbf {\bibinfo {volume} {102}},\ \bibinfo {pages} {123023} (\bibinfo {year} {2020})},\ \Eprint {https://arxiv.org/abs/2006.03789} {arXiv:2006.03789 [astro-ph.HE]} \BibitemShut {NoStop}%
\bibitem [{\citenamefont {{Springel}}(2010)}]{2010MNRAS.401..791S}%
  \BibitemOpen
  \bibfield  {author} {\bibinfo {author} {\bibfnamefont {V.}~\bibnamefont {{Springel}}},\ }\href {https://doi.org/10.1111/j.1365-2966.2009.15715.x} {\bibfield  {journal} {\bibinfo  {journal} {Mon. Not. Roy. Astron. Soc.}\ }\textbf {\bibinfo {volume} {401}},\ \bibinfo {pages} {791} (\bibinfo {year} {2010})},\ \Eprint {https://arxiv.org/abs/0901.4107} {arXiv:0901.4107 [astro-ph.CO]} \BibitemShut {NoStop}%
\bibitem [{\citenamefont {{Pakmor}}\ \emph {et~al.}(2016)\citenamefont {{Pakmor}}, \citenamefont {{Springel}}, \citenamefont {{Bauer}}, \citenamefont {{Mocz}}, \citenamefont {{Munoz}}, \citenamefont {{Ohlmann}}, \citenamefont {{Schaal}},\ and\ \citenamefont {{Zhu}}}]{2016MNRAS.455.1134P}%
  \BibitemOpen
  \bibfield  {author} {\bibinfo {author} {\bibfnamefont {R.}~\bibnamefont {{Pakmor}}}, \bibinfo {author} {\bibfnamefont {V.}~\bibnamefont {{Springel}}}, \bibinfo {author} {\bibfnamefont {A.}~\bibnamefont {{Bauer}}}, \bibinfo {author} {\bibfnamefont {P.}~\bibnamefont {{Mocz}}}, \bibinfo {author} {\bibfnamefont {D.~J.}\ \bibnamefont {{Munoz}}}, \bibinfo {author} {\bibfnamefont {S.~T.}\ \bibnamefont {{Ohlmann}}}, \bibinfo {author} {\bibfnamefont {K.}~\bibnamefont {{Schaal}}},\ and\ \bibinfo {author} {\bibfnamefont {C.}~\bibnamefont {{Zhu}}},\ }\href {https://doi.org/10.1093/mnras/stv2380} {\bibfield  {journal} {\bibinfo  {journal} {Mon. Not. Roy. Astron. Soc.}\ }\textbf {\bibinfo {volume} {455}},\ \bibinfo {pages} {1134} (\bibinfo {year} {2016})},\ \Eprint {https://arxiv.org/abs/1503.00562} {arXiv:1503.00562 [astro-ph.GA]} \BibitemShut {NoStop}%
\bibitem [{\citenamefont {Lioutas}\ \emph {et~al.}(2024)\citenamefont {Lioutas}, \citenamefont {Bauswein}, \citenamefont {Soultanis}, \citenamefont {Pakmor}, \citenamefont {Springel},\ and\ \citenamefont {R{\"o}pke}}]{Lioutas2022GeneralRM}%
  \BibitemOpen
  \bibfield  {author} {\bibinfo {author} {\bibfnamefont {G.}~\bibnamefont {Lioutas}}, \bibinfo {author} {\bibfnamefont {A.}~\bibnamefont {Bauswein}}, \bibinfo {author} {\bibfnamefont {T.}~\bibnamefont {Soultanis}}, \bibinfo {author} {\bibfnamefont {R.}~\bibnamefont {Pakmor}}, \bibinfo {author} {\bibfnamefont {V.}~\bibnamefont {Springel}},\ and\ \bibinfo {author} {\bibfnamefont {F.~K.}\ \bibnamefont {R{\"o}pke}},\ }\href {https://academic.oup.com/mnras/article/528/2/1906/7515304} {\bibfield  {journal} {\bibinfo  {journal} {Mon. Not. Roy. Astron. Soc.}\ }\textbf {\bibinfo {volume} {528}},\ \bibinfo {pages} {1906} (\bibinfo {year} {2024})}\BibitemShut {NoStop}%
\bibitem [{\citenamefont {Stergioulas}\ \emph {et~al.}(2011)\citenamefont {Stergioulas}, \citenamefont {Bauswein}, \citenamefont {Zagkouris},\ and\ \citenamefont {Janka}}]{Stergioulas:2011gd}%
  \BibitemOpen
  \bibfield  {author} {\bibinfo {author} {\bibfnamefont {N.}~\bibnamefont {Stergioulas}}, \bibinfo {author} {\bibfnamefont {A.}~\bibnamefont {Bauswein}}, \bibinfo {author} {\bibfnamefont {K.}~\bibnamefont {Zagkouris}},\ and\ \bibinfo {author} {\bibfnamefont {H.-T.}\ \bibnamefont {Janka}},\ }\href {https://doi.org/10.1111/j.1365-2966.2011.19493.x} {\bibfield  {journal} {\bibinfo  {journal} {Mon. Not. Roy. Astron. Soc.}\ }\textbf {\bibinfo {volume} {418}},\ \bibinfo {pages} {427} (\bibinfo {year} {2011})},\ \Eprint {https://arxiv.org/abs/1105.0368} {arXiv:1105.0368 [gr-qc]} \BibitemShut {NoStop}%
\bibitem [{\citenamefont {Hotokezaka}\ \emph {et~al.}(2013{\natexlab{a}})\citenamefont {Hotokezaka}, \citenamefont {Kiuchi}, \citenamefont {Kyutoku}, \citenamefont {Muranushi}, \citenamefont {Sekiguchi}, \citenamefont {Shibata},\ and\ \citenamefont {Taniguchi}}]{PhysRevD.88.044026}%
  \BibitemOpen
  \bibfield  {author} {\bibinfo {author} {\bibfnamefont {K.}~\bibnamefont {Hotokezaka}}, \bibinfo {author} {\bibfnamefont {K.}~\bibnamefont {Kiuchi}}, \bibinfo {author} {\bibfnamefont {K.}~\bibnamefont {Kyutoku}}, \bibinfo {author} {\bibfnamefont {T.}~\bibnamefont {Muranushi}}, \bibinfo {author} {\bibfnamefont {Y.-i.}\ \bibnamefont {Sekiguchi}}, \bibinfo {author} {\bibfnamefont {M.}~\bibnamefont {Shibata}},\ and\ \bibinfo {author} {\bibfnamefont {K.}~\bibnamefont {Taniguchi}},\ }\href {https://doi.org/10.1103/PhysRevD.88.044026} {\bibfield  {journal} {\bibinfo  {journal} {Phys. Rev. D}\ }\textbf {\bibinfo {volume} {88}},\ \bibinfo {pages} {044026} (\bibinfo {year} {2013}{\natexlab{a}})}\BibitemShut {NoStop}%
\bibitem [{\citenamefont {Bauswein}\ and\ \citenamefont {Stergioulas}(2015)}]{PhysRevD.91.124056}%
  \BibitemOpen
  \bibfield  {author} {\bibinfo {author} {\bibfnamefont {A.}~\bibnamefont {Bauswein}}\ and\ \bibinfo {author} {\bibfnamefont {N.}~\bibnamefont {Stergioulas}},\ }\href {https://doi.org/10.1103/PhysRevD.91.124056} {\bibfield  {journal} {\bibinfo  {journal} {Phys. Rev. D}\ }\textbf {\bibinfo {volume} {91}},\ \bibinfo {pages} {124056} (\bibinfo {year} {2015})}\BibitemShut {NoStop}%
\bibitem [{\citenamefont {Bauswein}\ \emph {et~al.}(2016)\citenamefont {Bauswein}, \citenamefont {Stergioulas},\ and\ \citenamefont {Janka}}]{Bauswein_2016}%
  \BibitemOpen
  \bibfield  {author} {\bibinfo {author} {\bibfnamefont {A.}~\bibnamefont {Bauswein}}, \bibinfo {author} {\bibfnamefont {N.}~\bibnamefont {Stergioulas}},\ and\ \bibinfo {author} {\bibfnamefont {H.-T.}\ \bibnamefont {Janka}},\ }\href {http://dx.doi.org/10.1140/epja/i2016-16056-7} {\bibfield  {journal} {\bibinfo  {journal} {Eur. Phys. J. A}\ }\textbf {\bibinfo {volume} {52}},\ \bibinfo {pages} {56} (\bibinfo {year} {2016})}\BibitemShut {NoStop}%
\bibitem [{\citenamefont {Takami}\ \emph {et~al.}(2015)\citenamefont {Takami}, \citenamefont {Rezzolla},\ and\ \citenamefont {Baiotti}}]{PhysRevD.91.064001}%
  \BibitemOpen
  \bibfield  {author} {\bibinfo {author} {\bibfnamefont {K.}~\bibnamefont {Takami}}, \bibinfo {author} {\bibfnamefont {L.}~\bibnamefont {Rezzolla}},\ and\ \bibinfo {author} {\bibfnamefont {L.}~\bibnamefont {Baiotti}},\ }\href {https://doi.org/10.1103/PhysRevD.91.064001} {\bibfield  {journal} {\bibinfo  {journal} {Phys. Rev. D}\ }\textbf {\bibinfo {volume} {91}},\ \bibinfo {pages} {064001} (\bibinfo {year} {2015})}\BibitemShut {NoStop}%
\bibitem [{\citenamefont {Maione}\ \emph {et~al.}(2017)\citenamefont {Maione}, \citenamefont {De~Pietri}, \citenamefont {Feo},\ and\ \citenamefont {L\"offler}}]{PhysRevD.96.063011}%
  \BibitemOpen
  \bibfield  {author} {\bibinfo {author} {\bibfnamefont {F.}~\bibnamefont {Maione}}, \bibinfo {author} {\bibfnamefont {R.}~\bibnamefont {De~Pietri}}, \bibinfo {author} {\bibfnamefont {A.}~\bibnamefont {Feo}},\ and\ \bibinfo {author} {\bibfnamefont {F.}~\bibnamefont {L\"offler}},\ }\href {https://doi.org/10.1103/PhysRevD.96.063011} {\bibfield  {journal} {\bibinfo  {journal} {Phys. Rev. D}\ }\textbf {\bibinfo {volume} {96}},\ \bibinfo {pages} {063011} (\bibinfo {year} {2017})}\BibitemShut {NoStop}%
\bibitem [{\citenamefont {Bauswein}\ and\ \citenamefont {Stergioulas}(2019)}]{Bauswein_2019}%
  \BibitemOpen
  \bibfield  {author} {\bibinfo {author} {\bibfnamefont {A.}~\bibnamefont {Bauswein}}\ and\ \bibinfo {author} {\bibfnamefont {N.}~\bibnamefont {Stergioulas}},\ }\href {https://doi.org/10.1088/1361-6471/ab2b90} {\bibfield  {journal} {\bibinfo  {journal} {J. Phys. G}\ }\textbf {\bibinfo {volume} {46}},\ \bibinfo {pages} {113002} (\bibinfo {year} {2019})}\BibitemShut {NoStop}%
\bibitem [{\citenamefont {Raithel}\ and\ \citenamefont {Paschalidis}(2023)}]{PhysRevD.108.083029}%
  \BibitemOpen
  \bibfield  {author} {\bibinfo {author} {\bibfnamefont {C.~A.}\ \bibnamefont {Raithel}}\ and\ \bibinfo {author} {\bibfnamefont {V.}~\bibnamefont {Paschalidis}},\ }\href {https://doi.org/10.1103/PhysRevD.108.083029} {\bibfield  {journal} {\bibinfo  {journal} {Phys. Rev. D}\ }\textbf {\bibinfo {volume} {108}},\ \bibinfo {pages} {083029} (\bibinfo {year} {2023})}\BibitemShut {NoStop}%
\bibitem [{\citenamefont {Fields}\ \emph {et~al.}(2023)\citenamefont {Fields}, \citenamefont {Prakash}, \citenamefont {Breschi}, \citenamefont {Radice}, \citenamefont {Bernuzzi},\ and\ \citenamefont {da~Silva~Schneider}}]{Fields:2023bhs}%
  \BibitemOpen
  \bibfield  {author} {\bibinfo {author} {\bibfnamefont {J.}~\bibnamefont {Fields}}, \bibinfo {author} {\bibfnamefont {A.}~\bibnamefont {Prakash}}, \bibinfo {author} {\bibfnamefont {M.}~\bibnamefont {Breschi}}, \bibinfo {author} {\bibfnamefont {D.}~\bibnamefont {Radice}}, \bibinfo {author} {\bibfnamefont {S.}~\bibnamefont {Bernuzzi}},\ and\ \bibinfo {author} {\bibfnamefont {A.}~\bibnamefont {da~Silva~Schneider}},\ }\href {https://doi.org/10.3847/2041-8213/ace5b2} {\bibfield  {journal} {\bibinfo  {journal} {Astrophys. J. Lett.}\ }\textbf {\bibinfo {volume} {952}},\ \bibinfo {pages} {L36} (\bibinfo {year} {2023})},\ \Eprint {https://arxiv.org/abs/2302.11359} {arXiv:2302.11359 [astro-ph.HE]} \BibitemShut {NoStop}%
\bibitem [{\citenamefont {Fernández}\ and\ \citenamefont {Metzger}(2016)}]{Fernandez2016}%
  \BibitemOpen
  \bibfield  {author} {\bibinfo {author} {\bibfnamefont {R.}~\bibnamefont {Fernández}}\ and\ \bibinfo {author} {\bibfnamefont {B.~D.}\ \bibnamefont {Metzger}},\ }\href {https://doi.org/https://doi.org/10.1146/annurev-nucl-102115-044819} {\bibfield  {journal} {\bibinfo  {journal} {Annu. Rev. Nucl. Part. Sci.}\ }\textbf {\bibinfo {volume} {66}},\ \bibinfo {pages} {23} (\bibinfo {year} {2016})}\BibitemShut {NoStop}%
\bibitem [{\citenamefont {Baiotti}\ and\ \citenamefont {Rezzolla}(2017)}]{Baiotti2017-vm}%
  \BibitemOpen
  \bibfield  {author} {\bibinfo {author} {\bibfnamefont {L.}~\bibnamefont {Baiotti}}\ and\ \bibinfo {author} {\bibfnamefont {L.}~\bibnamefont {Rezzolla}},\ }\href {https://doi.org/10.1088/1361-6633/aa67bb} {\bibfield  {journal} {\bibinfo  {journal} {Rept. Prog. Phys.}\ }\textbf {\bibinfo {volume} {80}},\ \bibinfo {pages} {096901} (\bibinfo {year} {2017})},\ \Eprint {https://arxiv.org/abs/1607.03540} {arXiv:1607.03540 [gr-qc]} \BibitemShut {NoStop}%
\bibitem [{\citenamefont {Shibata}\ and\ \citenamefont {Hotokezaka}(2019)}]{ShibataMasura2019}%
  \BibitemOpen
  \bibfield  {author} {\bibinfo {author} {\bibfnamefont {M.}~\bibnamefont {Shibata}}\ and\ \bibinfo {author} {\bibfnamefont {K.}~\bibnamefont {Hotokezaka}},\ }\href {https://doi.org/https://doi.org/10.1146/annurev-nucl-101918-023625} {\bibfield  {journal} {\bibinfo  {journal} {Annu. Rev. Nucl. Part. Sci.}\ }\textbf {\bibinfo {volume} {69}},\ \bibinfo {pages} {41} (\bibinfo {year} {2019})}\BibitemShut {NoStop}%
\bibitem [{\citenamefont {Metzger}(2019)}]{Metzger2019}%
  \BibitemOpen
  \bibfield  {author} {\bibinfo {author} {\bibfnamefont {B.~D.}\ \bibnamefont {Metzger}},\ }\href {https://doi.org/10.1007/s41114-019-0024-0} {\bibfield  {journal} {\bibinfo  {journal} {Living Rev. Relativ.}\ }\textbf {\bibinfo {volume} {23}},\ \bibinfo {pages} {1} (\bibinfo {year} {2019})}\BibitemShut {NoStop}%
\bibitem [{\citenamefont {Radice}\ \emph {et~al.}(2020)\citenamefont {Radice}, \citenamefont {Bernuzzi},\ and\ \citenamefont {Perego}}]{RadiceBernuzzi2020}%
  \BibitemOpen
  \bibfield  {author} {\bibinfo {author} {\bibfnamefont {D.}~\bibnamefont {Radice}}, \bibinfo {author} {\bibfnamefont {S.}~\bibnamefont {Bernuzzi}},\ and\ \bibinfo {author} {\bibfnamefont {A.}~\bibnamefont {Perego}},\ }\href {https://doi.org/https://doi.org/10.1146/annurev-nucl-013120-114541} {\bibfield  {journal} {\bibinfo  {journal} {Annu. Rev. Nucl. Part. Sci.}\ }\textbf {\bibinfo {volume} {70}},\ \bibinfo {pages} {95} (\bibinfo {year} {2020})}\BibitemShut {NoStop}%
\bibitem [{\citenamefont {Nakar}(2020)}]{NAKAR20201}%
  \BibitemOpen
  \bibfield  {author} {\bibinfo {author} {\bibfnamefont {E.}~\bibnamefont {Nakar}},\ }\href {https://doi.org/https://doi.org/10.1016/j.physrep.2020.08.008} {\bibfield  {journal} {\bibinfo  {journal} {Phys. Rep.}\ }\textbf {\bibinfo {volume} {886}},\ \bibinfo {pages} {1} (\bibinfo {year} {2020})}\BibitemShut {NoStop}%
\bibitem [{\citenamefont {Cowan}\ \emph {et~al.}(2021)\citenamefont {Cowan}, \citenamefont {Sneden}, \citenamefont {Lawler}, \citenamefont {Aprahamian}, \citenamefont {Wiescher}, \citenamefont {Langanke}, \citenamefont {Mart\'{\i}nez-Pinedo},\ and\ \citenamefont {Thielemann}}]{RevModPhys.93.015002}%
  \BibitemOpen
  \bibfield  {author} {\bibinfo {author} {\bibfnamefont {J.~J.}\ \bibnamefont {Cowan}}, \bibinfo {author} {\bibfnamefont {C.}~\bibnamefont {Sneden}}, \bibinfo {author} {\bibfnamefont {J.~E.}\ \bibnamefont {Lawler}}, \bibinfo {author} {\bibfnamefont {A.}~\bibnamefont {Aprahamian}}, \bibinfo {author} {\bibfnamefont {M.}~\bibnamefont {Wiescher}}, \bibinfo {author} {\bibfnamefont {K.}~\bibnamefont {Langanke}}, \bibinfo {author} {\bibfnamefont {G.}~\bibnamefont {Mart\'{\i}nez-Pinedo}},\ and\ \bibinfo {author} {\bibfnamefont {F.-K.}\ \bibnamefont {Thielemann}},\ }\href {https://doi.org/10.1103/RevModPhys.93.015002} {\bibfield  {journal} {\bibinfo  {journal} {Rev. Mod. Phys.}\ }\textbf {\bibinfo {volume} {93}},\ \bibinfo {pages} {015002} (\bibinfo {year} {2021})}\BibitemShut {NoStop}%
\bibitem [{\citenamefont {Rosswog}\ and\ \citenamefont {Korobkin}(2024)}]{Rosswog:2022tus}%
  \BibitemOpen
  \bibfield  {author} {\bibinfo {author} {\bibfnamefont {S.}~\bibnamefont {Rosswog}}\ and\ \bibinfo {author} {\bibfnamefont {O.}~\bibnamefont {Korobkin}},\ }\href {https://doi.org/10.1002/andp.202200306} {\bibfield  {journal} {\bibinfo  {journal} {Ann. Phys.}\ }\textbf {\bibinfo {volume} {536}},\ \bibinfo {pages} {2200306} (\bibinfo {year} {2024})},\ \Eprint {https://arxiv.org/abs/2208.14026} {arXiv:2208.14026 [astro-ph.HE]} \BibitemShut {NoStop}%
\bibitem [{\citenamefont {Janka}\ and\ \citenamefont {Bauswein}(2023)}]{Janka:2022krt}%
  \BibitemOpen
  \bibfield  {author} {\bibinfo {author} {\bibfnamefont {H.-T.}\ \bibnamefont {Janka}}\ and\ \bibinfo {author} {\bibfnamefont {A.}~\bibnamefont {Bauswein}},\ }\bibinfo {title} {{Dynamics and Equation of State Dependencies of Relevance for Nucleosynthesis in Supernovae and Neutron Star Mergers}},\ in\ \href {https://doi.org/10.1007/978-981-15-8818-1_93-1} {\emph {\bibinfo {booktitle} {{Handbook of Nuclear Physics}}}},\ \bibinfo {editor} {edited by\ \bibinfo {editor} {\bibfnamefont {I.}~\bibnamefont {Tanihata}}, \bibinfo {editor} {\bibfnamefont {H.}~\bibnamefont {Toki}},\ and\ \bibinfo {editor} {\bibfnamefont {T.}~\bibnamefont {Kajino}}}\ (\bibinfo  {publisher} {Springer Nature Singapore},\ \bibinfo {year} {2023})\ pp.\ \bibinfo {pages} {1--98},\ \Eprint {https://arxiv.org/abs/2212.07498} {arXiv:2212.07498 [astro-ph.HE]} \BibitemShut {NoStop}%
\bibitem [{\citenamefont {Margutti}\ and\ \citenamefont {Chornock}(2021)}]{Raffaella2021}%
  \BibitemOpen
  \bibfield  {author} {\bibinfo {author} {\bibfnamefont {R.}~\bibnamefont {Margutti}}\ and\ \bibinfo {author} {\bibfnamefont {R.}~\bibnamefont {Chornock}},\ }\href {https://doi.org/https://doi.org/10.1146/annurev-astro-112420-030742} {\bibfield  {journal} {\bibinfo  {journal} {Annu. Rev. Astron. Astrophys.}\ }\textbf {\bibinfo {volume} {59}},\ \bibinfo {pages} {155} (\bibinfo {year} {2021})}\BibitemShut {NoStop}%
\bibitem [{\citenamefont {Burns}(2020)}]{Burns2020}%
  \BibitemOpen
  \bibfield  {author} {\bibinfo {author} {\bibfnamefont {E.}~\bibnamefont {Burns}},\ }\href {https://doi.org/10.1007/s41114-020-00028-7} {\bibfield  {journal} {\bibinfo  {journal} {Living Rev. Relativ.}\ }\textbf {\bibinfo {volume} {23}},\ \bibinfo {pages} {4} (\bibinfo {year} {2020})}\BibitemShut {NoStop}%
\bibitem [{\citenamefont {Sarin}\ and\ \citenamefont {Lasky}(2021)}]{Sarin2021}%
  \BibitemOpen
  \bibfield  {author} {\bibinfo {author} {\bibfnamefont {N.}~\bibnamefont {Sarin}}\ and\ \bibinfo {author} {\bibfnamefont {P.~D.}\ \bibnamefont {Lasky}},\ }\href {https://doi.org/10.1007/s10714-021-02831-1} {\bibfield  {journal} {\bibinfo  {journal} {Gen. Relativ. Gravit.}\ }\textbf {\bibinfo {volume} {53}},\ \bibinfo {pages} {59} (\bibinfo {year} {2021})}\BibitemShut {NoStop}%
\bibitem [{\citenamefont {Hotokezaka}\ \emph {et~al.}(2013{\natexlab{b}})\citenamefont {Hotokezaka}, \citenamefont {Kiuchi}, \citenamefont {Kyutoku}, \citenamefont {Okawa}, \citenamefont {Sekiguchi}, \citenamefont {Shibata},\ and\ \citenamefont {Taniguchi}}]{Hotokezaka2013}%
  \BibitemOpen
  \bibfield  {author} {\bibinfo {author} {\bibfnamefont {K.}~\bibnamefont {Hotokezaka}}, \bibinfo {author} {\bibfnamefont {K.}~\bibnamefont {Kiuchi}}, \bibinfo {author} {\bibfnamefont {K.}~\bibnamefont {Kyutoku}}, \bibinfo {author} {\bibfnamefont {H.}~\bibnamefont {Okawa}}, \bibinfo {author} {\bibfnamefont {Y.-i.}\ \bibnamefont {Sekiguchi}}, \bibinfo {author} {\bibfnamefont {M.}~\bibnamefont {Shibata}},\ and\ \bibinfo {author} {\bibfnamefont {K.}~\bibnamefont {Taniguchi}},\ }\href {https://doi.org/10.1103/PhysRevD.87.024001} {\bibfield  {journal} {\bibinfo  {journal} {Phys. Rev. D}\ }\textbf {\bibinfo {volume} {87}},\ \bibinfo {pages} {024001} (\bibinfo {year} {2013}{\natexlab{b}})},\ \Eprint {https://arxiv.org/abs/1212.0905} {arXiv:1212.0905 [astro-ph.HE]} \BibitemShut {NoStop}%
\bibitem [{\citenamefont {Bauswein}\ \emph {et~al.}(2013{\natexlab{a}})\citenamefont {Bauswein}, \citenamefont {Goriely},\ and\ \citenamefont {Janka}}]{Bauswein2013}%
  \BibitemOpen
  \bibfield  {author} {\bibinfo {author} {\bibfnamefont {A.}~\bibnamefont {Bauswein}}, \bibinfo {author} {\bibfnamefont {S.}~\bibnamefont {Goriely}},\ and\ \bibinfo {author} {\bibfnamefont {H.~T.}\ \bibnamefont {Janka}},\ }\href {https://doi.org/10.1088/0004-637X/773/1/78} {\bibfield  {journal} {\bibinfo  {journal} {Astrophys. J.}\ }\textbf {\bibinfo {volume} {773}},\ \bibinfo {pages} {78} (\bibinfo {year} {2013}{\natexlab{a}})},\ \Eprint {https://arxiv.org/abs/1302.6530} {arXiv:1302.6530 [astro-ph.SR]} \BibitemShut {NoStop}%
\bibitem [{\citenamefont {Bauswein}\ \emph {et~al.}(2013{\natexlab{b}})\citenamefont {Bauswein}, \citenamefont {Baumgarte},\ and\ \citenamefont {Janka}}]{PhysRevLett.111.131101}%
  \BibitemOpen
  \bibfield  {author} {\bibinfo {author} {\bibfnamefont {A.}~\bibnamefont {Bauswein}}, \bibinfo {author} {\bibfnamefont {T.~W.}\ \bibnamefont {Baumgarte}},\ and\ \bibinfo {author} {\bibfnamefont {H.-T.}\ \bibnamefont {Janka}},\ }\href {https://doi.org/10.1103/PhysRevLett.111.131101} {\bibfield  {journal} {\bibinfo  {journal} {Phys. Rev. Lett.}\ }\textbf {\bibinfo {volume} {111}},\ \bibinfo {pages} {131101} (\bibinfo {year} {2013}{\natexlab{b}})}\BibitemShut {NoStop}%
\bibitem [{\citenamefont {{Bauswein}}\ \emph {et~al.}(2017)\citenamefont {{Bauswein}}, \citenamefont {{Just}}, \citenamefont {{Janka}},\ and\ \citenamefont {{Stergioulas}}}]{Bauswein2017}%
  \BibitemOpen
  \bibfield  {author} {\bibinfo {author} {\bibfnamefont {A.}~\bibnamefont {{Bauswein}}}, \bibinfo {author} {\bibfnamefont {O.}~\bibnamefont {{Just}}}, \bibinfo {author} {\bibfnamefont {H.-T.}\ \bibnamefont {{Janka}}},\ and\ \bibinfo {author} {\bibfnamefont {N.}~\bibnamefont {{Stergioulas}}},\ }\href {https://doi.org/10.3847/2041-8213/aa9994} {\bibfield  {journal} {\bibinfo  {journal} {\apjl}\ }\textbf {\bibinfo {volume} {850}},\ \bibinfo {eid} {L34} (\bibinfo {year} {2017})},\ \Eprint {https://arxiv.org/abs/1710.06843} {arXiv:1710.06843 [astro-ph.HE]} \BibitemShut {NoStop}%
\bibitem [{\citenamefont {Bauswein}\ and\ \citenamefont {Stergioulas}(2017)}]{10.1093/mnras/stx1983}%
  \BibitemOpen
  \bibfield  {author} {\bibinfo {author} {\bibfnamefont {A.}~\bibnamefont {Bauswein}}\ and\ \bibinfo {author} {\bibfnamefont {N.}~\bibnamefont {Stergioulas}},\ }\href {https://doi.org/10.1093/mnras/stx1983} {\bibfield  {journal} {\bibinfo  {journal} {Mon. Not. Roy. Astron. Soc.}\ }\textbf {\bibinfo {volume} {471}},\ \bibinfo {pages} {4956} (\bibinfo {year} {2017})}\BibitemShut {NoStop}%
\bibitem [{\citenamefont {Köppel}\ \emph {et~al.}(2019)\citenamefont {Köppel}, \citenamefont {Bovard},\ and\ \citenamefont {Rezzolla}}]{Köppel_2019}%
  \BibitemOpen
  \bibfield  {author} {\bibinfo {author} {\bibfnamefont {S.}~\bibnamefont {Köppel}}, \bibinfo {author} {\bibfnamefont {L.}~\bibnamefont {Bovard}},\ and\ \bibinfo {author} {\bibfnamefont {L.}~\bibnamefont {Rezzolla}},\ }\href {https://doi.org/10.3847/2041-8213/ab0210} {\bibfield  {journal} {\bibinfo  {journal} {Astrophys. J. Lett.}\ }\textbf {\bibinfo {volume} {872}},\ \bibinfo {pages} {L16} (\bibinfo {year} {2019})}\BibitemShut {NoStop}%
\bibitem [{\citenamefont {Bauswein}\ \emph {et~al.}(2021)\citenamefont {Bauswein}, \citenamefont {Blacker}, \citenamefont {Lioutas}, \citenamefont {Soultanis}, \citenamefont {Vijayan},\ and\ \citenamefont {Stergioulas}}]{PhysRevD.103.123004}%
  \BibitemOpen
  \bibfield  {author} {\bibinfo {author} {\bibfnamefont {A.}~\bibnamefont {Bauswein}}, \bibinfo {author} {\bibfnamefont {S.}~\bibnamefont {Blacker}}, \bibinfo {author} {\bibfnamefont {G.}~\bibnamefont {Lioutas}}, \bibinfo {author} {\bibfnamefont {T.}~\bibnamefont {Soultanis}}, \bibinfo {author} {\bibfnamefont {V.}~\bibnamefont {Vijayan}},\ and\ \bibinfo {author} {\bibfnamefont {N.}~\bibnamefont {Stergioulas}},\ }\href {https://doi.org/10.1103/PhysRevD.103.123004} {\bibfield  {journal} {\bibinfo  {journal} {Phys. Rev. D}\ }\textbf {\bibinfo {volume} {103}},\ \bibinfo {pages} {123004} (\bibinfo {year} {2021})}\BibitemShut {NoStop}%
\bibitem [{\citenamefont {{Agathos}}\ \emph {et~al.}(2020)\citenamefont {{Agathos}}, \citenamefont {{Zappa}}, \citenamefont {{Bernuzzi}}, \citenamefont {{Perego}}, \citenamefont {{Breschi}},\ and\ \citenamefont {{Radice}}}]{Agathos2020}%
  \BibitemOpen
  \bibfield  {author} {\bibinfo {author} {\bibfnamefont {M.}~\bibnamefont {{Agathos}}}, \bibinfo {author} {\bibfnamefont {F.}~\bibnamefont {{Zappa}}}, \bibinfo {author} {\bibfnamefont {S.}~\bibnamefont {{Bernuzzi}}}, \bibinfo {author} {\bibfnamefont {A.}~\bibnamefont {{Perego}}}, \bibinfo {author} {\bibfnamefont {M.}~\bibnamefont {{Breschi}}},\ and\ \bibinfo {author} {\bibfnamefont {D.}~\bibnamefont {{Radice}}},\ }\href {https://doi.org/10.1103/PhysRevD.101.044006} {\bibfield  {journal} {\bibinfo  {journal} {\prd}\ }\textbf {\bibinfo {volume} {101}},\ \bibinfo {eid} {044006} (\bibinfo {year} {2020})},\ \Eprint {https://arxiv.org/abs/1908.05442} {arXiv:1908.05442 [gr-qc]} \BibitemShut {NoStop}%
\bibitem [{\citenamefont {Tootle}\ \emph {et~al.}(2021)\citenamefont {Tootle}, \citenamefont {Papenfort}, \citenamefont {Most},\ and\ \citenamefont {Rezzolla}}]{Tootle_2021}%
  \BibitemOpen
  \bibfield  {author} {\bibinfo {author} {\bibfnamefont {S.~D.}\ \bibnamefont {Tootle}}, \bibinfo {author} {\bibfnamefont {L.~J.}\ \bibnamefont {Papenfort}}, \bibinfo {author} {\bibfnamefont {E.~R.}\ \bibnamefont {Most}},\ and\ \bibinfo {author} {\bibfnamefont {L.}~\bibnamefont {Rezzolla}},\ }\href {https://doi.org/10.3847/2041-8213/ac350d} {\bibfield  {journal} {\bibinfo  {journal} {Astrophys. J. Lett.}\ }\textbf {\bibinfo {volume} {922}},\ \bibinfo {pages} {L19} (\bibinfo {year} {2021})}\BibitemShut {NoStop}%
\bibitem [{\citenamefont {{K{\"o}lsch}}\ \emph {et~al.}(2022)\citenamefont {{K{\"o}lsch}}, \citenamefont {{Dietrich}}, \citenamefont {{Ujevic}},\ and\ \citenamefont {{Br{\"u}gmann}}}]{Koelsch2022}%
  \BibitemOpen
  \bibfield  {author} {\bibinfo {author} {\bibfnamefont {M.}~\bibnamefont {{K{\"o}lsch}}}, \bibinfo {author} {\bibfnamefont {T.}~\bibnamefont {{Dietrich}}}, \bibinfo {author} {\bibfnamefont {M.}~\bibnamefont {{Ujevic}}},\ and\ \bibinfo {author} {\bibfnamefont {B.}~\bibnamefont {{Br{\"u}gmann}}},\ }\href {https://doi.org/10.1103/PhysRevD.106.044026} {\bibfield  {journal} {\bibinfo  {journal} {\prd}\ }\textbf {\bibinfo {volume} {106}},\ \bibinfo {eid} {044026} (\bibinfo {year} {2022})},\ \Eprint {https://arxiv.org/abs/2112.11851} {arXiv:2112.11851 [gr-qc]} \BibitemShut {NoStop}%
\bibitem [{\citenamefont {{Kashyap}}\ \emph {et~al.}(2022)\citenamefont {{Kashyap}}, \citenamefont {{Das}}, \citenamefont {{Radice}}, \citenamefont {{Padamata}}, \citenamefont {{Prakash}}, \citenamefont {{Logoteta}}, \citenamefont {{Perego}}, \citenamefont {{Godzieba}}, \citenamefont {{Bernuzzi}}, \citenamefont {{Bombaci}}, \citenamefont {{Fattoyev}}, \citenamefont {{Reed}},\ and\ \citenamefont {{Schneider}}}]{Kashyap2022}%
  \BibitemOpen
  \bibfield  {author} {\bibinfo {author} {\bibfnamefont {R.}~\bibnamefont {{Kashyap}}}, \bibinfo {author} {\bibfnamefont {A.}~\bibnamefont {{Das}}}, \bibinfo {author} {\bibfnamefont {D.}~\bibnamefont {{Radice}}}, \bibinfo {author} {\bibfnamefont {S.}~\bibnamefont {{Padamata}}}, \bibinfo {author} {\bibfnamefont {A.}~\bibnamefont {{Prakash}}}, \bibinfo {author} {\bibfnamefont {D.}~\bibnamefont {{Logoteta}}}, \bibinfo {author} {\bibfnamefont {A.}~\bibnamefont {{Perego}}}, \bibinfo {author} {\bibfnamefont {D.~A.}\ \bibnamefont {{Godzieba}}}, \bibinfo {author} {\bibfnamefont {S.}~\bibnamefont {{Bernuzzi}}}, \bibinfo {author} {\bibfnamefont {I.}~\bibnamefont {{Bombaci}}}, \bibinfo {author} {\bibfnamefont {F.~J.}\ \bibnamefont {{Fattoyev}}}, \bibinfo {author} {\bibfnamefont {B.~T.}\ \bibnamefont {{Reed}}},\ and\ \bibinfo {author} {\bibfnamefont {A.~d.~S.}\ \bibnamefont {{Schneider}}},\ }\href {https://doi.org/10.1103/PhysRevD.105.103022} {\bibfield  {journal} {\bibinfo  {journal} {\prd}\ }\textbf {\bibinfo {volume}
  {105}},\ \bibinfo {eid} {103022} (\bibinfo {year} {2022})},\ \Eprint {https://arxiv.org/abs/2111.05183} {arXiv:2111.05183 [astro-ph.HE]} \BibitemShut {NoStop}%
\bibitem [{\citenamefont {{Sneppen}}\ \emph {et~al.}(2024)\citenamefont {{Sneppen}}, \citenamefont {{Just}}, \citenamefont {{Bauswein}}, \citenamefont {{Damgaard}}, \citenamefont {{Watson}}, \citenamefont {{Shingles}}, \citenamefont {{Collins}}, \citenamefont {{Sim}}, \citenamefont {{Xiong}}, \citenamefont {{Martinez-Pinedo}}, \citenamefont {{Soultanis}},\ and\ \citenamefont {{Vijayan}}}]{Sneppen2024}%
  \BibitemOpen
  \bibfield  {author} {\bibinfo {author} {\bibfnamefont {A.}~\bibnamefont {{Sneppen}}}, \bibinfo {author} {\bibfnamefont {O.}~\bibnamefont {{Just}}}, \bibinfo {author} {\bibfnamefont {A.}~\bibnamefont {{Bauswein}}}, \bibinfo {author} {\bibfnamefont {R.}~\bibnamefont {{Damgaard}}}, \bibinfo {author} {\bibfnamefont {D.}~\bibnamefont {{Watson}}}, \bibinfo {author} {\bibfnamefont {L.~J.}\ \bibnamefont {{Shingles}}}, \bibinfo {author} {\bibfnamefont {C.~E.}\ \bibnamefont {{Collins}}}, \bibinfo {author} {\bibfnamefont {S.~A.}\ \bibnamefont {{Sim}}}, \bibinfo {author} {\bibfnamefont {Z.}~\bibnamefont {{Xiong}}}, \bibinfo {author} {\bibfnamefont {G.}~\bibnamefont {{Martinez-Pinedo}}}, \bibinfo {author} {\bibfnamefont {T.}~\bibnamefont {{Soultanis}}},\ and\ \bibinfo {author} {\bibfnamefont {V.}~\bibnamefont {{Vijayan}}},\ }\href@noop {} {} (\bibinfo {year} {2024}),\ \Eprint {https://arxiv.org/abs/2411.03427} {arXiv:2411.03427 [astro-ph.HE]} \BibitemShut {NoStop}%
\bibitem [{\citenamefont {Blacker}\ and\ \citenamefont {Bauswein}(2024)}]{Blacker:2024tet}%
  \BibitemOpen
  \bibfield  {author} {\bibinfo {author} {\bibfnamefont {S.}~\bibnamefont {Blacker}}\ and\ \bibinfo {author} {\bibfnamefont {A.}~\bibnamefont {Bauswein}},\ }\href@noop {} {} (\bibinfo {year} {2024}),\ \Eprint {https://arxiv.org/abs/2406.14669} {arXiv:2406.14669 [astro-ph.HE]} \BibitemShut {NoStop}%
\bibitem [{\citenamefont {{Tsokaros}}\ \emph {et~al.}(2024)\citenamefont {{Tsokaros}}, \citenamefont {{Bamber}}, \citenamefont {{Ruiz}},\ and\ \citenamefont {{Shapiro}}}]{2024arXiv241100939T}%
  \BibitemOpen
  \bibfield  {author} {\bibinfo {author} {\bibfnamefont {A.}~\bibnamefont {{Tsokaros}}}, \bibinfo {author} {\bibfnamefont {J.}~\bibnamefont {{Bamber}}}, \bibinfo {author} {\bibfnamefont {M.}~\bibnamefont {{Ruiz}}},\ and\ \bibinfo {author} {\bibfnamefont {S.~L.}\ \bibnamefont {{Shapiro}}},\ }\href {https://doi.org/10.48550/arXiv.2411.00939} {} (\bibinfo {year} {2024}),\ \Eprint {https://arxiv.org/abs/arXiv:2411.00939} {arXiv:2411.00939 [gr-qc]} \BibitemShut {NoStop}%
\bibitem [{\citenamefont {Keller}\ \emph {et~al.}(2021)\citenamefont {Keller}, \citenamefont {Wellenhofer}, \citenamefont {Hebeler},\ and\ \citenamefont {Schwenk}}]{PhysRevC.103.055806}%
  \BibitemOpen
  \bibfield  {author} {\bibinfo {author} {\bibfnamefont {J.}~\bibnamefont {Keller}}, \bibinfo {author} {\bibfnamefont {C.}~\bibnamefont {Wellenhofer}}, \bibinfo {author} {\bibfnamefont {K.}~\bibnamefont {Hebeler}},\ and\ \bibinfo {author} {\bibfnamefont {A.}~\bibnamefont {Schwenk}},\ }\href {https://doi.org/10.1103/PhysRevC.103.055806} {\bibfield  {journal} {\bibinfo  {journal} {Phys. Rev. C}\ }\textbf {\bibinfo {volume} {103}},\ \bibinfo {pages} {055806} (\bibinfo {year} {2021})}\BibitemShut {NoStop}%
\bibitem [{\citenamefont {{Keller}}\ \emph {et~al.}(2023)\citenamefont {{Keller}}, \citenamefont {{Hebeler}},\ and\ \citenamefont {{Schwenk}}}]{Keller2023}%
  \BibitemOpen
  \bibfield  {author} {\bibinfo {author} {\bibfnamefont {J.}~\bibnamefont {{Keller}}}, \bibinfo {author} {\bibfnamefont {K.}~\bibnamefont {{Hebeler}}},\ and\ \bibinfo {author} {\bibfnamefont {A.}~\bibnamefont {{Schwenk}}},\ }\href {https://doi.org/10.1103/PhysRevLett.130.072701} {\bibfield  {journal} {\bibinfo  {journal} {\prl}\ }\textbf {\bibinfo {volume} {130}},\ \bibinfo {eid} {072701} (\bibinfo {year} {2023})},\ \Eprint {https://arxiv.org/abs/2204.14016} {arXiv:2204.14016 [nucl-th]} \BibitemShut {NoStop}%
\end{thebibliography}%
